\documentclass[fleqn,usenatbib,useAMS]{mnras}

\usepackage{graphicx}	% Including figure files
\usepackage{amsmath}	% Advanced maths commands
\usepackage{amssymb}	% Extra maths symbols
\usepackage{multicol}   % Multi-column entries in tables
\usepackage{bm}		    % Bold maths symbols, including upright Greek
\usepackage{pdflscape}	% Landscape pages
\usepackage{enumitem}	% Including figure files
\usepackage{multirow}   % Allow multi-row cells in tables
\usepackage{xspace}
\usepackage{CJK}
\usepackage{longtable}
\usepackage{xcolor}
\usepackage{etoolbox}
\newcommand{\Teff}{$T_\mathrm{eff}$\xspace}
\newcommand{\logg}{$\log g$\xspace}
\newcommand{\feh}{$\mathrm{[Fe/H]}$\xspace}
\newcommand{\fehatmo}{$\mathrm{[Fe/H]}_\text{atmo}$\xspace}
\newcommand{\alphafe}{$\mathrm{[\upalpha/Fe]}$\xspace}
\newcommand{\vmic}{$v_\mathrm{mic}$\xspace}
\newcommand{\lbol}{$L_\mathrm{bol}$\xspace}
\newcommand{\vmac}{$v_\mathrm{mac}$\xspace}
\newcommand{\vsini}{$v \sin i$\xspace}
\newcommand{\vbroad}{$v_\mathrm{broad}$\xspace}
\newcommand{\vrad}{$v_\mathrm{rad}$\xspace}
\newcommand{\numax}{$\nu_\mathrm{max}$\xspace}
\newcommand{\Gaia}{\textit{Gaia}\xspace}
\newcommand{\Hipparcos}{{\sc{Hipparcos}}\xspace}
\newcommand{\tess}{TESS}
\newcommand{\Ktwo}{K2}

\definecolor{C0}{rgb}{0.12156862745098039, 0.4666666666666667, 0.7058823529411765}
\definecolor{C1}{rgb}{1.0 0.4980392156862745 0.054901960784313725}
\definecolor{C2}{rgb}{0.17254901960784313 0.6274509803921569 0.17254901960784313}
\definecolor{C3}{rgb}{0.8392156862745098 0.15294117647058825 0.1568627450980392}
\definecolor{C4}{rgb}{0.5803921568627451 0.403921568627451 0.7411764705882353}

\usepackage[T1]{fontenc}
\usepackage{ae,aecompl}
\usepackage{newtxtext,newtxmath}

\begin{document}
\begin{CJK*}{UTF8}{gbsn}
\label{firstpage}
\pagerange{\pageref{firstpage}--\pageref{lastpage}}

%%%%%%%%%%%%%%%%%%% TITLE PAGE %%%%%%%%%%%%%%%%%%%

\title{The GALAH+ Survey: Third Data Release}

% The list of authors, and the short list which is used in the headers.
% If you need two or more lines of authors, add an extra line using \newauthor
\author[Buder et al.]{Sven Buder$^{1,2,3}$\thanks{Contact e-mail: \href{mailto:sven.buder@anu.edu.au}{sven.buder@anu.edu.au}},
Sanjib~Sharma$^{4,2}$, 
Janez~Kos$^{5}$,
Anish~M.~Amarsi$^{6}$,\newauthor
Thomas~Nordlander$^{1,2}$, 
Karin~Lind$^{7,3}$,
Sarah~L.~Martell$^{8,2}$, 
Martin~Asplund$^{9}$,\newauthor
Joss~Bland-Hawthorn$^{4,2}$,
Andrew~R.~Casey$^{10,11}$, 
Gayandhi~M.~De~Silva$^{12,13}$, \newauthor
Valentina~{D'Orazi}$^{14}$,
Ken~C.~Freeman$^{1,2}$,  
Michael~R.~Hayden$^{4,2}$,
Geraint~F.~Lewis$^{4}$,\newauthor
Jane~Lin$^{1,2}$, 
Katharine.~J.~Schlesinger$^{1}$,
Jeffrey~D.~Simpson$^{8,2}$, 
Dennis~Stello$^{8,15,4}$,\newauthor
Daniel~B.~Zucker$^{13,16}$,
Toma\v{z}~Zwitter$^{5}$,
Kevin~L.~Beeson$^{5}$,
Tobias~Buck$^{17}$,\newauthor
Luca~Casagrande$^{1,2}$,
Jake~T.~Clark$^{18}$,
Klemen~{\v C}otar$^{5}$,
Gary~S.~Da~Costa$^{1,2}$,\newauthor
Richard~de~Grijs$^{13,16,19}$,
Diane~Feuillet$^{20,3}$,
Jonathan~Horner$^{18}$,
Prajwal~R.~Kafle$^{21}$,\newauthor
Shourya~Khanna$^{22}$,
Chiaki~Kobayashi$^{23,2}$,
Fan~Liu$^{24}$,
Benjamin~T.~Montet$^{8}$,\newauthor
Govind~Nandakumar$^{1,2}$,
David~M.~Nataf$^{25}$,
Melissa~K.~Ness$^{26,27}$,
Lorenzo~Spina$^{11,2,28}$,\newauthor
Thor Tepper-Garc\'{i}a$^{4,2,29}$,
Yuan-Sen~Ting (丁源森)$^{30,31,32,1}$,
Gregor~Traven$^{20}$,\newauthor
Rok~Vogrin{\v c}i{\v c}$^{5}$,
Robert~A.~Wittenmyer$^{18}$,
Rosemary~F.~G.~Wyse$^{25}$,
Maru{\v s}a {\v Z}erjal$^{1}$,\newauthor
and~the~GALAH~collaboration
\\
\\
(Affiliations listed after the references)}

% These dates will be filled out by the publisher
\date{Accepted 2021 April 27. Received 2021 April 26; in original form 2020 November 05}

% Enter the current year, for the copyright statements etc.
\pubyear{2021}

% Don't change these lines
\maketitle
\end{CJK*}

\begin{abstract}
The ensemble of chemical element abundance measurements for stars, along with precision distances and orbit properties, provides high-dimensional data to study the evolution of the Milky Way. With this third data release of the Galactic Archaeology with HERMES (GALAH) survey, we publish 678\,423 spectra for 588\,571 mostly nearby stars (81.2\% of stars are within $<$ 2 kpc), observed with the HERMES spectrograph at the Anglo-Australian Telescope. This release (hereafter GALAH+ DR3) includes all observations from GALAH Phase 1 (bright, main, and faint survey, 70\%), K2-HERMES (17\%), TESS-HERMES (5\%), and a subset of ancillary observations (8\%) including the bulge and $>$ 75 stellar clusters. We derive stellar parameters \Teff, \logg, \feh, \vmic, \vbroad, and \vrad using our modified version of the spectrum synthesis code Spectroscopy Made Easy ({\sc sme}) and 1D {\sc marcs} model atmospheres. We break spectroscopic degeneracies in our spectrum analysis with astrometry from \Gaia~DR2 and photometry from 2MASS. We report abundance ratios [X/Fe] for 30 different elements (11 of which are based on non-LTE computations) covering five nucleosynthetic pathways. We describe validations for accuracy and precision, flagging of peculiar stars/measurements and recommendations for using our results. Our catalogue comprises 65\% dwarfs, 34\% giants, and 1\% other/unclassified stars. Based on unflagged chemical composition and age, we find 62\% young low-$\upalpha$, 9\% young high-$\upalpha$, 27\% old high-$\upalpha$, and 2\% stars with $\mathrm{[Fe/H]} \leq -1$. Based on kinematics, 4\% are halo stars. Several Value-Added-Catalogues, including stellar ages and dynamics, updated after \Gaia eDR3, accompany this release and allow chrono-chemodynamic analyses, as we showcase.
\end{abstract}

\begin{keywords}
Surveys -- 
the Galaxy --
methods: observational --
methods: data analysis --
stars: fundamental parameters -- 
stars: abundances
\end{keywords}

%%%%%%%%%%%%%%%%%%%%%%%%%%%%%%%%%%%%%%%%%%%%%%%%%%

%%%%%%%%%%%%%%%%% BODY OF PAPER %%%%%%%%%%%%%%%%%%

%________________________________________________________________
\section{Introduction} \label{sec:introduction}

% This figure was created with the notebook GALAH_DR3/input/galah_dr3_input.ipynb
\begin{figure*}
\centering
\includegraphics[width=0.95\textwidth]{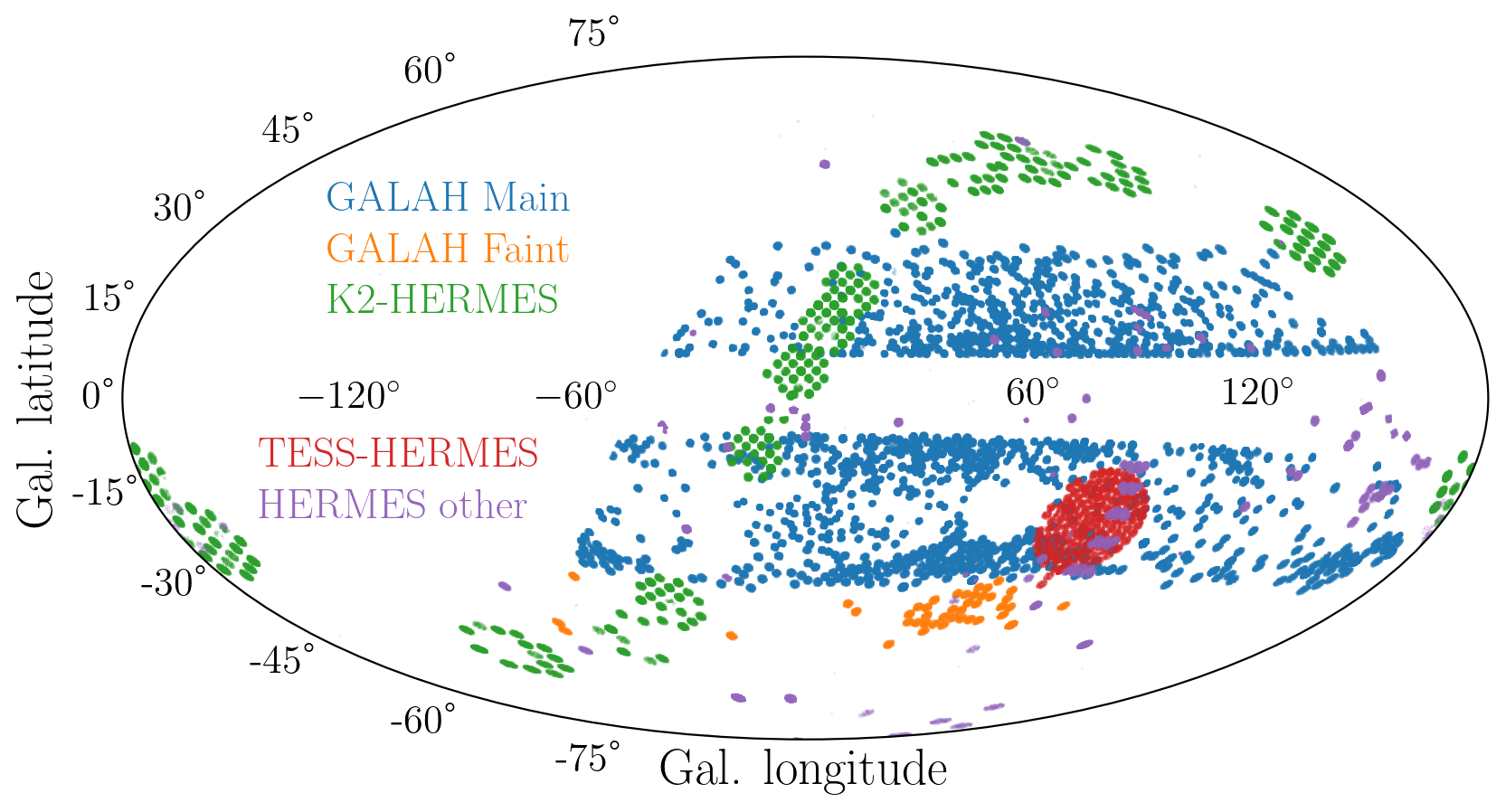}
\caption{\textbf{Overview of the distribution of stars included in this data release in Galactic coordinates with the centre of the Galaxy at the origin.} Shown are the GALAH main (blue) and faint (orange) targets, which avoid the Galactic plane. The targets of the \Ktwo-HERMES follow-up (green) fall within with the \Ktwo\ campaigns along the ecliptic and show the characteristic tile-pattern of the \textit{Kepler} telescope. The \tess-HERMES observations (red) are focused on the \tess\ Southern Continuous Viewing Zone. Other HERMES targets (purple) are distributed across the sky and were observed during independent programs.}
\label{fig:lb_overview_colored}
\end{figure*}

During the history of the Milky Way, the abundances of the different elements that make up the Galaxy's stars and planets have continually changed, as a result of the processing of the interstellar medium by successive generations of stars. As a result, the study of the elemental abundances in stars provides a direct record of the galaxy's history of star formation and evolution - a fact that has, in recent years, given birth to the science of Galactic Archaeology. 

Until recently, however, observational limitations meant that the data available to answer the questions of how the Milky Way formed and evolved was restricted to a few hundred or thousand stars with high-quality element abundances in our Solar neighbourhood \citep[see e.g.][]{Edvardsson1993, Nissen2010, Bensby2014}. 
In the last decade, advances in multi-object observations made by spacecraft (such as \Gaia) and ground-based facilities have brought about a revolution in the field of Galactic Archaeology. Where once the field was forced to focus on single-star population studies, it is now possible to carry out surveys that allow large-scale structural analyses.

Due to the intrinsic difficulty in determining the distances of stars, studies of the chemodynamical evolution of our Milky Way were previously restricted to nearby stars which were mapped by the \Hipparcos satellite \citep{ESA1997, Perryman1997, vanLeeuwen2007}. In the era of the \Gaia satellite \citep{Gaia-Collaboration2016, Brown2016, Brown2018}, we can now use astrometric and photometric observables and their physical relations with spectroscopic quantities to improve the analysis of spectra and thus the estimation of element abundances.

The connections between the chemical compositions and dynamics of stars across the vast populations in our Galaxy are a topic of significant ongoing research. Although we speak of the Milky Way in terms of the thin and thick disc \citep{Yoshii1982, Gilmore1983}, the bulge \citep{Barbuy2018}, and the stellar halo \citep{Helmi2020} as its main components \citep{BlandHawthorn_Gerhard2016}, we understand that the Galaxy is more than a superposition of independent populations. With the data now at hand, we can analyse the Galaxy from a chemodynamical perspective, and use stars of different ages as time capsules to trace back the formation history of our Galaxy \citep[see e.g.][]{Rix2013, BlandHawthorn2019}. As one example, the most recent data release from \Gaia\ has enabled significant leaps in our understanding of the enigmatic Galactic halo \citep[for an overview see e.g.][]{Helmi2020}. 6-d phase space information from \Gaia\ has revealed a large population of stars in the Solar neighbourhood that stand out against the smooth halo background as a coherent dynamical structure, pointing to a significant accretion event that is currently referred to as ``\Gaia-Enceladus-Sausage''  (GES) a combination of ``\Gaia-Enceladus'' \citep{Helmi2018} and ``\Gaia\ Sausage'' \citep{Belokurov2018}. Additionally, while we would expect the chemical composition of stars to be correlated with their ages and formation sites \citep[see e.g.][]{Minchev2017}, observations can now clearly demonstrate these connections \citep[see e.g.][]{Feuillet2018, Buder2019}, and can also demonstrate that stars within our Solar neighbourhood have experienced significant radial migration through their lifetimes \citep[see e.g.][]{Frankel2018, Hayden2020}.

Despite these significant advances, the full detail of our Galaxy's formation and history still elude us. Many of the pieces that make up that puzzle are presently missing, or remain contentious. As a result, a number of questions still remain to be answered. These include the discrete merger history of our Milky Way, the \mbox{(non-)existence} of an \textit{in situ} halo and the reason for the sharp transition from formation of stars with high $\upalpha$-element abundances in what has historically been called the ``thick disc'' to younger stars with Solar-like $\upalpha$-element abundances in the ``thin disc''. We have learnt a great deal about contributions of supernovae to element abundances, starting from the foundational work by \citet{B2FH1957}, and how we can use diagrams displaying element abundances, e.g. in \feh vs. \alphafe diagrams, as diagnostic tools of stellar and Galactic evolution. These advances in our understanding are largely thanks to the pioneering and seminal studies by \citet{Tinsley1979,Tinsley1980} and others, building on the trail-blazing observational achievements of \citet{Wallerstein1962} and others. To honour the fundamental contributions by Beatrice M. Tinsley and George Wallerstein, connecting the contributions of SNIa and SNII with element abundances in the \feh vs. \alphafe diagrams, we will hereafter refer to these as the ``Tinsley-Wallerstein diagrams''.

Previous and ongoing spectroscopic surveys by collaborations like RAVE \citep{Steinmetz2020a, Steinmetz2020b}, \Gaia-ESO \citep{Gilmore2012}, SDSS-IV APOGEE \citep{SDSSDR16}, and LAMOST \citep{Cui2012, Xiang2019} have certainly shed light on several of these outstanding questions. Answering them completely, requires more and/or better data to map out the correlations between stellar ages, abundances, and dynamics. Upcoming surveys like SDSS-V \citep{Kollmeier2017}, WEAVE \citep{WEAVE2018}, 4MOST \citep{4MOST2019}, and PFS \citep{Takada2014} will certainly continue to broaden our capabilities and understanding surrounding our galaxy's physical and chemical evolution. The data currently at hand, derived from spectroscopy, photometry, astrometry, and asteroseismology, provide high-dimensional information, and we must develop methods to extract the most accurate and precise information from them \citep[for reviews on this see e.g.][]{Nissen2018,Jofre2019}.

The recent growth in the quantity of available spectroscopic stellar data has delivered a new technique to galactic archaeologists - namely ``Chemical Tagging'', which allows the identification of stars that formed together using their chemical composition and an understanding of the astrophysics driving the dimensionality of chemical space. This technique is proving a vital tool, enabling us to observationally isolate and characterise the building blocks of our Galaxy. As a result, it remains a major science driver for the GALactic Archaeology with HERMES\footnote{High Efficiency and Resolution Multi-Element Spectrograph} (GALAH) collaboration\footnote{\url{https://www.galah-survey.org}} \citep{DeSilva2015}. With the large variety of nucleosynthetic channels that can enrich the birth material of stars \citep[see e.g.][]{Kobayashi2020}, the hypothesis is that we should be able to disentangle stars with different enrichment patterns, provided we observe enough elements with different enrichment origins. The success of some chemical tagging experiments \citep[see e.g.][]{Kos2018, PriceJones2020} is challenged by the broad similarities in chemical abundance in populations like the low-$\upalpha$ disc \citep[see e.g.][]{Ness2018}, and by the small but real inhomogeneities even within star clusters \citep{Liu2016,Liu2016b}. To put detailed chemical tagging into action, we will need a massive dataset \citep[see e.g.][]{Ting2016} consisting of measurements made with outstanding precision \citep{Ting2021}.

The publication of the previous second data release of the GALAH survey \citep{Buder2018}, entirely based on observations as part of GALAH Phase 1 with the HERMES spectrograph at the Anglo-Australian Telescope, has provided for the community abundance measurements of 23 elements based on 342\,682 spectra. Observations for this phase have continued and we are able to publish all 476\,863 spectra for 443\,843 stars of the now finished Phase 1 observations as part of this third data release. In parallel, spectroscopic follow-up observations of K2 and TESS targets have been performed with HERMES and we are able to also include these observations (112\,943 spectra for 99\,152 K2-HERMES stars and 34\,263 spectra for 26\,249 TESS-HERMES stars) in our release. We further include ancillary observations of 54\,354 spectra for 28\,205 stars in fields towards the bulge and more than 75 stellar clusters. Given the significant contribution from these programs to this GALAH release, we will hereafter refer to the release as GALAH+ DR3.

For the previous (second) data release of the GALAH survey \citep{Buder2018}, we made use of the data-driven tool \textit{The Cannon} \citep{Ness2015} to improve both the speed and the precision of the spectroscopic analysis. This was performed almost entirely without non-spectroscopic information for individual stars, using a ``training set'' of stars with careful by-hand analysis.
Although the data-driven approaches were successful for the majority of GALAH~DR2 stars, we know that these approaches can suffer from signal aliasing (e.g. moving outliers closer to the main trends), can learn unphysical correlations between the input data and the output stellar labels, and that the results are not necessarily valid outside the parameter space of the training set. As part of the present study, we aim to assess how accurately and precisely the stellar parameters and abundances were estimated by the data-driven approaches. We have therefore adjusted our approach to the analysis of the whole sample and now restrict ourselves to smaller wavelength segments of the four wavelength windows observed by HERMES per star with reliable line information for spectrum synthesis instead and include even more grids for an accurate computation of line strengths when the conditions depart from local thermodynamic equilibrium (LTE; e.g. \citealt{Mihalas1973,Asplund2005,Amarsi2020}).

The publication of \Gaia~DR2 \citep{Brown2018, Lindegren2018} provided phase space information up to 6 dimensions (coordinates, proper motions, parallaxes, and sometimes also radial velocities) for 1.3 billion stars, and having this information available for essentially all (99\%) stars in GALAH has allowed us to make major improvements to our stellar analysis. By combining our knowledge of the (absolute) photometry and spectroscopy of stars, we can break several of the degeneracies in our stand-alone spectroscopic analyses, which arise due to the fact that absorption lines do not always change to a detectable level as a function of stellar atmospheric parameters. The data analysis process for this third data release from the GALAH collaboration makes use of fundamental correlations, and this quantifiably improves the accuracy and precision of our measurements.

As large Galactic Archaeology-focused surveys continue to collect data (like GALAH in its ongoing Phase 2), the overlap between them increases. This enables us to compare results when analysing stars in the overlap, which have the same stellar labels (stellar parameters, abundances or other non-spectroscopic stellar information), and it also allows us to propagate labels from one survey onto another \citep[see e.g.][]{Casey2017, Ho2017, Xiang2019, Wheeler2020, Nandakumar2020}. This label propagation makes it possible to combine these complementary surveys for global mapping of stellar properties and abundances, and we show an example of this in Section \ref{sec:galah_in_context}, placing GALAH+~DR3 data in context with the APOGEE and LAMOST surveys.

This paper is structured as follows: We describe our target selection, observations, and reductions in Sec.~\ref{sec:selection_observation_reduction}. While the target selection and observation of the several projects like \Ktwo-HERMES and \tess-HERMES were slightly different from the main GALAH survey, we have reduced and analysed all data (combined under the term GALAH+) in a consistent and homogeneous way. The analysis of the reduction products is described in Sec.~\ref{sec:analysis}, focusing on the description of the general workflow of the analysis group and highlighting changes with respect to the previous release (GALAH~DR2). Secs.~\ref{sec:validation_sp} and \ref{sec:validation_ab} address the validation efforts for stellar parameters and element abundances, respectively. These address the accuracy and precision of these labels as well as our algorithms to identify and flag peculiar measurements or peculiar stars. Based on experience with the data set, we stress the importance of the flags, but also how complex the flagging estimates are, with several examples of peculiar abundance patterns. We also highlight possible caveats (and possibly peculiar physical correlations) of our analysis in Sec.~\ref{sec:caveats}. We present the contents of the main catalogue of this data release in Sec.~\ref{sec:catalogues}. In this section we also present the Value-Added-Catalogues (VACs) that accompany this release, namely a VAC (see Sec.~\ref{sec:vac_gaia}) created by crossmatching our targets with \Gaia eDR3 \citep{Brown2020} and the distance estimates by \citet{BailerJones2020}, a VAC (see Sec.\ref{sec:vac_age}) with estimates (such as stellar ages and masses) from isochrone fitting, a VAC (see Sec.~\ref{sec:vac_dynamics}) with stellar kinematic and dynamic estimates, a VAC (see Sec.~\ref{sec:rv_vac}) with radial velocity estimates based on different methods, and a VAC (see Sec.~\ref{sec:vac_binaries}) on parameters of binary systems. While we made use of data from \Gaia DR2 \citep{Brown2018} for our spectroscopic analysis, we also provide a second version of each of our catalogues with new crossmatches and VACs with updated data making use of \Gaia eDR3 \citep{Brown2020}, which was published shortly after our data release and supersedes \Gaia DR2. We describe all changes of the catalogues between version 1 (based on \Gaia DR2) and version 2 (with VACs now based on \Gaia eDR3) in Sec.~\ref{sec:version2}.
We highlight the scientific potential of the data in this release in context by using the combination of dynamic information and ages together with the element abundances of the main catalogue in Sec.~\ref{sec:galah_in_context}. We focus on Galactic Archaeology on a global scale and the chemodynamical evolution of our Galaxy. Along with the main and value-added-catalogues of this release, we publish the observed optical spectra for each of the arms of HERMES on the DataCentral\footnote{\url{https://docs.datacentral.org.au/galah/}} and provide the scripts used for the analysis as well as post-processing online in an open-source repository\footnote{\url{http://github.com/svenbuder/GALAH\_DR3}}
GALAH+ DR3 was timed to allow the scientific community to directly use abundances together with the latest \Gaia eDR3 information. We have not yet incorporated the latter into our abundance analysis, but plan to do so in future data releases, as we outline in Sect.~\ref{sec:conclusions}. In this section, we conclude and give an outlook to future observations as part of the ongoing observations of the GALAH survey (called phase 2 with an adjusted target selection) and our next data release.

% This figure was created with the notebook GALAH_DR3/input/galah_dr3_input.ipynb
\begin{figure*}
\centering
\includegraphics[width=0.99\textwidth]{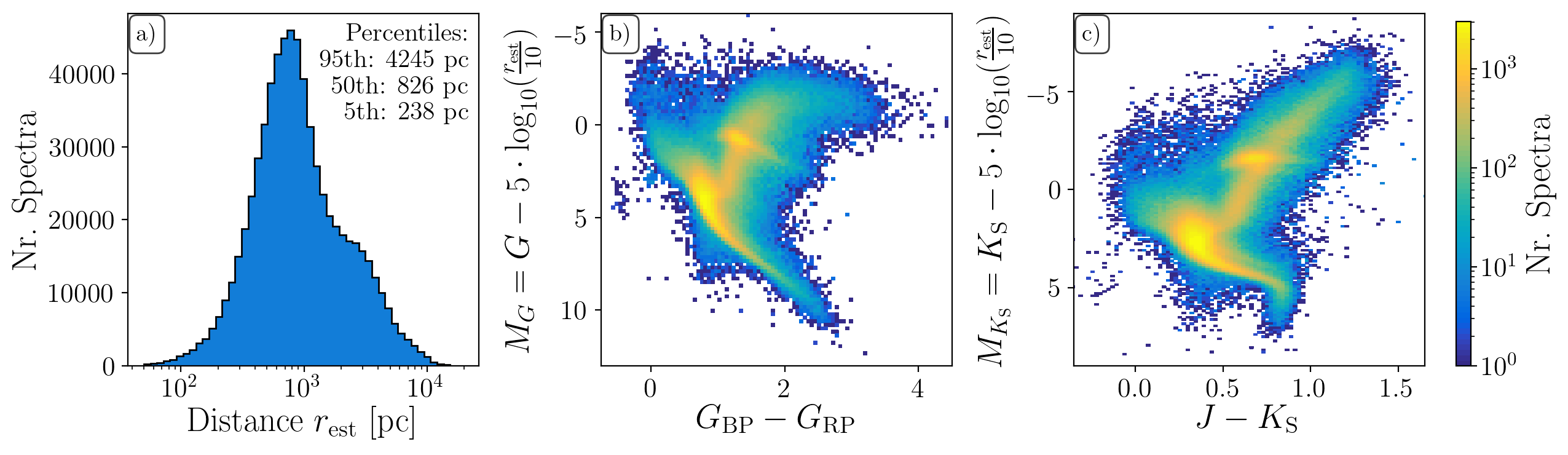}
\caption{
\textbf{Overview of distances and photometric information corresponding to the spectra (including repeats for some stars) observed as part of GALAH+~DR3 up to 25th February 2019.}
\textbf{Panel a)} shows the \citet{BailerJones2018} distances of stars in GALAH+~DR3. Due to the magnitude limited selection of stars, the majority of stars are not only dwarfs but also nearby; that is, within $1\,\mathrm{kpc}$. Only $5.8\%$ of stars are beyond $4\,\mathrm{kpc}$. 
\textbf{Panel b)} shows a colour-absolute magnitude diagram in the optical \Gaia passbands.
\textbf{Panel c)} shows an analogous diagram made with the infrared 2MASS passbands.
}
\label{fig:plot_distance_and_cmds}
\end{figure*}

%________________________________________________________________
\section{Target selection, observation, reduction} \label{sec:selection_observation_reduction}

While our previous data release \citep{Buder2018} contained only stars from the main GALAH survey, the current catalogue combines data from multiple projects with different science goals, all conducted with the HERMES spectrograph \citep{Sheinis2015} and the 2dF fibre positioning system \citep{Lewis2002} at the 3.9-metre Anglo-Australian Telescope. All the spectra have therefore been processed through the same data reduction pipeline. 
The collection into a single catalogue, which includes the K2-HERMES (S.~Sharma et al. in prep) and TESS-HERMES (S.~Sharma et al. in prep.) surveys, was chosen for ease of use. Full details of these additional surveys are presented in their corresponding data release papers and users are advised to refer to those when using data from these surveys.  The column \texttt{survey\_name} in the catalogue denotes the survey each star belongs to.
Data from four main projects, plus a number of smaller observing programs, are included in GALAH+~DR3. Fig.~\ref{fig:lb_overview_colored} shows their on-sky distribution. The majority of the stars are nearby, with a median distance of $826\,\mathrm{pc}$ (see Fig.~\ref{fig:plot_distance_and_cmds}a), and cover a large variety of stellar types and evolutionary stages, as can be seen in the colour-magnitude diagrams both with \Gaia (Fig.~\ref{fig:plot_distance_and_cmds}b) and 2MASS (Fig.~\ref{fig:plot_distance_and_cmds}c) bandpasses. Below, we describe the target selection for each of the four main projects.

\subsection{Target selection}

The GALAH input catalogue was made by combining the 2MASS \citep{Skrutskie2006} catalogue of infrared photometry with the UCAC4 \citep{Zacharias2013} proper motion catalogue. We only included stars with reliable 2MASS data, as captured in their data quality flags (Q=``A'', B=``1'', C=``0'', X=``0'', A=``0'', prox$\ge 6\arcsec$). We also rejected any star that had a nearby bright neighbour, with a rejection radius dependent on the bright star's $V$ magnitude, such that the potential target is rejected if the bright star is closer than $(130-[10\times V])$ arcseconds. The APASS photometric catalogue \citep{Henden2012} was not complete in the Southern sky at the start of GALAH observations in 2013, so we use a synthetic $V_\mathrm{JK}$ magnitude calculated from 2MASS photometry: $V_\mathrm{JK} = K+2(J-K+0.14)+0.382e^{((J-K-0.2)/0.5)}$. \citet{Sharma2018} demonstrate by using PARSEC isochrones \citep{Marigo2017} that this is a reasonable approximation for the $V$ magnitude for the types of stars observed in GALAH.

Four main projects are included in the GALAH+~DR3 catalogue (GALAH-main, GALAH-faint, \Ktwo-HERMES, and \tess-HERMES), each of which has its own selection function. We have attributed each possible pointing of the major sub surveys to a specific $\texttt{field\_id}$, as listed in Table~\ref{tab:field_ids}. The main GALAH survey takes as potential targets all stars with $12.0<V_\mathrm{JK}<14.0$, $\delta < +10^{\circ}$ and $|b|>10^{\circ}$ in regions of the sky that have at least 400 targets in $\pi$ square degrees (the 2dF field of view). We then segment this data set into 6546 ``fields'' with a fixed centre and radius between 0.7 and 1 degree. Fields containing more than 400 stars are observed multiple times with separate target lists. 
The GALAH-faint program was aimed at extending survey observations to regions with low target density. The target selection was shifted to $12<V_\mathrm{JK}<14.3$ as a way to maintain at least 400 stars per field. 
The GALAH survey also includes a few other extensions. The GALAH-bright program targets bright stars ($9.0<V_\mathrm{JK}<12.0$) to be observed in twilight or poor observing conditions. For bright stars, we use the same field centres as in regular survey observing, and require at least 200 stars per field. 
The GALAH-ultrafaint program targets very faint stars $14<V_\mathrm{JK}<16$. This was aimed at extending the survey into regions further away from the Sun. These fields were only observed under dark conditions.

The \Ktwo-HERMES survey leverages the excellent match between the two degree diameter of the 2dF fibre positioner and the five square degrees covered by each detector in the \textit{Kepler} spacecraft to create an efficiently observed spectroscopic complement for red giants in the \Ktwo\ campaign fields. The \Ktwo-HERMES program has both ``bright'' ($10<V_\mathrm{JK}<13$) and ``faint'' ($13<V_\mathrm{JK}<15$, $J-K_S>0.5$) target cohorts, to complement the asteroseismic targets that are the focus of the \Ktwo\ Galactic Archaeology Program \citep{Stello2015,Stello2017}. Analysis of asteroseismic and spectroscopic data together is key for GALAH+~DR3, and enables in-depth exploration of the structure and history of the Milky Way \citep[e.g.,][]{Sharma2016,Sharma2019}. The spectroscopic data also provide essential insights for the planet hosting stars identified in \Ktwo\ data \citep{Wittenmyer2018,Wittenmyer2020}.

The \tess-HERMES survey collected spectra for stars in the range $10.0<V_\mathrm{JK}<13.1$ in the \tess\ Southern Continuous Viewing Zone, within 12 degrees of the Southern ecliptic pole. \tess-HERMES aimed to provide accurate stellar parameters for candidate \tess\ input catalogue stars \citep{Stassun2019}, to better focus \tess\ target selection on the most promising asteroseismic targets. The results of the \tess-HERMES project are publicly available, and the project and outputs are described in \citet{Sharma2018}. 54\,354 in the ``HERMES other'' program are from targeted observations of stars in open clusters, the GALAH Pilot Survey \citep{Martell2017}, or targets from other HERMES observing that were not part of any of these surveys.

% Made with GALAH_DR3/validation/comparisons/Galactic_Archaeology_In_Context.ipynb
\begin{figure*}
\includegraphics[width=0.988\textwidth]{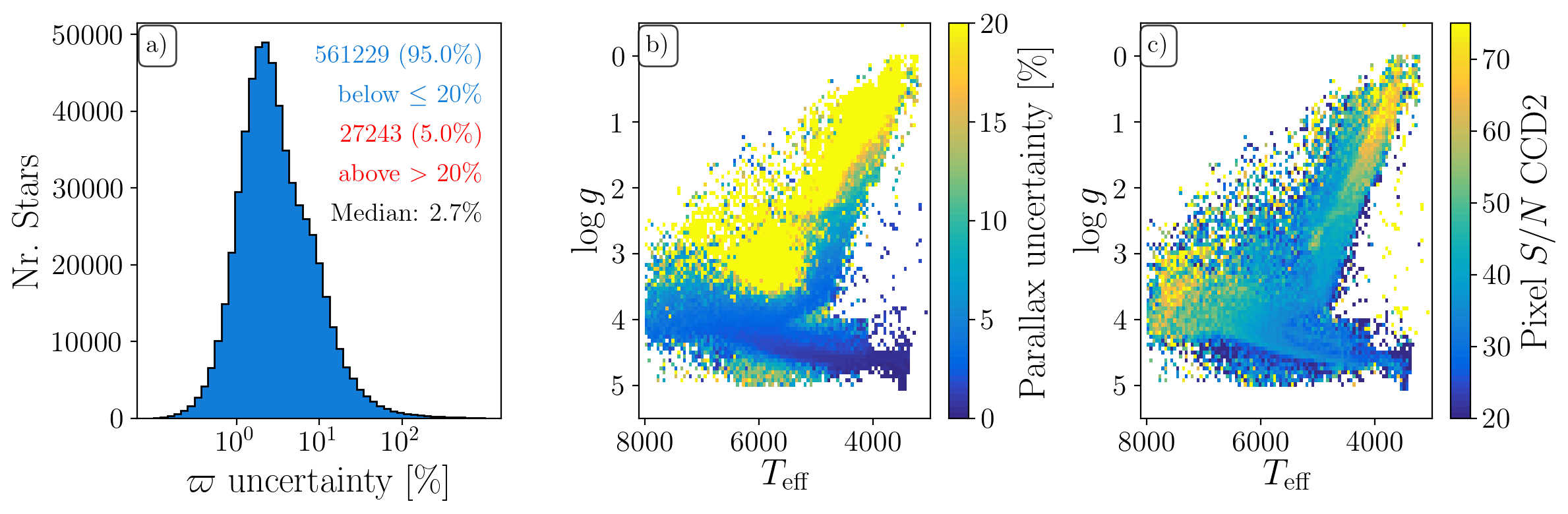}
  \caption{
\textbf{Overview and distribution of parallax uncertainty and $S/N$ for different types of stars (not spectra as in Fig.~\ref{fig:plot_distance_and_cmds}).}
\textbf{Panel a)} Parallax ($\varpi$) uncertainty provided by \Gaia~DR2. We note that the median uncertainty has decreased from 2.7\% to 1.5\% between \Gaia DR2 and \Gaia eDR3. 561\,229 (95\%) stars sit below 20\% in fractional uncertainty, and 27\,243 (5\%) stars fall above 20\%. 
\textbf{Panel b)} Distribution of \Gaia~DR2 fractional parallax uncertainty across the stellar parameters \Teff and \logg derived by GALAH+~DR3. Local cool dwarfs have the most reliable parallax information, while giants, and especially luminous giants have the worst.
\textbf{Panel c)} Distribution of $S/N$ per pixel for the green channel (CCD2) across the stellar parameters \Teff and \logg. Hot dwarfs (brighter than cool stars in the green channel) and luminous giants (brightest within the magnitude limited cohort) have the highest $S/N$ in the green channel. The $S/N$ for hot stars is typically better in the blue and green CCDs (relative to cool stars), whereas it is higher in the red and IR CCDs for the cool stars (relative to hot stars).}
  \label{fig:plx_snr_quality}
\end{figure*}

% Made with GALAH_DR3/input/galah_dr3_input.ipynb
\begin{table}
\centering
 \caption{\textbf{Field selection ($\texttt{field\_id}$) for the programs included in this data release.} Note the gaps between different TESS-HERMES fields are caused by other HERMES programs in between them.}
\label{tab:field_ids}
\begin{tabular}{ccrc}
\hline \hline
Program & $\texttt{field\_id}$  & Nr. Spectra & $\texttt{survey\_name}$\\
\hline
\textcolor{C0}{GALAH Main} & 0...6545 & 462045\\
\textcolor{C1}{GALAH Faint} & 6831...7116 & 14818\\
\textcolor{C2}{K2-HERMES} & 6546...6830 & 112943\\
\textcolor{C3}{TESS-HERMES} & 7117...7338 & 34263\\
 & 7358...7365 & \\
 & 7426...7431 & \\
\textcolor{C4}{HERMES other} & other & 54354 & \texttt{other}\\
\hline
Total &  & 678423\\

  \hline
 \end{tabular}
\end{table}

% Made with GALAH_DR3/input/galah_dr3_input.ipynb
\begin{figure*}
\centering
\includegraphics[width=0.988\textwidth]{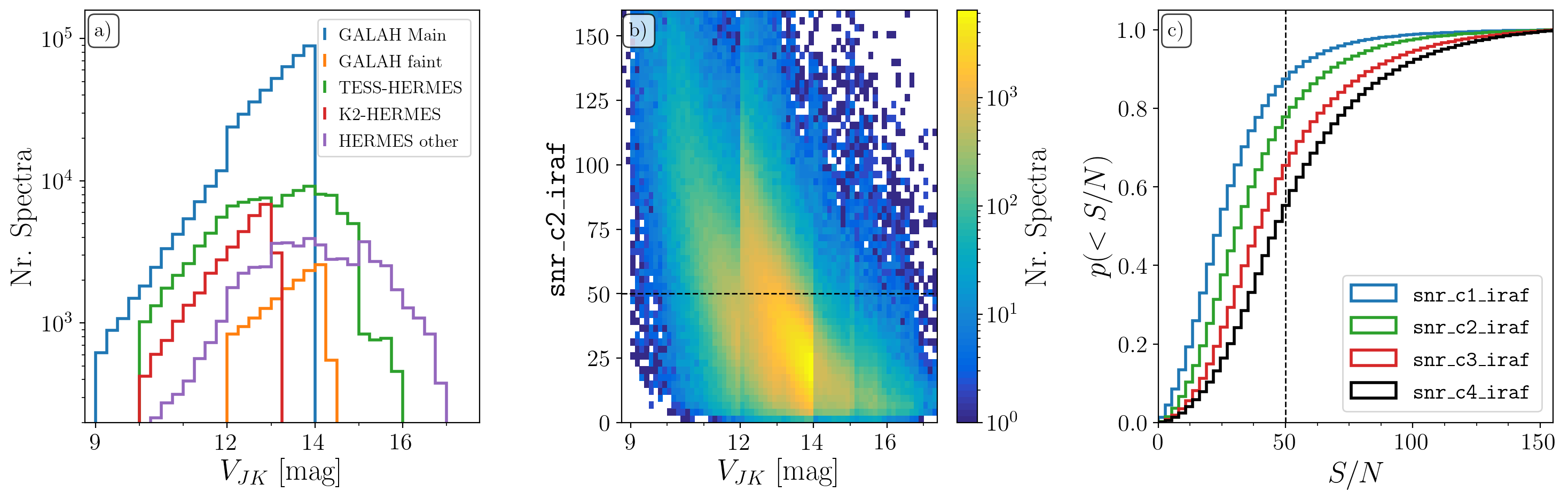}
\caption{
\textbf{Distributions of magnitudes and $S/N$ of GALAH+~DR3.}
\textbf{Panel a)} Distribution of $V$ magnitude calculated from 2MASS $J$ and $K_S$.
\textbf{Panel b)} Distribution of average achieved $S/N$ per pixel for the green band (CCD 2) as a function of $V_{JK}$.
\textbf{Panel c)} Cumulative distribution of the $S/N$ per pixel of the different bands/CCDs of HERMES for GALAH+~DR3. A black, dashed line indicates the overall $S/N$ of 50 that we initially aimed for for CCD2.}
\label{fig:hist_vjk_snr}
\end{figure*}

Since GALAH observes stars mainly nearby stars (81.2\% of the sample is within $2\,\mathrm{kpc}$, as shown in Fig.~\ref{fig:plot_distance_and_cmds}), almost all GALAH targets have well measured 5D (99\%) or even 6D (45\%) information from \Gaia \citep{Brown2018, Lindegren2018}. An overview of the astrometric and spectroscopic quality for the observed stars can be found in Fig.~\ref{fig:plx_snr_quality}a. The median fractional parallax error for GALAH stars is $2.7\,\%$, and $95\,\%$ (561\,229) of GALAH stars have parallax errors below $20\,\%$ (see panel a). A total of 588\,571 of our observations are of stars with matched \Gaia parallax measurements. When dividing the sample into giants (\Teff$ < 5500\,\mathrm{K}$ and $M_{K_S} < 2\,\mathrm{mag}$) and dwarfs (\Teff$ \geq 5500\,\mathrm{K}$ or $M_{K_S} \geq 2\,\mathrm{mag}$), $96\,\%$ (369\,227/383\,088) of the observed dwarf stars have parallax uncertainties below $10\,\%$ and $70\,\%$ (140\,840/200927) of the observed giant stars have parallax uncertainties below $10\,\%$. The inferred distance estimates from \citet{BailerJones2018}, used for the spectroscopic analysis in this release, are crucial for the small fraction of GALAH+~DR3 stars with parallax uncertainties above $20\,\%$.

Additionally, the available asteroseismic information is growing steadily as the analysis of data from the \Ktwo\ campaigns progresses. The overlap between GALAH targets and \Ktwo\ targets from campaign C1-C8 and C10-C18 has increased to more than 10\,000 stars with measured asteroseismic $\nu_\mathrm{max}$ values \citep{Zinn2020} and spectroscopic information, and covers almost the entire red giant branch ($\log g{\sim}1.5-3.0\,\mathrm{dex}$) and helium-core burning red clump.

The magnitude limited selection of the GALAH survey (see the magnitude distribution in  Fig.~\ref{fig:hist_vjk_snr}a) causes a strong correlation between increasing distance (and decreasing parallax quality) with increasing luminosity. This tradeoff between luminosity and parallax uncertainty was also visible for the stars in common between \Gaia~DR1 and GALAH~DR2 \citep{Buder2019} and is still present with the use of \Gaia~DR2, as we illustrate in Kiel diagrams in 
Fig.~\ref{fig:plx_snr_quality}b, 
showing that especially giants with larger distances suffer from large parallax uncertainties.

\subsection{Observations}

GALAH data are acquired with the 3.9-metre Anglo-Australian Telescope at Siding Spring Observatory. Up to 392 stars can be observed simultaneously using the 2dF robotic fibre positioner \citep{Lewis2002} that sits at the telescope's prime focus. The fibres run to the High Efficiency and Resolution Multi-Element Spectrograph \citep[HERMES;][]{DeSilva2015, Sheinis2015}, where the light is dispersed at $R{\sim}28\,000$ and captured by four independent cameras. HERMES records ${\sim}1000$~\AA\ of the optical spectrum across its four non-contiguous channels (4713-4903, 5648-5873, 6478-6737, and 7585-7887~\AA). Details of the instrument design and as-built performance of HERMES can be found in \citet{Barden2010}, \citet{Brzeski2011}, \citet{Heijmans2012}, \citet{Farrell2014} and \citet{Sheinis2015}.

Since HERMES was first commissioned, raw data it obtains has been contaminated by odd saturated points with vertical streaking, which was traced back to the choice of glass for the field flattening lens inside each of the four cameras \citep{Martell2017}. The original glass had been chosen for its high index of refraction, but uranium in the glass emitted $\upalpha$ particles that caused the saturated points and vertical readout streaks when they were captured by the HERMES CCDs \citep{Edgar2018}. In the first half of 2018, the original field flattening lenses were replaced with lenses made from a less radioactive glass, and the vertical streaks have almost stopped occurring in the data. The point spread function in the HERMES cameras changed as a result of changing the field flattening lenses, and is now larger and less symmetric in the corners of the detectors. As part of HERMES recommissioning, the GALAH team fed light from a Fabry-Perot interferometer into HERMES to characterise the new PSF across each detector, and this information has been incorporated into the data reduction procedure.

The observing procedure and targeting strategy for this data release are the same as for previous public GALAH data, including the selection of fields via the GALAH-internal $\texttt{obsmanager}$ (keeping track of already observed fields and suggesting fields with lowest airmass at a given observing time for a given program) and the assignment of targets onto 2dF fibres via $\texttt{configure}$ \citep{Miszalski2006}. For further information on the strategy of GALAH Phase 1, with the GALAH main and faint observations, we refer the reader to \citet{Buder2018}. For the K2-HERMES observing strategy, the reader is referred to \citet{Wittenmyer2018} and \citet{Sharma2019}, and for TESS-HERMES to \citet{Sharma2018}.

GALAH+~DR3 expands the number of targets from DR2 significantly and includes all data taken between November 2013 and February 2019. The distribution of GALAH+~DR3 stars across $V_{JK}$ and signal-to-noise ratio (S/N) is shown in Fig.~\ref{fig:hist_vjk_snr}b and adds another perspective on the complex correlation of luminosity (or surface gravity $\log g$) with $S/N$ for the observed stars, as shown in Fig.~\ref{fig:plx_snr_quality}c.

GALAH, \Ktwo-HERMES, and \tess-HERMES observers choose from a database of available fields depending on conditions, limiting the hour angle to within $\pm 2$ hours whenever possible. The standard observing procedure for regular GALAH survey fields is to take three 1200s exposures, with an arc lamp and flat lamp exposure taken at the same sky position as each field to enable proper extraction and calibration of the data. Bright-star fields are observed in evening and morning twilight, and when the seeing is too poor for the regular survey fields. They receive three 360s exposures and the same calibration frames as for the regular fields.

The median seeing at the AAT is $1\farcs 5$, and the exposure time is extended by $33\,\%$ if the seeing is between $2\farcs 0$ and $2\farcs 5$ and by $100\,\%$ if the seeing is between $2\farcs 5$ and $3\farcs 0$. This exposure time was chosen to achieve a S/N of 50 per pixel (equivalent to 100 per resolution element) in the HERMES green channel (CCD 2).  This is accomplished in nominal seeing when a star has an apparent magnitude of 14 in the photometric band matched to the camera (B=14 and CCD 1, V=14 and CCD 2, etc.) Mismatches between predicted and actual data quality are mainly caused by a combination of seeing, cloud and airmass. We show the distribution for the actual S/N per pixel as a cumulative distribution for all four HERMES channels in $V_{JK}$ in Fig.~\ref{fig:hist_vjk_snr}c. Depending on the spectral type, the S/N achieved for a given star in each CCD varies (i.e., a hot star will be brighter in the blue and green passbands and fainter in the red and infrared passbands).

\subsection{Reductions}\label{sec:reductions}

Since the release of GALAH~DR2, we have improved our reduction pipeline \citep{Kos2017}, and as a result, all spectra included in DR3 have been reduced using the new, improved pipeline. As in GALAH~DR2, raw images are corrected for bias level and flat field, and cosmic rays are removed with a modified LaCosmic algorithm \citep{vandokkum2001}. Scattered light and fibre-cross talk signals are removed. The wavelength solution for the extracted spectra is found via fitting of ThXe arc lamp observations. Sky spectra are modelled from the 25 sky fibres included in each field and subtracted, and synthetic telluric lines are computed using \texttt{molecfit} \citep{Kausch2015, Smette2015} and removed from observed spectra. The reduction pipeline runs a cross-correlation with \texttt{AMBRE} spectra \citep{DeLaverny2012} to provide a first estimate of the stellar parameters effective temperature \Teff, surface gravity \logg, iron abundance \feh, as well as radial velocity \vrad, and to normalise the spectra.

The main improvement is the wavelength solution, which is now more stable at the edges of the green and red CCDs, where we lack arc lines. This has been achieved by monitoring the solution and fixing the polynomial describing the pixel-to-wavelength transformation, if deviations from a typical or average solution are detected. The solution is described by a 4th order Chebyshev polynomial. We use \textsc{IRAF}'s \texttt{identify} function to find the positions of arc lines in each image and match them with our linelist. Fitting the solution, however, is now done in a more elaborate way. Initially, all spectra from the same image are allowed to have an independent solution. Then the four coefficients of the Chebyshev polynomial are compared. The first coefficient defines the zero-point. Because the 2dF fibres are not arranged monotonically in the pseudo-slit, the first coefficient is truly independent of the spectrum number (spectra being numbered 1 to 400 in each image). The values of the other three coefficients should be a smooth function of the spectrum number. If a coefficient for a specific spectrum deviates by more than $3\upsigma$ from a smooth function, it is corrected to lie on the smooth function. This successfully fixes the previous problems with incorrect wavelength solutions at the edge of the image. 

Our improved reduction pipeline also features an improved parameterization of cross-talk. It can only be measured in larger gaps between every 10th spectrum. Cross-talk was previously represented as a function of the position in the image, but now each batch of 10 spectra (from one slitlet) is assigned the measured cross-talk without any interpolation. The cross-talk is still a function of the direction along the dispersion axis. The normalisation has been improved with a new identification of continuum sections (regions of a spectrum where the continuum is measured) and optimised polynomial orders. The pipeline has been actively maintained and adapted to perform well with the recommissioned instrument following the replacement of the field flattening lenses in 2018 May. Other minor improvements and computing optimisations were made.

As we lay out in the next Section, with the analysis approach via spectrum synthesis chosen for GALAH+~DR3, we will not make use of the full spectra of GALAH, but restrict ourselves to absorption features with reliable line information for spectrum synthesis, when estimating stellar parameters and abundances.

%________________________________________________________________
\section{Data analysis} \label{sec:analysis}

In this section, we describe how the outputs from the data reduction process are used to estimate the final stellar parameters for each spectrum as well as up to 30 element abundances. The starting points for the analysis are the products of the reduction pipeline, described in Section \ref{sec:reductions}, that is reduced spectra, initial estimates of radial velocity \vrad, as well as initial estimates of the stellar parameters \Teff, \logg, and \feh. Contrary to GALAH~DR2, we do not use the reduction-pipeline based, \vrad-shifted, normalised spectra for the spectrum analysis of this release.

\subsection{Changes from GALAH~DR2 to GALAH+~DR3:}

\begin{figure*}
\centering
\includegraphics[width=0.996\textwidth]{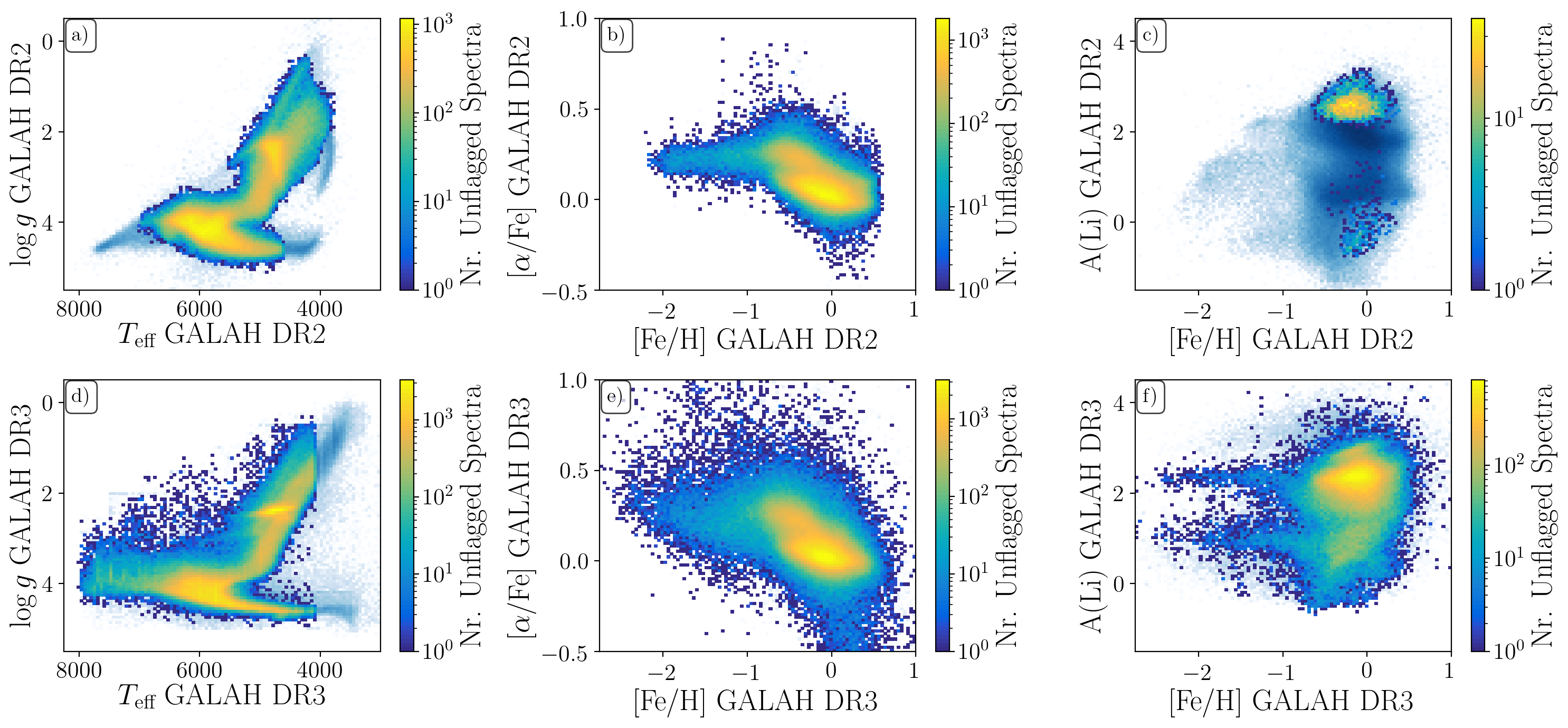}
\caption{
\textbf{Comparison of GALAH~DR2 (upper panels) and GALAH+~DR3 (lower panels, this release).}
The light-blue background indicates all measurements, whereas the colourmap shows the number of successful (unflagged) measurements at each point. \textbf{Left panels:} Kiel diagrams, i.e. \Teff versus \logg, for stars of DR2 \textbf{(a)} and DR3 \textbf{(d)}. \textbf{Middle panels:} Abundance pattern of iron vs. $\upalpha$-process elements, i.e. Tinsley-Wallerstein diagrams displaying \feh versus \alphafe, for DR2 \textbf{(b)} and DR3 \textbf{(e)}. \textbf{Right panels: } Absolute Li abundance as a function of iron abundance, i.e. \feh versus A(Li), for DR2 \textbf{(c)} and DR3 \textbf{(f)}. The stellar parameters and abundances from GALAH~DR2 appear more tightly constrained, but we note that this is an artefact of the data-driven approach, which tends to find solutions closer to the mean parameter/abundance patterns. We include all DR2 and DR3 stars in these panels, and not just the stars common to both to highlight the increase in observations, accuracy of stellar parameters, and coverage of abundances, rather than the improvement in precision for the same spectra.}
\label{fig:galah_dr3_comparison_dr2}
\end{figure*}

The two most important differences to the workflow of our analysis are the following: First, we are using astrometric information from the \Gaia mission to break spectroscopic degeneracies. Secondly, we do not use data-driven approaches for the spectrum analysis in GALAH+~DR3, but only the spectrum synthesis code Spectroscopy Made Easy \citep[][hereafter {\textsc{sme}}]{Valenti1996, Piskunov2017}, which had only been used for the training set analysis in DR2. We visualise the reasons for this step with the comparison of GALAH~DR2 and DR3 in Fig.~\ref{fig:galah_dr3_comparison_dr2}. We found in DR2 that stars at the periphery in stellar label space, e.g. high temperature (compare panels a) and d) or low metallicity (compare panels b) and e)) did not receive optimal labels from the data driven process.

Data-driven models that also use astrometric information may likely perform equally well as, or possibly better than, our DR3 analysis for many aspects. In DR3, we chose to apply the more traditional method to the full sample to assess the limitations of the data-driven approach. This includes testing the flexibility of the model we used, as we found that quadratic models (as used for GALAH~DR2) are too inflexible to be applicable across the entire stellar parameter space. Training by using $\chi^2$-optimisation may give too much weight to outliers. Furthermore, we want the results to be independent of the exact selection criteria used to define the training set, since data-driven models can struggle to inter- and extrapolate for spectra which are not sufficiently represented and modelled in the training step. Limited by the scope of the training set, we had to flag several abundance measurements for a vast majority of elements, where we suspected extrapolation. The flagged results are shown as the lighter blue background in Fig.~\ref{fig:galah_dr3_comparison_dr2}, where we find in panel a) that some of the inferred stellar parameters are unphysical, such as the upturn in the low-mass main sequence and the correlation between \Teff and \logg for hot stars. The effect of flagging on the number of inferred stellar abundances can best be seen in the drastic increase in Li detections in DR3 (compare panels (c) and (f)), where detections in DR2 were limited to warm dwarfs and Li-rich giants. This was a direct result of the choice of training set stars, with the numbers of detections in DR2 being further lowered by our use of more conservative criteria of detections for lines.

Being able to estimate reliable stellar parameters for hot stars (see panel d) has also enabled the determination of several of their abundance patterns, which was not possible in DR2. Intriguingly, some of the A- and F-type main sequence stars exhibit under-abundant \alphafe (see lowest measurements in panel e) and overabundant iron-peak and neutron-capture elements, which is the peculiar chemical compositions of Am/Fm stars \citep[see e.g.][]{Xiang2020}. In DR3, we are also able to estimate more accurate element abundances for metal-poor stars ($\mathrm{[Fe/H]} < -1\,\mathrm{dex}$), in particular those with the previously identified low-$\upalpha$\footnote{These stars have lower abundances in the $\upalpha$-elements at fixed \feh when compared to the high-$\upalpha$ disc population.} ``outer'' halo pattern \citep[see e.g.][]{Nissen2010}.

% This figure was created with the notebook GALAH_DR3/validation/comparisons/comparison_galah_dr2/galah_dr3_comparison_dr2.ipynb

For this data release, we run the analysis pipeline only for spectra that have initial radial velocity estimates either as part of previous unflagged GALAH data releases, unflagged measurements from the DR3 reduction pipeline, or from Gaussian fits to the Balmer lines with less than $5\,\mathrm{km\,s^{-1}}$ uncertainty. We therefore excluded 81\,007 spectra\footnote{We note that these observations include spectra with low $S/N$ as well as a few calibration observations like sky or dome flats falsely labelled as stellar observations over the course of 5 years of observations.}). Further we restrict this release to stars with external information on parallaxes from \Gaia \citep{Lindegren2018}, thus excluding 9,080 spectra. For a few tens of bright stars that are not in \Gaia~DR2, we take distances and parallaxes from \Hipparcos \citep{vanLeeuwen2007}, which can be identified via missing extended astrometric information.

\subsection{The general workflow}

Our general workflow follows the same approach as the spectrum synthesis analysis for DR2, with the aim to homogeneously and automatically analyse a large number of spectra that intrinsically look very different. The analysis is divided into two fundamental steps: first, we estimate the stellar parameters; and second, we keep the stellar parameters fixed while only fitting one abundance at a time for the different lines/elements in the GALAH wavelength range.
For the stellar parameter estimation (first step), we first perform a normalisation and a first rough stellar parameter fit with one iteration, followed by a final normalisation and finer parameter fit that is iterated on until convergence. For this, we limit ourselves to 46 selected segments of the GALAH spectra which include the lines with reliable line data for the stellar parameter analysis, as described in detail in Sec.~\ref{sec:spec_details}.
For the abundance analysis (second step), we perform only one normalisation and iteratively optimise the abundance based on those data points of the lines/elements that we estimate to be unblended enough after comparing a synthetic spectrum with all lines with another one that only has the lines of the element in question. To achieve most accurate results, our analysis is based on the most recent line data, atmosphere grids, and grids to compute departures from the common LTE assumption during the line formation, as described in detail in Sec.~\ref{sec:spec_details}. Below we describe this workflow in more detail, which illustrates the challenges of homogeneously analysing very different spectra:
\begin{enumerate}
\item Initialise {\sc sme} (version 536) with choices of line data, atmosphere grid, non-LTE departure grids, observed spectrum (limited to the 46 segments\footnote{These can be found in our \href{https://github.com/svenbuder/GALAH_DR3/tree/master/analysis/stellar\_parameters}{online documentation}.} used for the parameter estimation) including selection of continuum and line masks, initial parameters for $\chi^2$ optimisation. Check if all external information is provided and then update the initial \logg with this external information and the initial stellar parameters as outlined in the explanation of \logg (Sec.~\ref{sec:spec_details}).
\item Normalise all 46 segments individually with the chosen initial setup by fitting linear functions first to the observed spectrum (iteratively and with sigma-clipping) and then to the difference of the observed and synthetic spectrum.
\item Optimise the stellar parameters \Teff, \feh, \vbroad (\vsini with \vmac set to 0\footnote{At the resolution of GALAH, \vsini and \vmac are degenerate broadening influences. We thus fit them with {\sc sme} by setting \vmac to 0 and only fit \vsini.}), and global \vrad with 2 major {\sc sme} update loops (calculating double-sided partial derivatives and exploring the local $\chi^2$ surface with up to 5 different parameter choices). Consistently update \logg and \vmic from physical and empirical relations, respectively, with every change of \Teff or \feh. In our test, this already led to updated parameters close to the $\chi^2$ global minima.
\item Normalise all 46 segments again individually as in step 2, but with updated stellar parameters.
\item Optimise the stellar parameters \Teff, \feh, \vbroad, and \vrad with up to 20 major {\sc sme} update loops as in step 3 until the fractional change of $\chi^2$ is below 0.001.
\item Collect stellar parameters for validation. Save covariance uncertainties, based on the statistical $\chi^2$ uncertainties given the uncertainties of the normalized flux, in addition to the uncertainties delivered by {\sc sme} \citep[see][for more details]{Piskunov2017}. The validation of stellar parameters (see Sec.\ref{sec:validation_sp}) led to an adjustment of the estimated atmospheric \feh ({\sc sme}.feh) by adding $0.1\,\mathrm{dex}$\footnote{This is not the final $\mathrm{[Fe/H]} = \textsc{fe\_h}$ as reported in this data release, but a pseudo iron abundance $\textsc{sme}.\mathrm{feh} = \textsc{fe\_h\_atmo}$, estimated from H, Sc, Ti, and Fe lines.}.
\item Initialise {\sc sme} for the element abundance estimation with choices of line data, atmosphere grid, non-LTE departure grids, observed spectrum (limited to line segement(s) used for the element abundance estimation) including selection of continuum and line masks, final global atmosphere parameters for $\chi^2$ optimisation. Contrary to steps 3 and 5, hereafter the aforementioned global parameters, including \vrad, are kept fixed\footnote{For the Li line, at the end of CCD3, we have found that for roughly 10\% of the spectra, the wavelength solution is not reliable enough and therefore simultaneously fitted [Li/Fe] and \vrad.}.
\item Normalise the segment(s) for the particular line (for the line-by-line analyses, e.g. Sr6550) or for all lines of the particular chemical species (e.g. Ca) with the chosen initial setup by fitting linear functions first to the observed spectrum. Improve this normalisation by fitting a linear function to the difference between the observed and synthetic spectrum to create a ``full'' synthetic spectrum.
\item Because the same line exhibits different degrees of blending in different stars, which are complex and difficult to predict ab-initio, perform a blending test by creating a ``clean'' synthetic spectrum only based on the lines of the element to be fitted. Then compare the ``full'' and ``clean'' spectra for elements other than Fe for the chosen line mask pixels and neglect those which deviate more than $\Delta\chi^2>0.005$.
\item Optimise the relevant element abundance entry in the abundance table ({\sc sme}.abund) with up to 20 major {\sc sme} update loops until fractional change of $\chi^2$ is below 0.001. The atmosphere is updated with each change of chemical composition to stay consistent, but we note that for the sake of computation cost with {\sc sme}, the abundances, that are not fitted, are kept at scaled-solar, with the exception of Li with A(Li) = 2.3, an enhancement of $0.4\,\mathrm{dex}$ for N in giants, and the precomputed $\upalpha$-enhancement for $\upalpha$-process elements.
\item Collect stellar parameters and element abundances for validation and post-processing.
\item Calculate upper limits for each element/line for non-detections by estimating the lowest abundance that would lead to a line flux depression of 0.03 below the normalised continuum (see more detailed explanations at the end of this section).
\item Post-processing: apply flagging algorithms, calculate final uncertainties from accuracy and precision estimates, combine line-by-line measurements of element abundances weighted by their uncertainties.
\end{enumerate}

For each star, the computational costs amount to between 50 CPU minutes (for the hottest stars with few lines), 2 CPU hours for the Sun, up to 6 CPU hours (for the coolest stars with most lines), with around 30-50\% of that used for the stellar parameter step and the rest for abundance estimation for all lines. The total computational costs amount to 1.2 Mio CPU hours for the stellar parameter and abundance fitting, that is, neglecting data collection and post-processing.

\subsection{Details of the spectroscopic analysis} \label{sec:spec_details}

\paragraph*{The line data} are based on the corresponding compilation for the \Gaia-ESO survey \citep[][Heiter et al. in press]{Heiter2015b, Heiter2020} with updated oscillator strengths ($\log gf$ values) for some elements, in particular \ion{V}{I} \citep{2014ApJS..215...20L}, \ion{Cr}{I} \citep{2017ApJS..228...10L}, \ion{Co}{I} \citep{2015ApJS..220...13L}, \ion{Ni}{I} \citep{2014ApJS..211...20W} and \ion{Y}{II} \citep{2017MNRAS.471..532P}. In addition, we ``astrophysically'' tuned (based on solar abundances and observations) the $\log gf$-values for approximately 100 lines that were not used for abundance measurements, but affected the continuum placement and blending fraction for the main diagnostic lines. The final compilation of the lines used for stellar parameter and element abundance estimation together with the most important line data are listed in Table~\ref{tab:linelist}.

\paragraph*{The 46 segments and masks for stellar parameter estimation} are based on selected neutral and ionised Sc, Ti, and Fe lines as well as the two Balmer lines $\mathrm{H_\alpha}$ and $\mathrm{H_\beta}$. We chose these lines based on the high quality of their experimental or theoretical line data and limit ourselves to the least blended lines or parts of lines. The masks used for parameter and abundance optimisation were selected based on the line shapes of several thousand randomly selected spectra (including those of crowded spectra of cool stars with dominant molecular absorption bands). The masks used for continuum placement were selected on-the-fly as the regions with smallest amount of line absorption, ensuring a sufficient number of (pseudo-)continuum points on either side of the line mask.

\paragraph*{The model atmospheres} used for our analysis are theoretical 1D hydrostatic models taken from the {\sc marcs} grid \citep[][{\sc marcs}2014]{Gustafsson2008}. The adopted grid is the same as in GALAH~DR2 \citep[][Sect. 3.2]{Buder2018}. In brief, they cover $2500 \leq T_\mathrm{eff} \leq 8000\,\mathrm{K}$, $-0.5 \leq \log g \leq 5.5\,\mathrm{dex}$ with the exclusion of the hottest and lowest surface gravity regions, $-5 \leq \mathrm{[Fe/H]} \leq 1$, and were computed with the Solar chemical composition of \citet{Grevesse2007}, scaled by \feh and with $\upalpha$-enhancements as laid out later in this section. Plane-parallel models were adopted for $\log g \geq 4$, and spherically-symmetric models for $\log g < 4$.

\paragraph*{The non-LTE grids} of departure coefficients that we use for the on-the-fly synthesis of 1D non-LTE spectra are described in \citet{Amarsi2020}. In brief, new grids of departure coefficients were constructed by adopting the non-LTE model atoms presented for H \citep{Amarsi2018}, Li \citep{Lind2009, Wang2020}, C \citep{Amarsi2019}, O \citep{Amarsi2018b}, Na \citep{Lind2011}, Mg \citep{Osorio2015}, Al \citep{Nordlander2017}, Si \citep{Amarsi2017}, K \citep{Reggiani2019}, Ca \citep{Osorio2019}, Mn \citep{Bergemann2019b}, and Ba \citep{Gallagher2020}, and running on the {\sc marcs} model atmosphere grid using the non-LTE radiative transfer code {\sc balder} \citep{Amarsi2018}, a modified version of {\sc multi3d} \citep{Leenaarts2009}.  For Fe, the same non-LTE grids of departure coefficients that were used in GALAH~DR2 were adopted here \citep{Amarsi2016a, Lind2017}. As we demonstrated in \citet{Amarsi2020}, relaxing LTE reduces the dispersion in the [X/Fe] versus \feh plane significantly, for example by $0.1\,\mathrm{dex}$ for Mg and Si, and it can remove spurious differences between the dwarfs and giants by up to $0.2\,\mathrm{dex}$. Recent progress in this field will allow the implementation of non-LTE also for other large surveys \citep{Amarsi2020, Osorio2020}. The use of non-LTE grids is unique to GALAH, whereas most other current major surveys, like APOGEE \citep{Joensson2020}, RAVE \citep{Steinmetz2020a} and \Gaia-ESO \citep{Smiljanic2014} report 1D LTE results in their public data releases.

\paragraph*{Initial stellar parameters and abundances} are chosen depending on the quality of reduction products and their availability in GALAH~DR2 \citep{Buder2018}. If the stellar parameters of GALAH~DR2 (and non-published spectra of K2-HERMES, TESS-HERMES and other spectra analysed in the same way via \textit{The Cannon}) are not flagged, we use those. Otherwise, we use initial rough stellar parameters provided as part of the reduction pipeline during its radial velocity estimation with grid interpolation, if they are unflagged. Otherwise we use a set of fiducial stellar parameters ($T_\text{eff} = 5500\,\mathrm{K}$, $\log g = 3.0\,\mathrm{K}$, and $\mathrm{[Fe/H]} = -0.5\,\mathrm{dex}$ as well as the result of Gaussian fits to the two Balmer lines for \vrad). We initialise the abundance pattern as scaled-solar, but adjust the alpha-enhancement as described later in this section.

\paragraph*{Surface gravities} are updated self-consistently with the other stellar parameters for each synthesis step via the fundamental relation of \logg, stellar mass $\mathcal{M}$, and bolometric luminosity $L_\text{bol}$
\begin{equation}
\log g = \log g_\odot + \log \frac{\mathcal{M}}{\mathcal{M_\odot}} + 4 \log \frac{T_\mathrm{eff}}{T_\mathrm{eff,\odot}} - \log \frac{L_\mathrm{bol}}{L_\mathrm{bol,\odot}} \label{eq:logg}
\end{equation}

\paragraph*{Bolometric luminosities} are estimated via
\begin{equation}
\log \frac{L_\mathrm{bol}}{L_\mathrm{bol,\odot}} = -0.4 \cdot \left(K_S - 5\cdot \log \frac{D_\varpi}{10} + BC(K_S) - A(K_S) - M_\mathrm{bol,\odot} \right) \label{eq:lbol}
\end{equation}
from the 2MASS \citep{Skrutskie2006} $K_S$ band, a consistently calculated bolometric correction $BC(K_S)$ for this band using stellar parameters for each synthesis step, distances $D_\varpi = \texttt{r\_est}$ from \citet{BailerJones2018} as well as extinctions $A_{K_S}$ in the $K_S$ band. If both 2MASS $H$ \citep{Skrutskie2006} and WISE $W2$ \citep{Cutri2013} band information have quality A (98\% of our sample) we used the RJCE method \citep{Majewski2011} to compute $A_{K_S}  = 0.917 \cdot \left( H - W2 - 0.08 \right)$. For the remaining 2\% of our sample we used $A_{K_S} = 0.38 \cdot \texttt{E(B-V)}$ \citep{Savage1979}.

\paragraph*{Bolometric corrections} were estimated via interpolation of the grids from \citet{Casagrande2014, Casagrande2018}. The interpolation needs stellar parameters and extinction and is performed whenever a stellar parameter is adjusted during its optimisation step to ensure a self-consistent set of bolometric corrections and stellar parameters at any time. The upper limit of the extinction $E(B-V)$ for 95\% of our sample is below $0.48\,\mathrm{mag}$ based on the maps from \citet{Schlegel1998}, and to speed up calculations, we have therefore truncated the extinction input for the interpolation to this value.

\paragraph*{Stellar masses} are needed to estimate the surface gravities according to Eq.~\ref{eq:logg}, but also depend on the surface gravity, luminosities or absolute magnitudes, when estimated via isochrone interpolation. We therefore estimate those masses iteratively and self-consistently together with $\log g$ via isochrone interpolation whenever a stellar parameter $O_i \in$ [\Teff, \logg, \feh, and \lbol] is updated during the parameter optimisation. We assume that these parameters have Gaussian uncertainties and no covariances. This is a bold assumption, given that we use both \logg and \lbol, which convey very similar information. However, we use large uncertainties for \logg, to limit its influence to extreme cases and can then write a likelihood for each isochrone point with model parameters $S_i$
\begin{equation}
\mathcal{L} \sim \prod_i \frac{1}{\sqrt{2 \pi} \sigma_i} \cdot \exp \left( - \frac{\left( O_i - S_i \right)^2}{2 \sigma_i^2} \right)
\end{equation}

{As we do not have final uncertainties for the stars at this stage of the analysis, we assume that the parameter uncertainties $\sigma_i$ are $100\,\mathrm{K}$, $0.5\,\mathrm{dex}$, $0.2\,\mathrm{dex}$, and $0.1\cdot L_\text{bol}$ for \Teff, \logg, \feh, and \lbol, respectively. We want to stress that these are not the final average uncertainties, but that these values were chosen after extensive tests of ensuring enough isochrone points to be considered for the mass interpolation within the uncertainties. For the final uncertainties of this release, we use a more sophisticated implementation (see Sec.~\ref{sec:validation_sp}). We convert the iron abundances into a measurement of metallicity $Z$ by assuming the $\upalpha$ enhancement to follow the stellar parameter relation laid out later in this section and combine this $\mathrm{[\upalpha/Fe]}$ and the atmospheric iron abundance to [M/H] via the correlation by \citet{Salaris2006} and into $Z$ with the Solar value from the {\sc parsec+colibri} isochrones \citep{Bressan2012, Marigo2017}. We then use these {\sc parsec+colibri} isochrones on a grid with ages of $0.5...(0.5)...13.5\,\mathrm{Gyr}$ and $\mathrm{[Fe/H]} = -2.4...(0.1)...0.6\,\mathrm{dex}$. to estimate maximum likelihood masses on-the-fly.

\paragraph*{Microturbulence velocities} $v_\mathrm{mic}$ were computed consistently from the empirical relations estimated for the GALAH survey. For cool main sequence stars ($T_\mathrm{eff} \leq 5500\,\mathrm{K}$ and $\log g \geq 4.2\,\mathrm{dex}$) we use
\begin{equation}
v_\mathrm{mic} = 1.1 + 1.6 \cdot 10^{-4} \cdot (T_\mathrm{eff}-5500\,\mathrm{K})
\end{equation}
and for hotter or evolved stars ($T_\textrm{eff} \geq 5500\,\mathrm{K}$ or $\log g \leq 4.2\,\mathrm{dex}$) we use
\begin{equation}
1.1 + 1.0\cdot 10^{-4} \cdot (T_\mathrm{eff}-5500\,\mathrm{K}) + 4 \cdot 10^{-7} \cdot (T_\mathrm{eff}-5500\,\mathrm{K})^2, 
\end{equation}
where \vmic is given in $\mathrm{km\,s^{-1}}$. In Sec.~\ref{sec:caveats}, we elaborate on the possible systematic trends that this simplified function could introduce.

\paragraph*{Element abundances} are computed during the analysis with the {\sc sme}-internal notation of relative abundances for the first 99 elements, such that their sum amounts to 1. These are initialised consistently with the {\sc marcs} pattern from the Solar abundances of \citet{Grevesse2007}. This notation is different from the usual $A(\mathrm{X}) = \texttt{A\_x} = \log \epsilon (\mathrm{X})$ and we thus convert them when reading out the final abundance pattern. In our final notations of element X, we report $A(\mathrm{X})$ on the customary astronomical scale for logarithmic abundances, where H is defined to be $A(H) = 12.00$, that is, $A(\mathrm{X})= \log \frac{N_\mathrm{X}}{N_\mathrm{H}} + 12$, where $N_\mathrm{X}$ and $N_\mathrm{H}$ are the number densities of elements X and H, respectively. We further report relative abundances as $\mathrm{[X/H]} = A(\mathrm{X}) - A(\mathrm{X})_\odot$ and $\mathrm{[X/Fe]} = \mathrm{[X/H]} - \mathrm{[Fe/H]}$. For the explanation of our chosen values of $A(\mathrm{X})_\odot$ see Sec.~\ref{sec:accuracy_abundances} and for their values see Tab.~\ref{tab:solar_reference_values2}. This table also lists the lines used for the line-by-line analysis, which were later combined for the final element abundances reported as {\sc x\_fe} for element X.

\paragraph*{Line-by-line vs. combined abundance analysis} was selected based on the time and computation resources available. 
{We have therefore measured only the following elements line-by-line (and report a value based on error-weighted means): 
$\upalpha$ (see next paragraph), Li, C, K, Ti, V, Co, Ni, Cu, Zn, Rb, Sr, Y, Zr, Mo, Ru, La, Ce, Nd, Sm, Eu. To estimate the abundance of the following elements, we ran combined all lines of the particular elements to fit a global abundance at the same time: O, Na, Mg, Al, Si, Ca, Sc, Cr, Mn, Fe, Ba. We want to stress that the use of non-LTE grids does not affect the computation time. We chose to run several elements in a combined way because of the ongoing developments of their line selection or non-LTE grids. During the development of the pipeline we have tested all individual lines for the elements run with non-LTE and only selected those with similar trends and absolute abundances to run combined. By using individual lines, we are less prone to unreliable line data, such as unreliable $\log gf$ values. Incorrect oscillator strengths introduce a bias in the absolute abundance for each line. When the Solar abundance for these lines are however estimated independently from the others, they can still be used for the combined [X/Fe] abundance, after applying individual Sun-based corrections to the absolute abundances (see Table~\ref{tab:solar_reference_values2}).

\paragraph*{Alpha-enhancement $\mathrm{[\upalpha/Fe]}$} is treated differently during the stellar parameter estimation step and the abundance determination step for each of the alpha-elements. In all cases, we initialise the abundances with the scaled-Solar pattern. We then adjust the alpha-enhancement for the elements O, Ne, Mg, Si, S, Ar, Ca, and Ti with the common enhancement pattern of $\mathrm{[\upalpha/Fe]} = 0.4\,\mathrm{dex}$ for $\mathrm{[Fe/H]} \leq -1.0\,\mathrm{dex}$ as well as $\mathrm{[\upalpha/Fe]} = 0.0\,\mathrm{dex}$ for $\mathrm{[Fe/H]} \geq 0.0\,\mathrm{dex}$ and a linear function between both iron abundances. We update this value consistently whenever \feh changes. For the individual lines of O, Mg, Si, Ca, and Ti as part of GALAH+~DR3, we then update their actual abundances while keeping the other abundances fixed. The final reported global $\mathrm{[\upalpha/Fe]} = \texttt{alpha\_fe}$ is then an error-weighted combination of selected Mg, Si, Ca, and Ti lines (Mg5711, combined Si, combined Ca, Ti4758, Ti4759, Ti4782, Ti4802, Ti4820, and Ti5739). We stress however, that this combination is dependent on the detection of these lines and might come down to a single measurement, whereas other estimates are a combination of up to 9 measurements.

\paragraph*{Upper limits} are calculated for all measured lines/elements if no detection was possible. In this case we estimate the smallest abundance needed to explain the strength of a line, that is the difference of line to continuum flux in the normalised spectrum of at least 0.03 or at least $1.5/(S/N)$ in the line mask. We interpolate these values from precomputed estimates of line strengths for a set of stellar parameters and abundances. This approach was chosen and tested to estimate a larger number of upper limits for Li, but we want to caution the users to not blindly use them because we have not performed extensive tests for the other elements. They should only be used when essential for the science case and after thorough inspection of the observed and synthetic spectra. 

%________________________________________________________________
\section{Validation of stellar parameters} \label{sec:validation_sp}

In this section, we describe the tests that we perform to validate the stellar parameters we obtain in terms of their accuracy (systematic uncertainties) and precision. In addition, we then describe several other algorithms that we have developed in order to identify peculiar stars or spectra - cases for which our standard pipeline might fail. The result of the performed quality tests are summarised in the stellar parameter flag \texttt{flag\_sp} with flags that can be raised explained in Sec.~\ref{sec:flagging_sp}. We do strongly recommend that all users take these flags into account, and make use of them unless the flags are explicitly not advisable for their particular science case. By default we recommend to use quantities with \texttt{flag\_sp} 0{, as this indicates that the stellar parameter estimates have passed all quality tests. Several influences on the accuracy, like unresolved binarity, as well as some possible systematics / caveats that we have not been able to quantify and therefore not flag, are addressed in Sec.~\ref{sec:caveats}.

To assess the quality of the stellar parameters we obtain, we resort to the commonly used comparison samples for accuracy, that is, the Sun (see our results for sky flat observations compared to literature in Table~\ref{tab:solar_reference_values1}) and other \Gaia FGK Benchmark stars \citep[GBS][]{Heiter2015, Jofre2014, Jofre2015, Hawkins2016, Jofre2018a}, photometric temperatures from the Infrared Flux Method \citep[IRFM][]{Casagrande2010}, stars with asteroseismic information, and open as well as globular cluster stars. For the precision assessment we use the internal uncertainty estimates and repeat observations of the same stars.  We calculate the final stellar parameter errors for a given parameter $P$ via 
\begin{equation}
e_\text{final}^2 (P) = e_\text{accuracy}^2(P) + e_\text{precision}^2(P). \label{eq:usual_final_error}
\end{equation}

The precision uncertainty $e_\text{precision}^2(P)$ is usually quantified by either fitting uncertainty ($e_\text{fit}^2(P)$)\footnote{In our case we use the square root of diagonal elements of the fitting covariance matrix to trace the fitting uncertainty \citep{Piskunov2017}.} or uncertainty from repeated measurements ($e_\text{repeats}^2(P)$), which are typically expected to be of the same order. We compare and rescale these values to match in Sec.~\ref{sec:precision_sp}, so that we can use their maximum values for the individual measurements. Our repeat precision estimates are based on the behaviour with respect to our reference $S/N$, that is \texttt{snr\_c2\_iraf}, and lead to our applied uncertainty estimation of
\begin{equation}
e_\text{final}^2 (P) = e_\text{accuracy}^2(P) + \text{max} \left(e_\text{fit}^2(P), e_\text{repeats}^2(P, \texttt{snr\_c2\_iraf}) \right). \label{eq:final_error}
\end{equation}

For the repeat observations, we make use of the 51\,539 spectra of explicit repeat observations (typically at different nights) of stars, not the three individual observations scheduled for each star. Such repeat observations were mainly performed for the TESS-HERMES as well as bulge and cluster observations, but a smaller contribution comes from repeat observations of GALAH fields with bad seeing. Checks of the parameter distribution of the repeat observations and the overall sample suggest that they are representative of the sample.

The uncertainties in terms of accuracy and mean expected precision at $S/N = 40$ for the stellar parameters are listed in Tab.~\ref{tab:accuracy_sp}. We explain how we estimate the accuracy in Sec.~\ref{sec:accuracy_sp}\footnote{For \vbroad, we used the comparison with the \Gaia FGK Benchmark Stars and estimate the accuracy via the scatter of $2\,\mathrm{km\,s^{-1}}$ with respect to the square sum of the rotational and macroturbulence velocity as accuracy limit.} and elaborate on the choice of uncertainty combination when we assess the precision of the stellar parameters in Sec.~\ref{sec:precision_sp}. To identify the stars and spectra that have less reliable or unreliable information, we have implemented a combination of the flagging algorithms already applied to GALAH~DR2 \citep[see][]{Buder2018} and new algorithms, which we will present in Sec.~\ref{sec:flagging_sp}.

\begin{table}
\centering
 \caption{Accuracy values and expected precision at $S/N = \textsc{snr\_c2\_iraf} = 40$ per pixel for the stellar parameters. The stated precision value for \logg is the mean precision of the whole sample.}
 \label{tab:accuracy_sp}
 \begin{tabular}{lcc}
  \hline \hline
Parameter [Unit] & Accuracy Value & Precision ($S/N=40$)\\
\hline
\Teff [K] & 67 & 49 \\
\logg [$\mathrm{cm\,s^{-2}}$] & 0.12 & 0.07 \\
\feh [dex] & 0.034 & 0.055 \\
\fehatmo [dex] & 0.059 & 0.041\\
\vbroad [$\mathrm{km\,s^{-1}}$] & 2.0 & 0.83 \\
\vrad [$\mathrm{km\,s^{-1}}$] & 0.1 & 0.34 \\
\hline
\end{tabular}
\end{table}

\subsection{Accuracy of stellar parameters} \label{sec:accuracy_sp}

\subsubsection{Effective temperature}

Our effective temperatures are estimated from our spectra rather than photometry and because they correspond to the best-fit spectroscopic solution, we do report them rather than values calibrated to the photometric scale, but assess their accuracy.

We see typically good agreement with the GBS that are representative of the stars in this data release, as well as with the general trends from the IRFM method within the uncertainties, as laid out below. We therefore do not correct biases or trends for \Teff and use the scatter with respect to the GBS as accuracy measure for our \Teff. For purposes that need the temperatures to be tied to the photometric scale, we report however also IRFM temperatures to allow users to (re-)assess the temperatures and possible uncertainties on a star-by-star basis.

\paragraph*{\Gaia FGK benchmark stars (GBS).}

We have observed the GBS \citep{Heiter2015, Jofre2014, Jofre2015, Hawkins2016, Jofre2018a} in the Southern hemisphere as reference stars with external non-spectroscopic measurements of stellar parameters. Their reference \Teff are based on angular diameter measurements \citep[e.g.][]{Karovicova2018,Karovicova2020} and when we compare with the GALAH+~DR3 results (blue error bars in upper panel of Fig.~\ref{fig:gbs_performance}), we find an excellent agreement with these temperatures for most of the stars between $3500$ and $6250\,\mathrm{K}$. We note, however, significant differences for the two massive (${\sim}3\,\mathrm{M_\odot}$) giant stars $\upxi$~Hya, $\upepsilon$~Vir, and the subgiant $\upepsilon$~For. For these three stars, both \logg and \feh agree with the benchmark values within the uncertainties, however. We also notice an increasing disagreement for F stars (hotter than $6250\,\mathrm{K}$), that is Procyon, HD~84937, HD~49933. Nonetheless, our estimated values of \logg and \feh also agree within the uncertainties. We note, however, that the majority of stars of the GALAH sample have significantly lower masses (on average $1.08\pm0.28\,\mathrm{M_\odot}$) than these stars.

% Made with GALAH_DR3/validation/stellar_parameters/gaia_fgk_benchmark_stars/gbs_performance.ipynb
\begin{figure}
\centering
\includegraphics[width=\columnwidth]{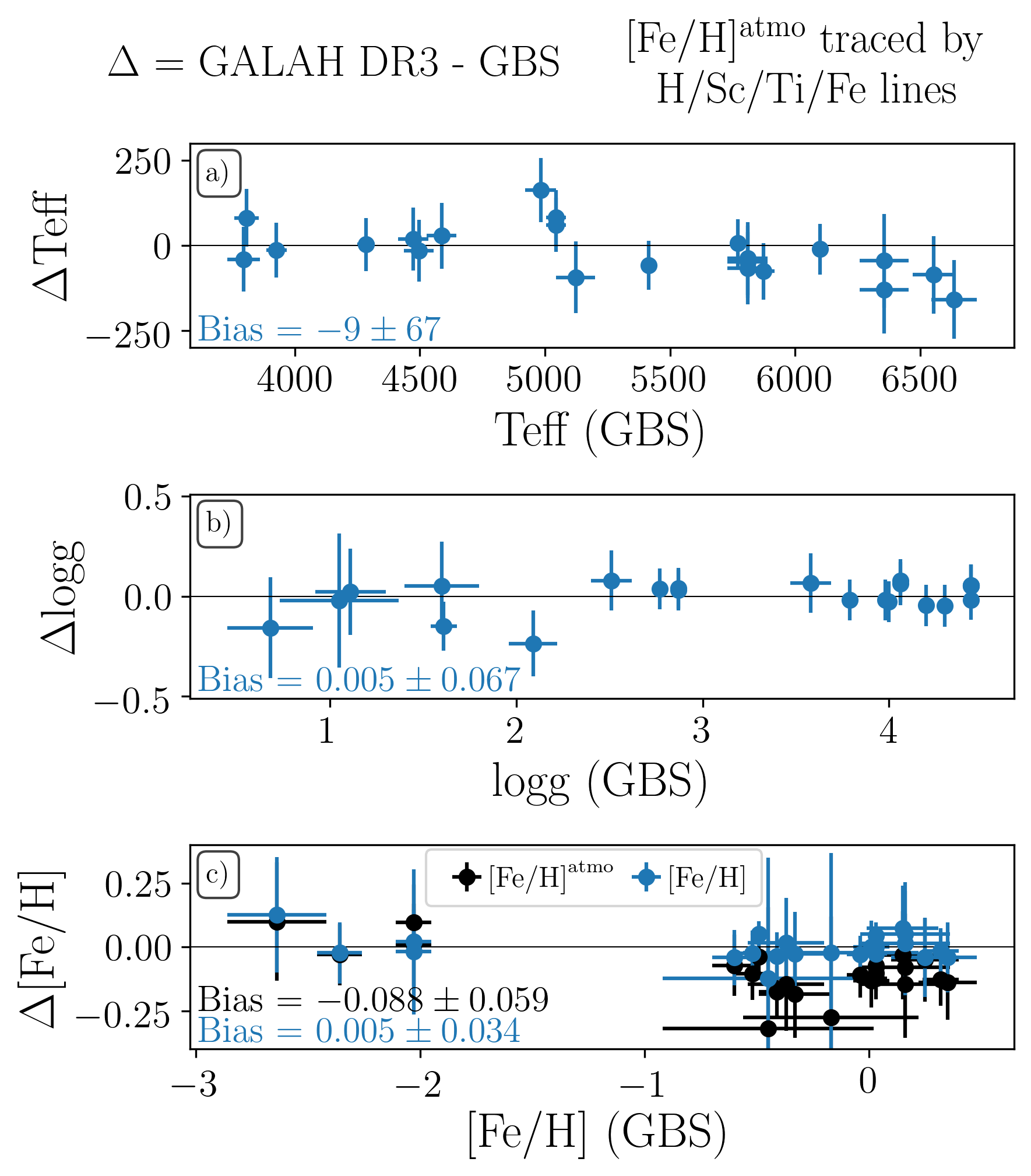}
\caption{
\textbf{Comparison of the stellar parameters \Teff (top), \logg (middle), and \feh (bottom) for the observed \Gaia FGK Benchmark stars.}
Differences are stated as GALAH+~DR3 - GBS \citep{Jofre2018a} and biases are error-weighted. The biases of \Teff are small but show similar to previous data releases, systematic deviations for F stars. The biases of \logg are small thanks to the improved \logg estimation. The disagreement between the GBS $\log g$ values and ours has decreased significantly from DR2 ($-0.06\pm0.16\,\mathrm{dex}$). During the stellar parameter estimation, the atmospheric iron abundance (black error bars) is estimated from mask regions of well selected H, Ti, Sc, and Fe lines and underestimates the true iron abundance. For the abundance fits, we have thus increased the atmospheric iron abundance by $+0.1\,\mathrm{dex}$. The final reported iron abundance (blue error bars) is only based on Fe lines and shows no bias. GBS with $T_\text{eff} > 6000\,\mathrm{K}$ were observed with $S/N{\sim}60$, whereas the other stars all cover $S/N$ between 150 and 800.}
\label{fig:gbs_performance}
\end{figure}

\paragraph*{Infrared Flux Method (IRFM) temperatures.}

% Based on GALAH_DR3/validation/stellar_parameters/irfm/galah_dr3_irfm.ipynb
\begin{figure*}
\centering
\includegraphics[width=\textwidth]{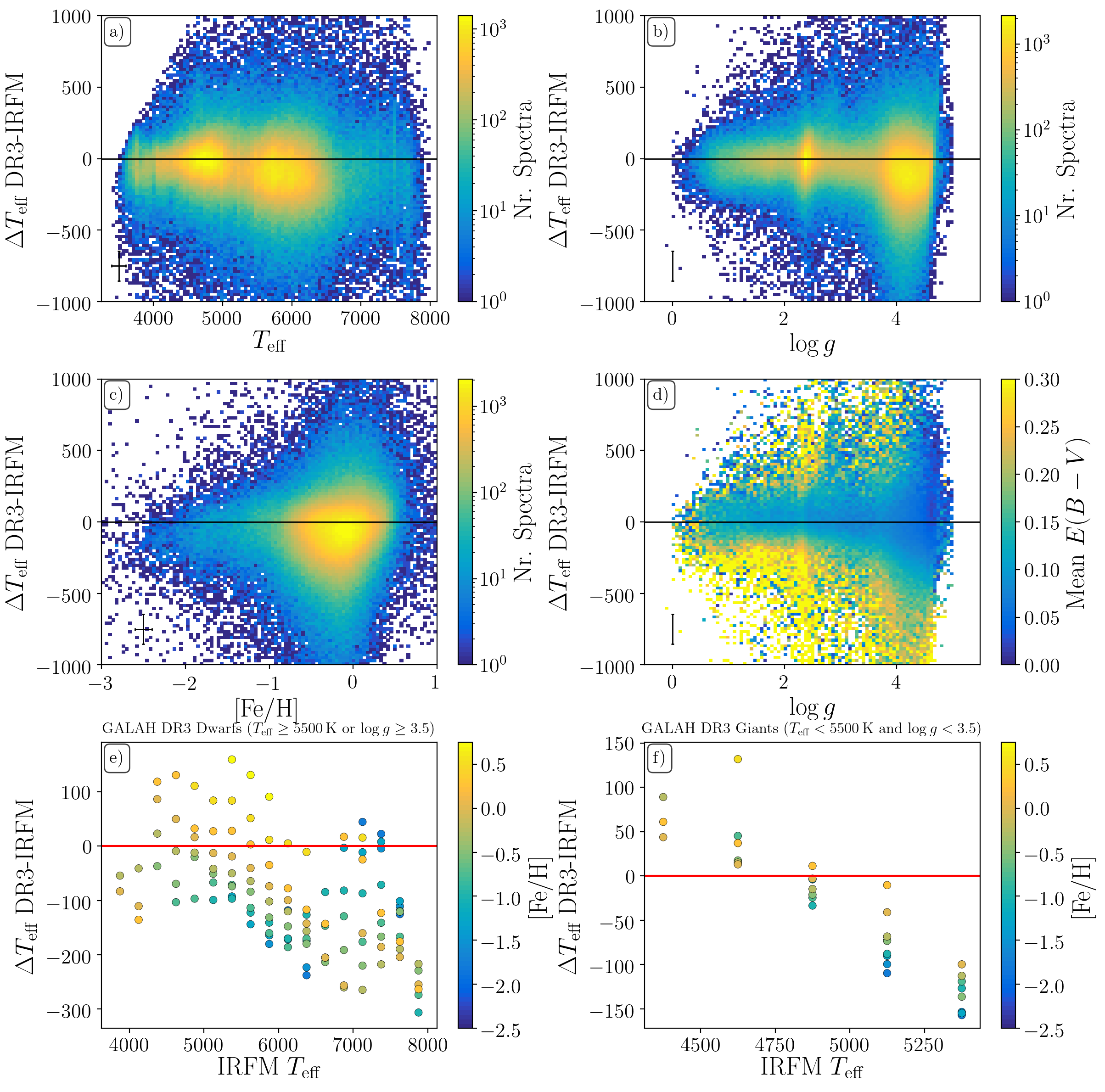}
\caption{
\textbf{Comparisons of spectroscopically determined $T_\text{eff}$ with $T_\text{eff}$ estimated via the Infrared Flux method following \citet{Casagrande2010, Casagrande2020}.} 
\textbf{Panel a-c)} Density distributions of the deviation of GALAH+~DR3 vs. IRFM \Teff as a function of GALAH+~DR3 \Teff, \logg, and \feh respectively. \textbf{Panel d)} Density distributions of the deviation of \Teff as function of GALAH+~DR3 \Teff coloured by the mean extinction $E(B-V)$ per bin. \textbf{Panels e) and f)} Distributions of deviations of \Teff ($3875..(250)..7875\,\mathrm{K}$) as a function of IRFM \Teff for different \feh bins ($-2.50..(0.25)..0.75\,\mathrm{dex}$) for dwarfs ($T_\text{eff} \geq 5500\,\mathrm{K}$ or $\log g \geq 3.5\,\mathrm{dex}$) and giants (i.e. not dwarfs), respectively. Points are coloured by the \feh bin and represent the median deviation for bins with at least 50 stars.}
\label{fig:IRFM}
\end{figure*}

\begin{figure*}
\centering
\includegraphics[width=\textwidth]{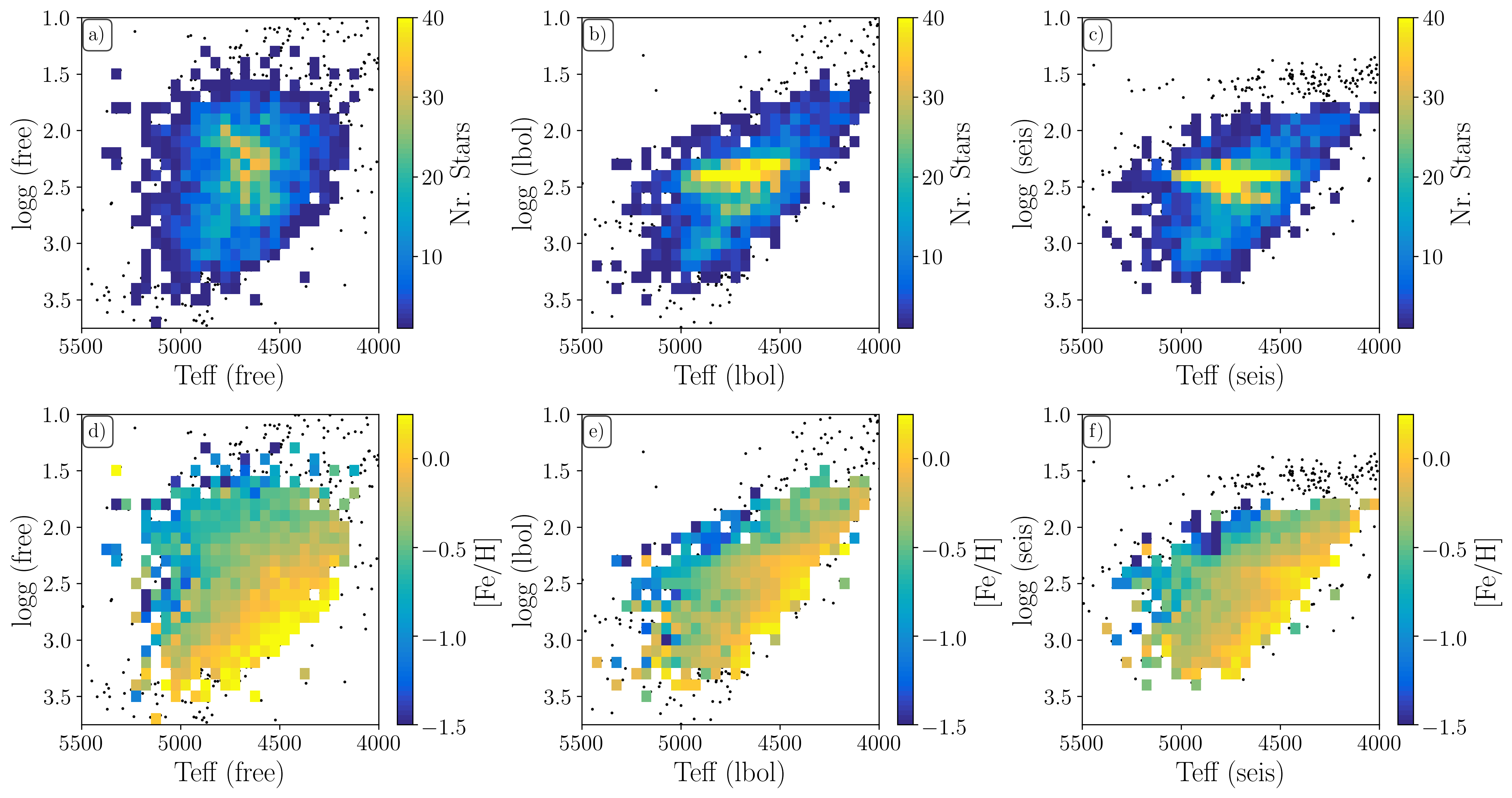}
\caption{
\textbf{Results of testing different pipeline versions with `free' (left panels), `bolometric' (middle panels), and `seismic' (right panels) estimates of \logg for the stars with both asteroseismic and parallax information available.}
Shown are the number density (upper panels) as well as the mean iron abundance (lower panels) in binned distributions.
The plots show that with good astrometric and photometric information, the `bolometric' pipeline (chosen for GALAH+~DR3) delivers accurate results similar to the `seismic' pipeline. The `bolometric' pipeline further shows much smaller scatters and biases than the `free' pipeline.
}
\label{fig:seis_comparison_3setups}
\end{figure*}

We apply the IRFM \citep{Casagrande2010} to estimate photometric \Teff. We use the 2MASS and \Gaia photometry to estimate photometric temperatures as described by \citet{Casagrande2020} and compare the differences between these temperatures  in Fig.~\ref{fig:IRFM}. Because the IRFM is sensitive to extinction, we subsequently limit the quantitative comparison (stating 16th, 50th, and 84th percentiles) to stars with small extinction $E(B-V) < 0.15\,\mathrm{mag}$ (see panel d). Most of the outliers can be explained by high extinction values (compare panel b and d).

The overall agreement is good for stars with lower temperatures ($T_\text{eff} < 5500\,\mathrm{K}$, see panel a) as well as stars with lower surface gravities ($\log g < 3.5\,\mathrm{dex}$, see panel b). We see a trend towards underestimated \Teff for hotter dwarfs, similar to previous GALAH analyses as well as the trend of the few benchmark stars.

For giants ($T_\text{eff} < 5500\,\mathrm{K}$ and $\log g < 3.5\,\mathrm{dex}$) we find a very good agreement for their whole temperature range of $-6_{-78}^{+80}\,\mathrm{K}$. For stars in the red clump region ($T_\text{eff} = 4800\pm400\,\mathrm{K}$, $\log g = 2.4\pm0.2\,\mathrm{dex}$), we find a difference of $2_{-75}^{+74}\,\mathrm{K}$.

When inspecting dwarfs ($T_\text{eff} \geq 5500\,\mathrm{K}$ or $\log g \geq 3.5\,\mathrm{dex}$) in bins of $4125..(250)..7250\,\mathrm{K}$ 
(covering $97\,\%$ of the dwarfs), we find an increasing differences from $-8_{-133}^{+138}\,\mathrm{K}$ at $4500\,\mathrm{K}$ towards $-125_{-176}^{+184}\,\mathrm{K}$ at $6750\,\mathrm{K}$. For Solar twins, that is stars similar to the Solar \Teff, \logg, and \feh within $100\,\mathrm{K}$, $0.1\,\mathrm{dex}$, $0.1\,\mathrm{dex}$ following the definition by \citet{Bedell2018}, we find a typical difference of $-95_{-119}^{+128}\,\mathrm{K}$.

Because the distribution of overall \Teff difference as a function of \feh (panel c) is not clear enough for a diagnostic of \feh trends, we analyse the difference as a function of different \feh bins for dwarfs (panel e) and giants (panel f). We find that our estimated \Teff best agrees for stars with Solar \feh (coinciding with the peak of the GALAH metallicity distribution function) but we tend to overestimate \Teff for stars with super-Solar \feh, while we tend to underestimate them for stars with sub-Solar \feh.

We note that discrepancies between spectroscopic and photometric temperatures, similar to the offsets that we find of $-1.3_{-2.2}^{+2.4}\,\%$ on average, are common \citep[see e.g.][]{Meszaros2013} and it is contentious if they should be corrected or not. Given that our spectroscopic temperatures correspond to the best spectroscopic fit, we choose to not correct our spectroscopic temperatures, unlike, for example, the approach followed by the APOGEE collaboration \citep{Joensson2020}. We do, however, provide IRFM temperatures along with adopted reddening values in our main catalogue. We note that we have not included the results of the IRFM \Teff comparisons for our accuracy estimates of our spectroscopic \Teff and therefore caution the user to decide which temperatures might be more useful for their science case and decide if they want to adjust the uncertainties by a systematic factor, for example a quadratic increase of accuracy uncertainty estimated from the difference of IRFM and spectroscopic \Teff.

\begin{figure*}
\centering
\includegraphics[width=\textwidth]{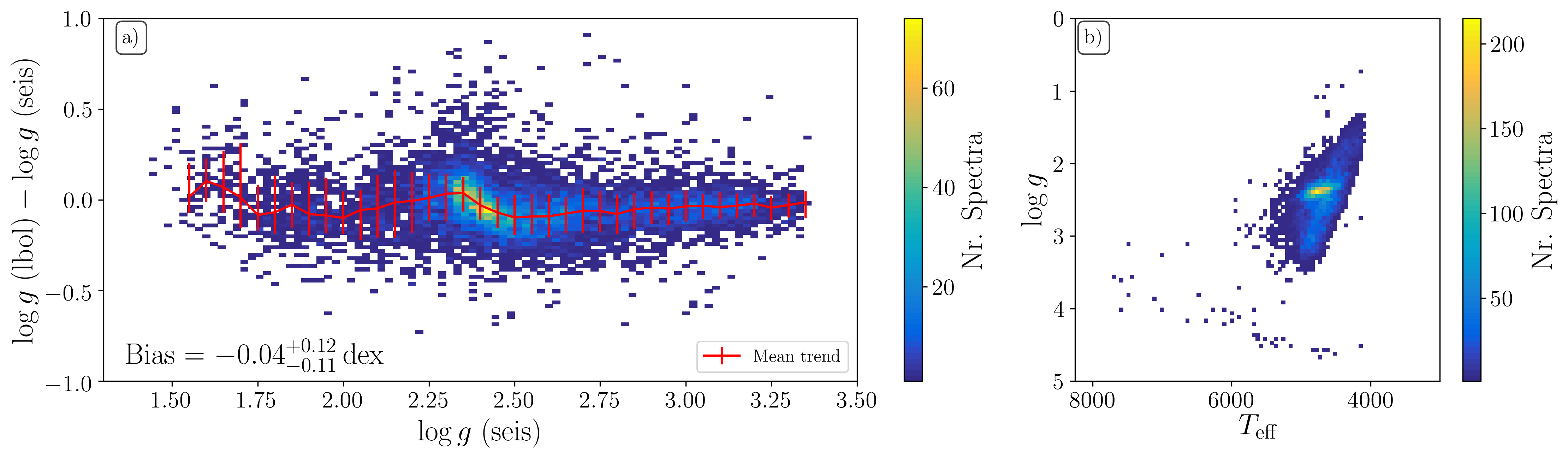}
\caption{
\textbf{Comparison of surface gravities of the stars with asteroseismic information from the K2 asteroseismic analyses (SYD pipeline).}
\textbf{Panel a)} shows the density distribution of the deviation between our surface gravities (lbol) and those estimated from scaling relations (seis). 
\textbf{Panel b)} showing the same stars, but in the Kiel diagram. Most stars are giants with $\log g$ between 1.5 and $3.5\,\mathrm{dex}$, but a few tens of dwarfs are visible. These are likely cases of photometric blends from dwarfs by giants (see text for further details).
}
\label{fig:all_seismic}
\end{figure*}

\subsubsection{Surface gravity}

We find excellent agreement and negligible biases between our derived surface gravities and those from the  GBS, as well as those obtained for stars with asteroseismic information. Due to the larger sample size of the stars with asteroseismic information, we apply the estimated scatter of this sample as an accuracy estimate for our \logg.

\paragraph*{GBS.}

The surface gravities we obtain are in excellent agreement with the accepted values for the GBS \citep{Heiter2015, Jofre2018a}, as expected given that both studies used the same approach to estimate these via bolometric relations. Due to the different implementations of this method, it is however reassuring to see the excellent agreement and low scatter (second panel of Fig.~\ref{fig:gbs_performance}). We note a slight disagreement for the highest bolometric luminosities and masses, which cancel each other out and lead to a good agreement in \logg. The only outlier of these measurements is the giant star HD~107328 (which has the largest relative mass and \logg uncertainty of the GBS and a significant change from \Hipparcos to \Gaia parallaxes); however, both \Teff and \feh are in excellent agreement with the GBS values.

\paragraph*{Stars with asteroseismic information.}

To test the GALAH+~DR3 pipeline, we analyse a subset of 3175 spectra, for which asteroseismic \numax estimates were available from the seismic SYD pipeline \citep{Huber2009} as part of the K2 Galactic Archaeology Program (GAP) data release 3 (J.~Zinn et al., in prep.). We compare the GALAH+~DR3 pipeline (`bolometric' or `lbol' pipeline) with an adjusted version (`asteroseismic' or `seis' pipeline) that uses the empirical (metallicity-independent) asteroseismic scaling relations of solar-like oscillators \citep[see e.g.][]{Kjeldsen1995, Bedding2010}:
\begin{equation}
\log g = \log g_\odot + \log \frac{\nu_\mathrm{max}}{\nu_\mathrm{max,\odot}} + \log \sqrt{\frac{T_\mathrm{eff}}{T_\mathrm{eff,\odot}}} \label{eq:seis}
\end{equation}
with $\nu_\mathrm{max,\odot} = 3090\,\mathrm{\upmu Hz}$ \citep{Huber2017} and $T_\mathrm{eff,\odot} = 5772\,\mathrm{K}$ (see Tab.~\ref{tab:solar_reference_values1}).

The difference in estimated \Teff of $-20_{-26}^{+25}\,\mathrm{K}$ and \fehatmo of $-0.02_{-0.03}^{0.03}\,\mathrm{dex}$ are both very small. Additionally, we have used the pipeline where \logg is a free parameter for the spectrum fit (``{free'') to assess the improvement of our parameter estimation thanks to the use of external information (see Fig.~\ref{fig:seis_comparison_3setups}). The ``free'' pipeline can only estimate \logg from the spectra and shows a significant scatter (especially for the red clump stars) for this stellar parameter, which propagates into larger scatter for \Teff and \feh as well. With the new constraints on \logg from external information from astrometry and photometry, the scatter of all parameters decreases significantly and the red clump stars show a tight distribution around $\log g{\sim}2.4\,\mathrm{dex}${. This is consistent with the most reliable measurements, which take into account asteroseismic information (right panels), although even finer structure within this small sample such as the separation between the red clump and the RGB bump is only seen when seismology is included.

We assess the final accuracy of \logg (and not the initial performance test described above) with all SYD-pipeline K2 GAP measurements from J.~Zinn et al. (in prep) overlapping with GALAH+~DR3. When comparing the difference of the final values for \logg estimated via Eq.~\ref{eq:seis} (seis) and GALAH+~DR3 (lbol) in Fig.~\ref{fig:all_seismic}a, we see that both the difference and the scatter of the \logg values has decreased from $-0.06\pm0.29\,\mathrm{dex}$ in GALAH~DR2 \citep[see Fig.~17 from][]{Buder2018} to $-0.04_{-0.11}^{+0.12}\,\mathrm{dex}$ on average and we see a good agreement with the majority of asteroseismic values (colourbar) for a large parameter range of $2\,\mathrm{dex}$. We note that the raw measurements of the K2 GAP overlap included 7.5\% K2 dwarf observations that were blended by giants in the K2 data, a slightly higher number than the 4\% blends found for the Kepler field \citep{Hon2019}. In the final K2 GAP sample, only a few tens of dwarfs (see high \logg stars in Fig.~\ref{fig:all_seismic}b) are likely blends.

\subsubsection{Metallicity and iron abundance}

For GALAH+~DR3, we strictly separate the notation of metallicity [M/H] and iron abundance \feh. We also refer to the atmosphere iron abundance \textsc{fe\_h\_atmo} (of the {\sc marcs} grids and {\sc sme}.feh). The last quantity is estimated mainly from Fe lines, but we also included Sc and Ti lines and thus would refer to it as pseudo-iron abundance. Only when we talk about the abundance estimated solely using Fe lines do we refer to \feh or \textsc{fe\_h}. We report the scatter of our measurements with those of the GBS as accuracy measures for both atmosphere (\textsc{fe\_h\_atmo}) and pure iron abundance \textsc{fe\_h}.

\paragraph*{GBS.}

After the collection of results from the stellar parameter estimations, we compare the atmosphere iron abundance to the values from \citet{Jofre2018a} and find a significant bias (see black errors bars in bottom panel of Fig.~\ref{fig:gbs_performance}). We have thus decided to shift the atmosphere value \textsc{sme}.feh by $+0.1\,\mathrm{dex}$ for the later abundance estimations.

From the observation of the sky flat as Solar reference, we estimate a final zero point value of $A(\mathrm{Fe})_\odot = 7.38$. This value is significantly smaller than the literature values of 7.45 and 7.50 from \citet{Grevesse2007} and \citet{Asplund2009}, respectively, and confirms that the absolute iron abundances would be estimated too low without zero point shifts. When using this value for the computation of the final \feh values, however, we find not only the Solar values, but also the GBS stars to be in agreement with the literature. We furthermore see that the scatter of this pure \feh value (black) is smaller than that of the atmosphere values (blue) in Fig.~\ref{fig:gbs_performance}c.

We note, however, that the coverage of the GBS in terms of iron abundance is very sparse. This is easily visible in the bottom panel of Fig.~\ref{fig:gbs_performance} for the iron abundances around $-1.5\,\mathrm{dex}$, but also concerns the most metal-rich stars, especially giants, for which we have to assume that the general agreement also applies.

\begin{table*}
    \centering
    \caption{Comparison of GALAH DR3 radial velocities methods with \Gaia eDR3 and APOGEE DR16. We have fitted 2 Gaussian distributions to the data and report their mean values and standard deviations as well as the their amplitude ratios (AR). Note that all methods except \texttt{rv\_obst} do not apply a gravitational redshift (GR), leading to significant shifts between these different methods. For quality cuts applied, see text.}  \label{tab:rv_comparison}

    \begin{tabular}{c|ccc|ccc|ccc}
    \hline \hline
    Method          &  \multicolumn{3}{c}{\texttt{rv\_obst} (GR)} & \multicolumn{3}{c}{\texttt{rv\_nogr\_obst} (no GR)} & \multicolumn{3}{c}{\texttt{rv\_sme\_v2} (no GR)} \\
    Component   & Narrow  & Broad & AR & Narrow & Broad & AR & Narrow & Broad & AR \\
    & [$\mathrm{km\,s^{-1}}$] & [$\mathrm{km\,s^{-1}}$] & & [$\mathrm{km\,s^{-1}}$] & [$\mathrm{km\,s^{-1}}$] & & [$\mathrm{km\,s^{-1}}$] & [$\mathrm{km\,s^{-1}}$] & \\
    \hline
    \Gaia eDR3     & $-0.16 \pm 0.45$ & $-0.46 \pm 1.13$ & 1:1.7 & $-0.02 \pm 0.48$ & $-0.13 \pm 1.22$ & 1:2.0 & $-0.09 \pm 0.45$ & $-0.22 \pm 1.18$ & 1:1.8 \\
    APOGEE DR16     &   $-0.07 \pm 0.18$ & $-0.37 \pm 0.44$ & 1:0.4 &  $-0.02 \pm 0.25$ & $-0.02 \pm 0.57$ & 1:3.7 & $-0.13 \pm 0.25$ & $-0.07 \pm 0.68$ & 1:5.3 \\
    \hline \hline
    \end{tabular}
\end{table*}

\subsubsection{Radial velocities} \label{sec:rv}

{In contrast to the approach taken in GALAH~DR2, we have estimated the radial velocities as a free parameter in the stellar parameter estimation and have thus been able to overcome a systematic trend of the reduction pipeline, which would otherwise have been overestimating the positive and underestimating the negative radial velocity by $1\%$, respectively.

{In version 2 of our data catalogues, we provide several different estimates for the community in a value-added-catalogue and add the values that we recommend to use for each spectrum in the column \texttt{rv\_galah}\footnote{We have updated this notation for convenience to allow the user to easily use our best estimate of radial velocities. In version 1 of the main catalogues, we reported only the \textsc{SME} based radial velocities, which used a erroneous barycentric correction, as we outline in this section.}.

{For each spectrum, we at least try to fit the radial velocity as part of the spectrum synthesis comparison, reported as \texttt{rv\_sme\_v2}. We have identified that a wrong barycentric correction was used for shifting the reduced spectra. The first version of the {\sc sme} radial velocity measurements (\texttt{rv\_sme\_v1}), solely based on these spectra were therefore on average lower by $0.35\pm0.19\,\mathrm{km\,s^{-1}}$, with 14 estimates shifted by more than $1\,\mathrm{km\,s^{-1}}$. We have uploaded a second version based on correct barycentric corrections.

{In addition to the \vrad provided by {\sc sme} as part of the stellar parameter pipeline, our VAC for \vrad also provides measurements which are done with the method described by \citet{Zwitter2018, Zwitter2020} via template spectra (stacked from observed GALAH spectra), rather than synthetic spectra, as well as the radial velocities reported by \Gaia eDR3 \citep{Brown2020}.

{We recommend users to consider the choice of radial velocity measurement for their specific science case. Our most accurate measurements are reported via \texttt{rv\_obst} and are recommended (if available) for Galactic stellar dynamics studies. If the user wants to compare radial velocities with other studies, we recommend the use of \texttt{rv\_nogr\_obst} (if available) or \texttt{rv\_sme\_v2} otherwise, as these do not correct for gravitational redshifts (and are currently the most commonly reported measurements by stellar spectroscopic surveys).

{To assess the accuracy of our radial velocity estimates, we therefore focus on the comparison of measurements from \texttt{rv\_nogr\_obst} and \texttt{rv\_sme\_v2} with those from \Gaia eDR3 and APOGEE DR16 \citep{Joensson2020}. Here, we limit ourselves to the unflagged GALAH spectra with sufficient signal-to-noise ratio in the second CCD, $S/N (C2) > 40$, and \Gaia eDR3 as well as APOGEE DR16 measurements with less than $2\,\mathrm{km\,s^{-1}}$ uncertainty. The differences with respect to \Gaia eDR3 and APOGEE DR16 can be best approximated with 2 Gaussian distributions, with the values listed in Table.~\ref{tab:rv_comparison}.

{For the difference of the template measurements (\texttt{rv\_nogr\_obst}) with respect to \Gaia eDR3, we find a narrow Gaussian peaking at $-0.02 \pm 0.48\,\mathrm{km\,s^{-1}}$ and a broader Gaussian (with amplitude ratio 1:2.0) at $-0.13 \pm 1.22\,\mathrm{km\,s^{-1}}$. For the stars overlapping with APOGEE DR16, the difference of \texttt{rv\_sme\_v2} and \texttt{VHELIO\_AVG} can best be described by a narrow Gaussian peaking at $-0.02 \pm 0.25\,\mathrm{km\,s^{-1}}$ and a broader Gaussian (with amplitude ratio 1:3.7) at $-0.02 \pm 0.57\,\mathrm{km\,s^{-1}}$.

{For the difference of the {\sc sme} measurements (\texttt{rv\_sme\_v2}) with respect to \Gaia eDR3, we find a narrow Gaussian peaking at $-0.09 \pm 0.45\,\mathrm{km\,s^{-1}}$ and a broader Gaussian (with amplitude ratio 1:1.8) at $-0.22 \pm 1.18\,\mathrm{km\,s^{-1}}$. For the stars overlapping with APOGEE DR16, the difference of \texttt{rv\_sme\_v2} and \texttt{VHELIO\_AVG} can best be described by a narrow Gaussian peaking at $-0.02 \pm 0.25\,\mathrm{km\,s^{-1}}$ and a broader Gaussian (with amplitude ratio 1:3.7) at $-0.02 \pm 0.57\,\mathrm{km\,s^{-1}}$.

{These comparisons confirm that \texttt{rv\_nogr\_obst} is very accurate and on the same scale as both \Gaia eDR3 and APOGEE DR16 with only $-0.02\,\mathrm{km\,s^{-1}}$ bias. The {\sc sme} measurements (\texttt{rv\_sme\_v2}) are slightly underestimated, with a bias of $-0.09$ to $-0.13\,\mathrm{km\,s^{-1}}$. We therefore add an accuracy uncertainty of $0.1\,\mathrm{km\,s^{-1}}$ to our \texttt{rv\_sme\_v2} measurements and prefer the template measurements where available.

{The source of the recommended, that is most accurate, values \texttt{rv\_galah} is indicated via the flag \texttt{use\_rv\_flag}, as we outline in the catalogue description in Sec.~\ref{sec:main_catalogue}.

\subsection{Precision of stellar parameters} \label{sec:precision_sp}

To estimate the precision of our stellar parameters, we use both internal {\sc sme} covariance errors and repeat observations of the same star for all stellar parameters except for \logg, for which we Monte Carlo sample the uncertainties.

\paragraph*{Stellar parameters except \textit{log}\,g}

We have estimated the standard deviations of repeat observations for three situations: star in the same fibre for each observation, star in different fibres and lastly all repeats irrespective of fibre. The results are shown in Fig.~\ref{fig:precision_sp}, where we plot the standard deviations standard deviations as a function of $S/N$ in CCD2\footnote{For the repeat observations we use the $S/N$ of the higher quality observation.} together with the median {\sc sme} covariance errors for the fitted stellar parameters $T_\text{eff}$, $\log g$, iron line abundance \feh, the atmosphere iron abundance \feh, rotational broadening $v_\text{broad}$, and radial velocity $v_\text{rad}$. 

% Made with GALAH_DR3/validation/repeat_observations/estimate_repeat_uncertainty.ipynb
\begin{figure*}
\centering
\includegraphics[width=0.915\textwidth]{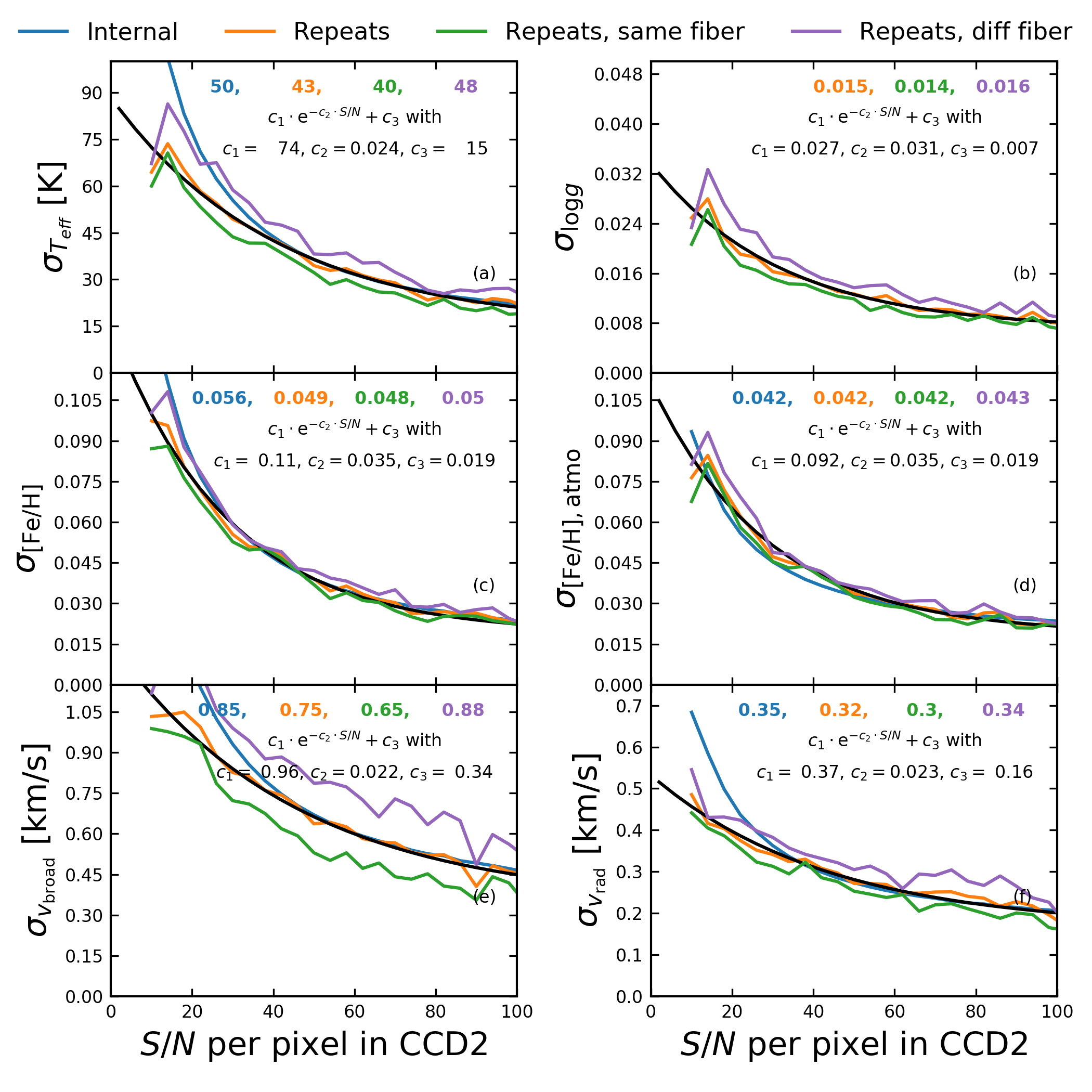}
\caption{
\textbf{Precision estimates from internal {\sc sme} covariance uncertainties (blue) as well as standard deviations from all (orange), same fibre (green), and different fibre (violet) repeat observations.} Black lines indicate the exponential fit to the orange lines. The functional form and best-fit coefficients are annotated for each panel and were used for the final assessment of precision. The numbers given across the top of each panel indicate the uncertainties estimated for $S/N = 40$ per pixel, similar to Fig. 15 from \citet{Buder2018}. Note that the internal precision was already adjusted for each parameter with the scaling relations outlined in the text. The internal precision for \logg is not given, because it was not fitted, but estimated via external information, as outlined in the text on the precision of \logg.}
\label{fig:precision_sp}
\end{figure*}

The trends of internal and repeat precision are expected to be similar, but we find that, for the stellar parameters, the uncertainties from the internal {\sc sme} covariance uncertainties, based on $\chi^2$ optimisation tend to overestimate the absolute quality of fit and are typically significantly lower than those from repeat observations, although tracing them well in a relative sense, when rescaled. As discussed when introducing the final error estimation with Eq.~\ref{eq:final_error}, the two precision estimates should be the same and we thus rescale the internal {\sc sme}-based uncertainties with a combination of factors and shifts, noted as ({factor,shift) with (3,7.5) for \Teff,(4,0.01) for \feh, (2,0.0125) for \fehatmo, (1.75,0.3) for \vbroad, and (2.0,0.15) for \vrad towards a minimum difference with respect to the exponential fit for the repeat observations (black curve in Fig.~\ref{fig:precision_sp}). For the estimation of these linear rescaling functions, we have focused on the $S/N$ interval of 40 to 200, which typically leads to larger internal uncertainties for those stars below $S/N < 20$, for which we believe a more conservative uncertainty estimate is justifiable. We use all repeats (orange lines), because we find typically a good agreement between same (green) and different (purple) fibre repeats. Only in the case of \vbroad, we see significant differences between the different repeat types, which is likely caused by unaccounted resolution variations which translate into a different broadening estimate of the same star in different fibres.

This rescaled internal precision now allows for a combination of the individual estimate of the fit quality (through the internal {\sc sme}-based uncertainty) with the general precision expected for a given $S/N$, which could otherwise be underestimated when only using to the raw internal uncertainty.

We further note that we have changed our definition of $S/N$ in these figures compared to DR2 \citep{Buder2018} to show the $S/N$ of the higher quality observation (DR3) instead of the lower one (DR2). The quantitative improvement of the precision from GALAH~DR2 to GALAH+~DR3 is thus not necessarily an indicator of the decreasing precision, but of a more reliable precision estimate.

\paragraph*{Precision of surface gravities \textit{log}\,g.}

We stress that \logg values are not optimised from the $\chi^2$-determination of the spectra like the other parameters, but from Eq.~\ref{eq:logg}. Instead of the internal {\sc sme} uncertainties, we sample the parameters used for Eq.~\ref{eq:logg} via Monte Carlo (MC) sampling. For computational reasons, we assume the uncertainties for the formula to be Gaussian and sample the parameters with uncertainties $\sigma (M) = 0.1\cdot M$, $\sigma (BC) = 0.1\,\mathrm{mag}$, $\sigma (T_\text{eff})$, $\sigma (D_\varpi)$, $\sigma (K_s)$, and $\sigma (A_{K_s})$. With this approach we estimate a mean internal uncertainty for \logg of $0.07\,\mathrm{dex}$. With this implementation, the uncertainty is driven by the mass uncertainty (contributing $0.044\,\mathrm{dex}$ for star with Solar mass) and the combination of the photometric uncertainties. For stars with precise parallaxes, the parallax uncertainty is contributing only a small fraction (the median parallax uncertainty of the sample of 2.7\% translates into roughly $0.024\,\mathrm{dex}$ uncertainty in \logg through Eqs.~\ref{eq:lbol} and \ref{eq:logg}), it is dominating the \logg uncertainty for the 5\% stars with parallax uncertainties above 20\% (see Fig.~\ref{fig:plx_snr_quality}).

By construction and due to the exquisite astrometric and photometric external information available, this internal precision is significantly better than the previous spectroscopic estimates from GALAH~DR2. We note, however, that these estimates do not take external influences like binarity or correlations of uncertainties into account.

% Created with GALAH_DR3/validation/comparisons/comparison_clusters/GALAH_DR3_Comparison_Clusters.ipynb
\begin{figure*}
\centering
\includegraphics[width=\textwidth]{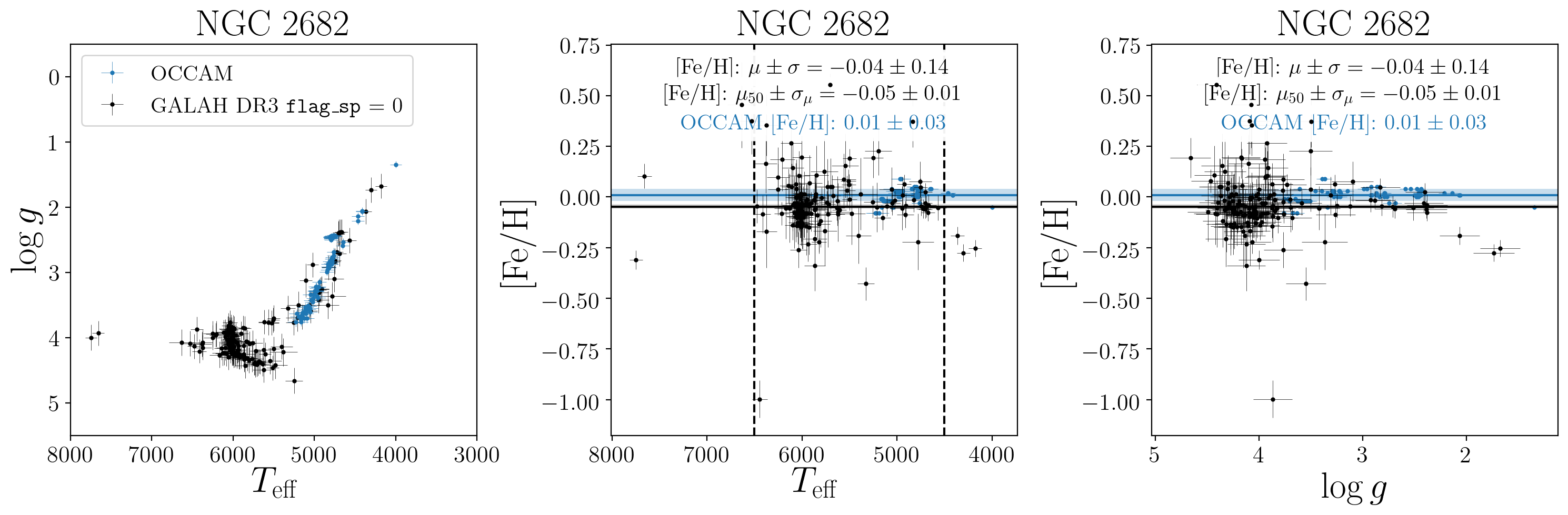}
\includegraphics[width=\textwidth]{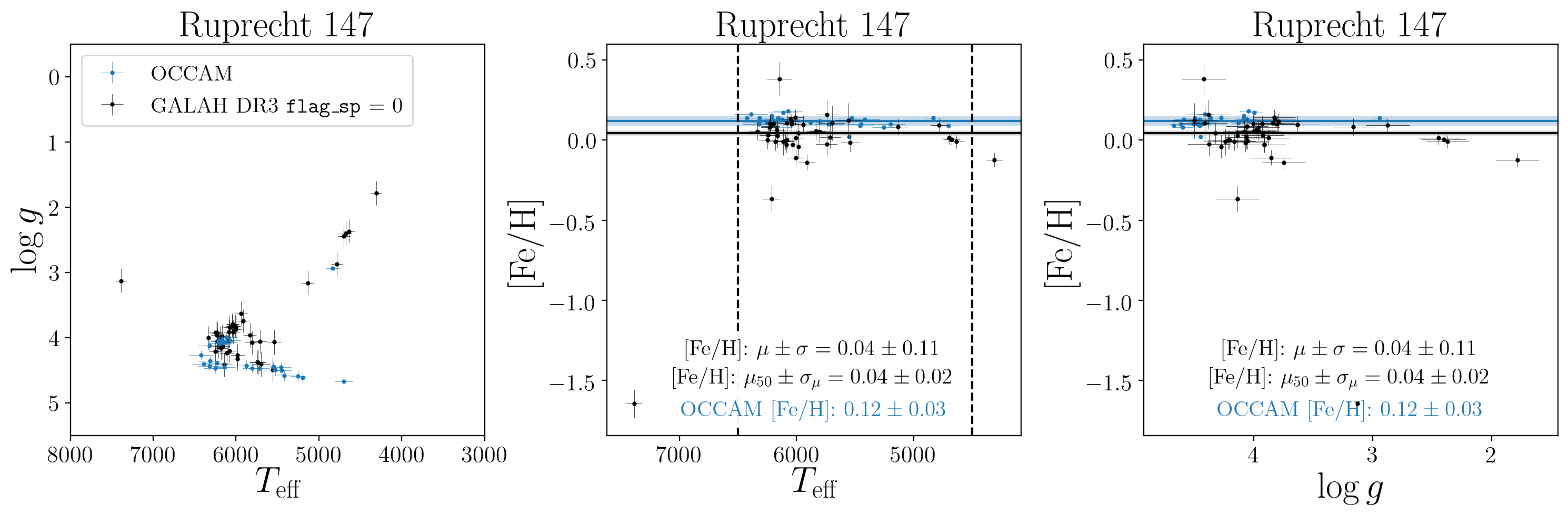}
\includegraphics[width=\textwidth]{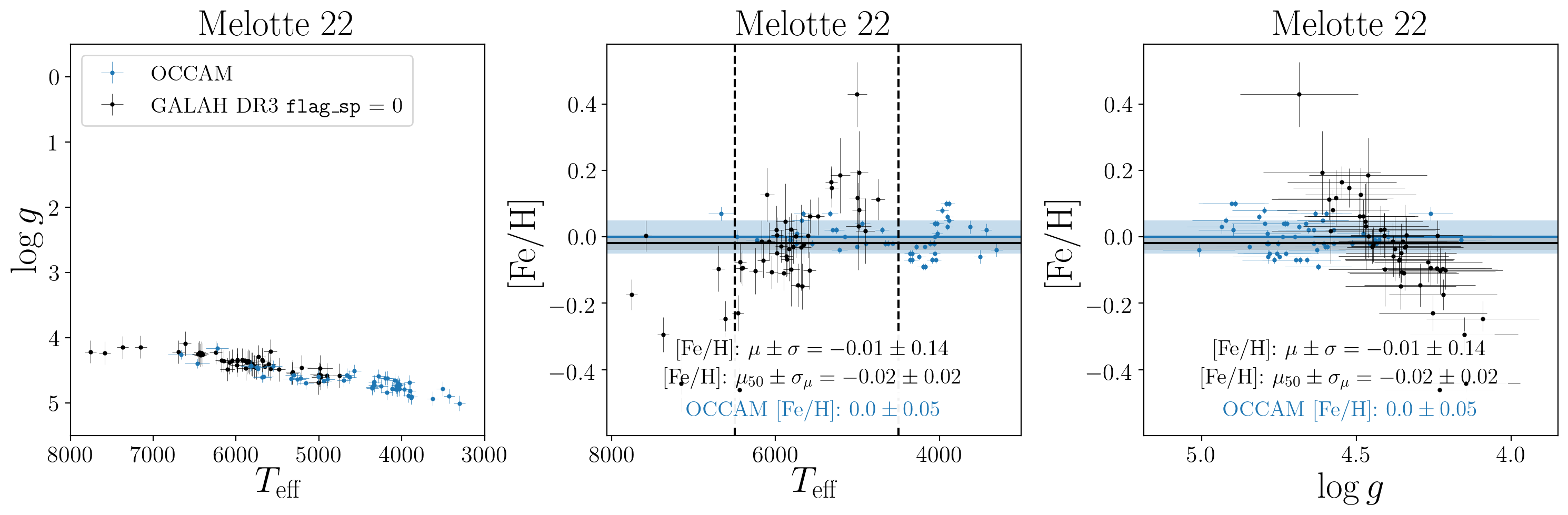}
\caption{
\textbf{Stellar parameters (combinations of \Teff, \logg, and \feh) of the three open clusters NGC~2682 (278 spectra, M~67), Ruprecht~147 (80), and Melotte~22 with data from GALAH+~DR3 (unflagged in black).} Unflagged data from the OCCAM survey \citep{Donor2020} is plotted in blue. Horizontal bars indicate the mean abundances of the clusters from GALAH in grey (estimated from unflagged measurements of for stars with $4500 < T_\mathrm{eff} < 6500\,\mathrm{K}$) and the OCCAM survey (blue).}
\label{fig:oc_stellar_params}
\end{figure*}

% Created with GALAH_DR3/validation/comparisons/comparison_clusters/GALAH_DR3_Comparison_Clusters.ipynb
\begin{figure*}
\centering
\includegraphics[width=\textwidth]{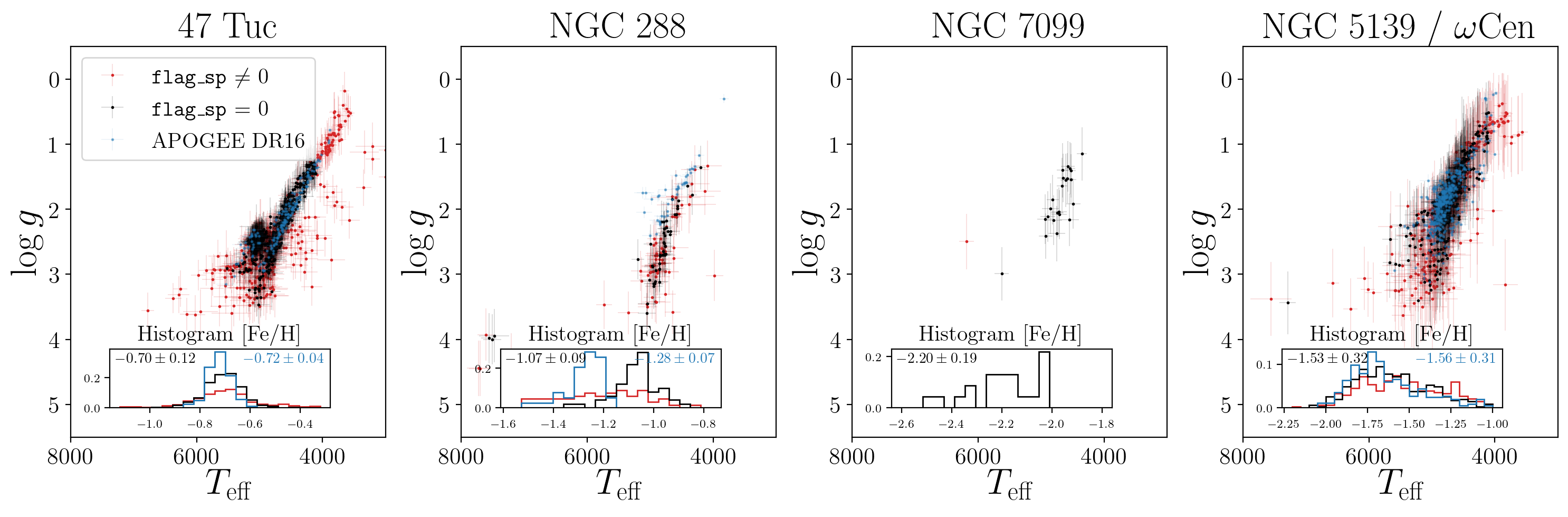}
\caption{
\textbf{Kiel diagrams (\Teff vs. \logg) for four globular clusters observed by GALAH.}
Unflagged measurements are shown in black, flagged ones in red. When available, unflagged data from APOGEE DR16 survey \citep{SDSSDR16} is plotted in blue. Inset plots in each panel indicate the normalised \feh distribution of the plotted stars with text annotating the simple mean and standard deviations of the these stars.}
\label{fig:gc_feh}
\end{figure*}

\paragraph*{Iron abundances of cluster stars.}

{Based on the open cluster membership analysis by \citet{CantatGaudin2020}, we estimate that we have intentionally and unintentionally observed members of 75 stellar clusters. The eight open clusters with most observations are NGC~2682 (278 spectra, M~67), NGC~2632 (117, M~44, Praesepe), NGC~2516 (83), NGC~2204 (81), Ruprecht~147 (80), Melotte~22 (74), Blanco~1 (67), and NGC~6253 (50). Furthermore we have observed 10 of the 128 open clusters\footnote{These are ASCC~16 (22 spectra), ASCC~16 (22), ASCC~21 (11), Berkeley~33 (8), Melotte~22 (74), NGC~2204 (81), NGC~2232 (20), NGC~2243 (8), NGC~2318 (2), NGC~2682 (278), and Ruprecht~147 (80).} of the OCCAM survey \citep{Donor2020}, included as VAC from SDSS DR16 \citep{SDSSDR16}. The analysis of all open clusters observed with GALAH is addressed in the dedicated paper by \citet{Spina2020b}{. Here we only focus on three open clusters which cover a large range of stellar evolutionary stages and are also reported by the OCCAM survey\footnote{The same plots as in Fig.~\ref{fig:oc_stellar_params} are available in our \href{https://github.com/svenbuder/GALAH_DR3/tree/master/validation/comparisons/comparison_clusters}{online documentation} for all our clusters observed by GALAH.}.

We show the coverage of evolutionary stages for these clusters in the left panels of Fig.~\ref{fig:oc_stellar_params} for both GALAH and OCCAM, which cover dwarfs for all clusters and giants for both NGC~2682 (M~67) and Ruprecht~147. When looking at the average values of \feh for these clusters as a function of \Teff (middle panels) as well as \logg (right panels), we firstly see very good agreement for the average values of Melotte~22 between GALAH and OCCAM. The \feh values for Ruprecht~147 and NGC~2682 disagree within the standard error of the mean, but agree within the standard deviation. We have limited the stars used for this averaging to those stars with $4500 < T_\mathrm{eff} < 6500\,\mathrm{K}$. We implement these cuts to avoid systematic trends on either side of the range, where either GALAH or APOGEE/OCCAM underperform{s. The parallax uncertainties of the cluster members are on average well below 12\%, suggesting that these observations should be reliable and representative for validation purposes.

\begin{figure*}
\centering
\includegraphics[width=\textwidth]{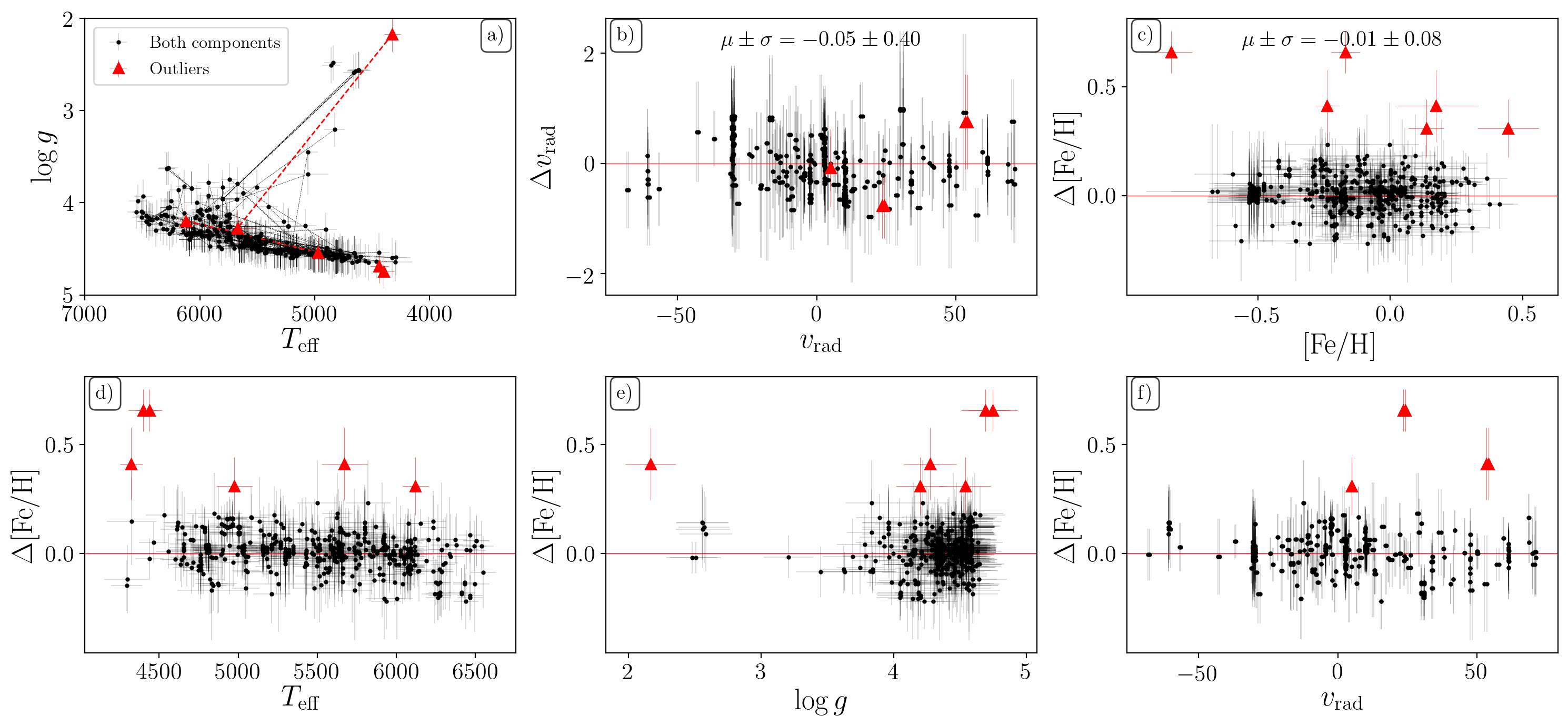}
\caption{
\textbf{Comparison of stellar parameters \Teff, \logg, \feh, and \vrad for wide binaries identified with the algorithm by \citet{ElBadry2018c}.}
We plot different combinations of these parameters as references (black) and assess the difference in \feh and \vrad. Red dots show significant outliers in $\Delta$\feh, that is, two very cool dwarfs as well as two stars with more than $1000\,\mathrm{K}$ difference in \Teff (see panel d). We include stars with \texttt{flag\_sp} up to 128.}
\label{fig:wide_binary_sp}
\end{figure*}

We also have observed several globular clusters, and plot the Kiel diagrams as well as \feh histograms for four of them, namely 47~Tuc, NGC~288, NGC~7099~(M30), and NGC~5139~($\omega$\,Cen) in the four panels of Fig.~\ref{fig:gc_feh}. For each of them we show the unflagged (black) and flagged (red) measurements from GALAH+~DR3 and where possible also the likely members observed as part of APOGEE~DR16\footnote{We have selected likely members via selecting stars within the mean cluster estimates by \citet{Baumgardt2019}.}. In particular, 47~Tuc and $\omega$\,Cen show excellent agreement between APOGEE~DR16 and GALAH+~DR3 in their mean \feh. The \feh distribution looks sharper for APOGEE in 47~Tuc due to that survey's higher $S/N$ observations and the spurious trends between \feh and \Teff, and \feh and \logg, found in the GALAH data. For $\omega$\,Cen they look similar and show a large spread in \feh. When comparing the literature compilation by \citet[][H96]{Harris1996} with our mean, standard deviation, and standard error \feh values (subsequently $\mu\pm\sigma\pm\sigma_\mu$), we obtain
\begin{description}
\item $\mathrm{[Fe/H]} = -0.70\pm0.12\pm0.01\,\mathrm{dex}$ (47~Tuc, -0.76 H96), 
\item $\mathrm{[Fe/H]} = -0.95\pm0.18\pm0.07\,\mathrm{dex}$ (NGC~6362, -1.06 H96), 
\item $\mathrm{[Fe/H]} = -1.99\pm0.28\pm0.07\,\mathrm{dex}$ (NGC~6397, -1.91 H96), 
\item $\mathrm{[Fe/H]} = -2.20\pm0.19\pm0.04\,\mathrm{dex}$ (NGC~7099, -2.12, H96), 
\item $\mathrm{[Fe/H]} = -1.53\pm0.32\pm0.02\,\mathrm{dex}$ (NGC~5139, -1.57 H96),
\item $\mathrm{[Fe/H]} = -0.97\pm0.06\pm0.02\,\mathrm{dex}$ (NGC~1851, -1.26 H96), 
\item $\mathrm{[Fe/H]} = -1.07\pm0.09\pm0.01\,\mathrm{dex}$ (NGC~288, -1.24 H96), and
\item $\mathrm{[Fe/H]} = -1.00\pm0.10\pm0.01\,\mathrm{dex}$ (NGC~362, -1.16 H96).
\end{description}
{Thus we find good agreement for high and low \feh, that is for 47~Tuc, NGC~6362, NGC~6397, NGC~7099, and NGC~5139. However, for the intermediate \feh clusters NGC~1851, NGC~288, and NGC~362 we find disagreement with \citet{Harris1996}, for which we have no explanation. While APOGEE~DR16 also has higher \feh for NGC~1851 ($-1.08\pm0.07\,\mathrm{dex}$) and NGC~362 ($-1.09\pm0.05\,\mathrm{dex}$), their \feh for NGC~288 agrees with \citet{Harris1996}. We note that the parallax uncertainties of stars in each of the three clusters is 30-40\%, which is significant and significantly higher than the uncertainties for 95\% of GALAH's targets. Taking also into account the parallax uncertainties of NGC~7099 and NGC~5139 of on average 60 and 46\% respectively, we conclude that these clusters are not suitable to reliably validate our pipeline.

\paragraph*{Stellar parameters of wide binaries.}

We used the approach by \citet{ElBadry2018c} to select wide binaries using \Gaia and further limit the selection to those with similar GALAH+~DR3 \vrad (within $1\,\mathrm{km\,s^{-1}}$). We find 268 pairs, including dwarf-giant pairs. In Fig.~\ref{fig:wide_binary_sp} we plot the stellar parameters \Teff, \logg, \feh, and \vrad to illustrate the difference of \feh and \vrad for these stars with sometimes quite different stellar parameters. We want to stress that we include also stars with \texttt{flag\_sp} up to 128 (for example the very cool, flagged stars as well as apparent photometric binary stars with unreliable \logg). As previous studies have shown \citep{ElBadry2018b, ElBadry2018c}, we expect very similar abundances for these pairs and indeed can confirm that their \feh and \vrad are consistent within the uncertainties for almost all cases. The average differences of $\Delta v_\text{rad} = -0.05 \pm 0.41\,\mathrm{km\,s^{-1}}$ and $\Delta \mathrm{[Fe/H]} = -0.01 \pm 0.08\,\mathrm{dex}$ show excellent agreement over large scales (when neglecting the 8 outliers of 268 pairs, shown in red). We furthermore do not see significant trends of the differences of \feh with \Teff, \logg, \feh, or \vrad, which lends confidence that our analysis is reliable within the stellar parameter range of the observed wide binaries. Even most of the dwarf-giant pairs show a very good agreement.

\begin{table}
\centering
 \caption{Flags used for GALAH+~DR3 to estimate the final bit-flag \texttt{flag\_sp} (stellar parameters), \texttt{flag\_X\_fe} (final reported element abundances), and  \texttt{ind\_flag\_X1234} (individual raw line/element abundances measurements) via addition of the individual bitmask flags.}
 \label{tab:flag_sp_galah_dr3}
 \begin{tabular}{rl}
  \hline \hline
		&	\texttt{flag\_sp}	\\
\hline
   1	&	\Gaia RUWE > 1.4 \\
   	&	\citep[unreliable astrometric solution, see][]{Lindegren2018b} \\
   2	&	Unreliable broadening \\
   4	&	Low S/N (below 10 for CCD 2) \\
   8	&	Reduction issues \\
	&	a) Wavelength solution (propagating of \texttt{red\_flag}), \\
	&	b) t-SNE projected reduction issues, \\
	&	c) Negative/positive fluxes, spikes, etc. \\
  16	&	t-SNE projected emission features \\
  32	&	t-SNE projected binaries \\
  64	&	Binary sequence/pre-main sequence flag \\
 128	&	SNR-dependent high {\sc sme} chi2 (bad fit) \\
 256	&	Problems with Fe: line flux is not between \\
 	&	0.03 and 1.00, \feh unreliable, or blending suspected\\
 512	&	{\sc sme} did not finish \\
     	&	a) No convergence == non-finite stellar parameters \\
   	&	b) Gaussian RV fit failed \\
1024	&	{\sc marcs} grid limit reached or \\
	&	outside of reasonable parameter range \\
  \hline
		&	\texttt{flag\_X\_fe}	\\
  \hline
  1	&	Upper Limit \\
  32	&	No reliable measurement reported \\
  \hline
		&	\texttt{ind\_flag\_X1234}	\\
  \hline
  1	&	Upper Limit \\
  2	&	Bad fit / large $\chi^2$ \\
  4	&	Uncertain measurement / saturation \\
  8	&	Bad wavelength solution / rv for Li6708 \\
  16	&	Bad stellar parameter flag (>= 128) \\
  32	&	No measurement available \\
  \hline
\end{tabular}
\end{table}

\subsection{Flagging of stellar parameters} \label{sec:flagging_sp}

After the stellar parameters have been estimated, we raise flags according to the individual criteria listed in Table~\ref{tab:flag_sp_galah_dr3}. Fig.~\ref{fig:hrd_galah_dr3} shows the derived parameter values associated with all flagged spectra. The most used flags are 8 (8.6\%), 1 (8.5\%), 256 (8.0\%), 4 (5.6\%), 512 (2.4\%), and 1024 (2.2\%). Less than 2\% of spectra have raised flags 2, 16, 32, 64, or 128.

As for GALAH~DR2 \citep{Buder2018}, we have applied the algorithm developed by \citet{Traven2017}, which combined the dimensionality reduction method t-SNE \citep{vanderMaaten2008} with the clustering algorithm \textsc{DBSCAN} \citep{Ester1996} to arrange similar looking spectra close to each other. With these techniques we have been able to identify clusters of spectra with reduction issues, emission features, as well as clear line-splitting binaries. We have further identified possible astrometric binaries or pre-main sequence stars (\texttt{flag\_sp} = 64) by selecting the oldest {\sc parsec} isochrones for the particular iron abundance of each star and selecting all stars with surface gravity lower by $\Delta \log g = 0.15$ and cooler by $\Delta T_\text{eff} = 150\,\mathrm{K}$. This selection is most effective for the identification of binaries on the secondary main-sequence (with slightly lower $\log g$). For stars with equal bolometric luminosity, for example a binary system with the same stellar parameters, the estimated $\log g$ can be smaller by up to $\sim 0.3$. This deviation can be approximated via Eq.~\ref{eq:logg} when assuming that the bolometric luminosity of the system is twice that of a single star and the mass is estimated to be that of a single star, so that $\Delta \log g{\sim}- \log L_\text{bol, binary} + \log L_\text{bol, single} = - \log \left( 2 \cdot L_\text{bol, single} \right) + \log L_\text{bol, single} = - \log 2 $. We have also identified unreliable parameter estimates for the coolest bright giants, for which unreasonably low iron abundances have been estimated (see tip of the RGB in Fig.~\ref{fig:hrd_galah_dr3}(d) with many stars with incorrectly low \feh). We elaborate on this in Sec.~\ref{sec:caveats} when addressing \vmic and metallicity trends. Based on the overall distribution of stars in the $S/N$ vs. $\chi^2$ plane (median $\chi^2 = 0.748$), we have implemented a $\chi^2$ flag (\texttt{flag\_sp} = 128) for $\chi^2 > 0.1 \cdot S/N + \exp(0.08 \cdot S/N)$

\section{Validation of element abundances} \label{sec:validation_ab}

We validate element abundances in terms of accuracy and precision  and discuss the two validations separately. Unlike for GALAH~DR2, we are now not limited by the influence of the training set for this data release{. We have also tried to estimate more upper limits and outline our approach in this section, followed by the description of our flagging algorithms with the aim to allow the community to make informed choices on the use of abundance measurements. \textbf{As for the previous data releases, we want to stress that we discourage the use of flagged element abundances without consideration of the possible systematics that the inclusion of these flagged measurements can introduce.}

{For our accuracy studies (Sec.~\ref{sec:accuracy_abundances}), the techniques we could adopt for comparisons with accurate benchmark are limited. Contrary to the stellar parameters, where multiple methods, and especially those which are independent of spectroscopy, are available for accuracy estimations, the available benchmarks for abundance accuracy are based on spectrocopy and - with exception of the Sun and Solar twins - also strongly limited in terms of accuracy (e.g. due to neglected 3D and non-LTE effects). We are therefore only using the Sun and Solar twins to validate the accuracy of our measurements to zeroth order (abundance zero points). We do not assess our abundance accuracy quantitatively beyond this, but we do also investigate systematic trends with respect to other spectroscopic studies presented in the literature. We want to stress that a proper estimation of the accuracy uncertainties would have to involve the systematic influence of the individual stellar parameters within their uncertainties, the uncertainties of the absolute abundances / zero points in terms of $\log gf$ values and additional uncertainties from the fit to the sky flat, Arcturus, the comparison with the Solar circle sample, as well as the Solar twin comparison. For computational reasons we have not been able to quantify all of these influences, but report them if possible (see e.g. Tab.~\ref{tab:solar_reference_values2}).

{Our precision estimates for abundances show a similar behaviour to that of the fitting uncertainties $e_\text{fit}(X)$ and the $S/N$-dependent repeat uncertainties $ e_\text{repeats}(X,S/N)$, as we will explore in Sec.~\ref{sec:precision_ab}.
{Our reported final uncertainties are thus limited to a formula depending on element/line X and $S/N$ of CCD2
\begin{equation}
e_\text{final}^2 (X,S/N) = \text{max} \left[ e_\text{fit}(X),e_\text{repeats}(X,S/N) \right]. \label{eq:final_error_abundances}
\end{equation}

\subsection{Accuracy of element abundances} \label{sec:accuracy_abundances}

We estimate our element abundances by changing the absolute abundance for each element that is measured in the initially scaled-Solar chemical composition by \citet{Grevesse2007} of the model atmosphere. We convert the abundances to the customary astronomical scale for logarithmic abundances and report the raw values of these measurements, A(X1234) for the 1234\,\AA~ line of element X in the \texttt{allspec} catalogue (see Sec.~\ref{sec:main_catalogue}). We subtract the Solar value A(X1234), that we define for this data release (see Tab.~\ref{tab:solar_reference_values2}), from this measurement to estimate the ratio [X/H] and for elements other than Fe, we further subtract the iron abundance to estimate the ratio [X/Fe].

In addition to the definition of the abundance zero points, we also validated the accuracy of our element abundances by comparison to measurements of the GBS and Solar twin stars, as well as members of open and globular clusters.

\paragraph*{Abundance zero points.}

Following the definition of the bracket notation, the Solar value A(X1234) should be strictly estimated from the measurement of the particular line in the Solar spectrum. For several lines within the GALAH wavelength range, however, we are facing difficulties in estimating the Solar A(X1234). Firstly, via 2df-HERMES, we can only perform sky (flat) observations rather than observing the Sun directly. Secondly, our observation setup is much shorter as for the normal setup of our observations. Thirdly, some lines are either not detectable (even within high $S/N$ spectra) or their equivalent width or line strength does not increase significantly with increasing A(X), that is, we perform a measurement at a plateau on the curve of growth. Contrary to many other studies or surveys, we choose to report the absolute abundances, and only use laboratory oscillator strengths ($\log gf$) rather than tuning these astro-physically based on the Solar spectrum with literature abundances. There are thus several solutions available to still estimate abundance zero points, which we will discuss subsequently:
\begin{enumerate}
\item Measure A(X) from the same line in Solar / sky flat / asteroid spectrum.
\item Use a different line in the Solar spectrum (because A(X) has to be the same).
\item Use another benchmark star (like Arcturus) via bridge measurements. To do this, one would measure the line that is weak in the Sun in Arcturus as well as another line of the element that is strong enough in both stars. In that case the difference in A(X) for the lines in Arcturus can be used to transfer them onto lines in the Sun.
\item Use the element abundance ratios of stars in the Solar circle, for studies suggest that the abundances should be Solar. APOGEE has followed this approach to estimate their zero points since their DR14 \citep[see][]{Holtzman2018, SDSSDR16}.
\item Compare with a literature study, e.g. via the estimates for Arcturus \citep[e.g.][]{Ramirez2011}, Solar twins \citep[e.g.][]{Bedell2018} or the overlap with a different survey, e.g. APOGEE~DR16 \citep{SDSSDR16}.
\end{enumerate}

For GALAH+~DR3 we try to use the first method whenever reliable and validate it using the other approaches. Whenever this approach was not advisable or the differences to the other methods were too significant (pointing to issues in the sky flat spectra), we used the fourth and fifth option to ensure the best overall consistency.

% This figure was created with GALAH_DR3/validation/stellar_parameters/overview/stellar_parameter_overview.ipynb
\begin{figure*}
\centering
\includegraphics[width=\textwidth]{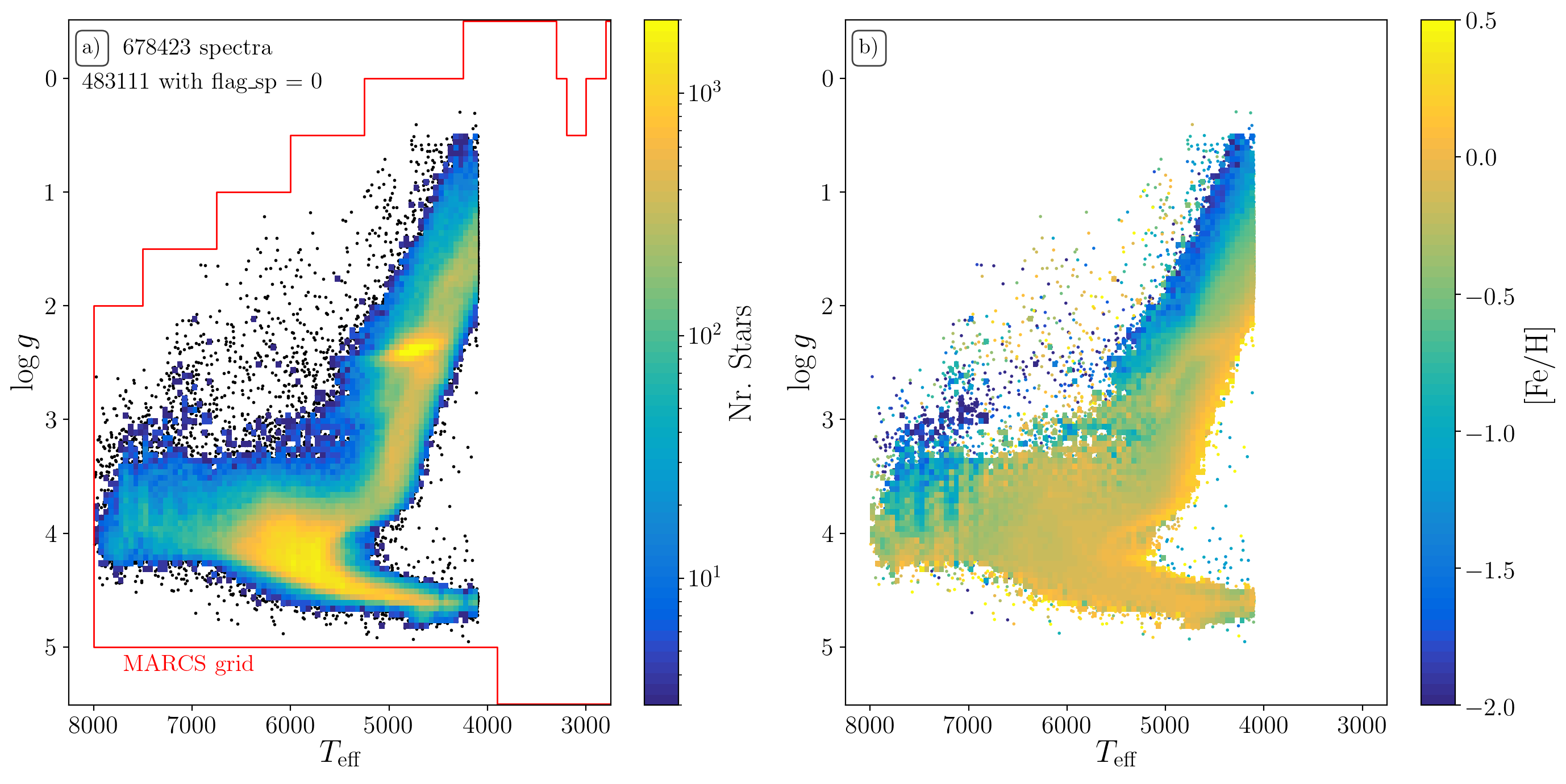}
\includegraphics[width=\textwidth]{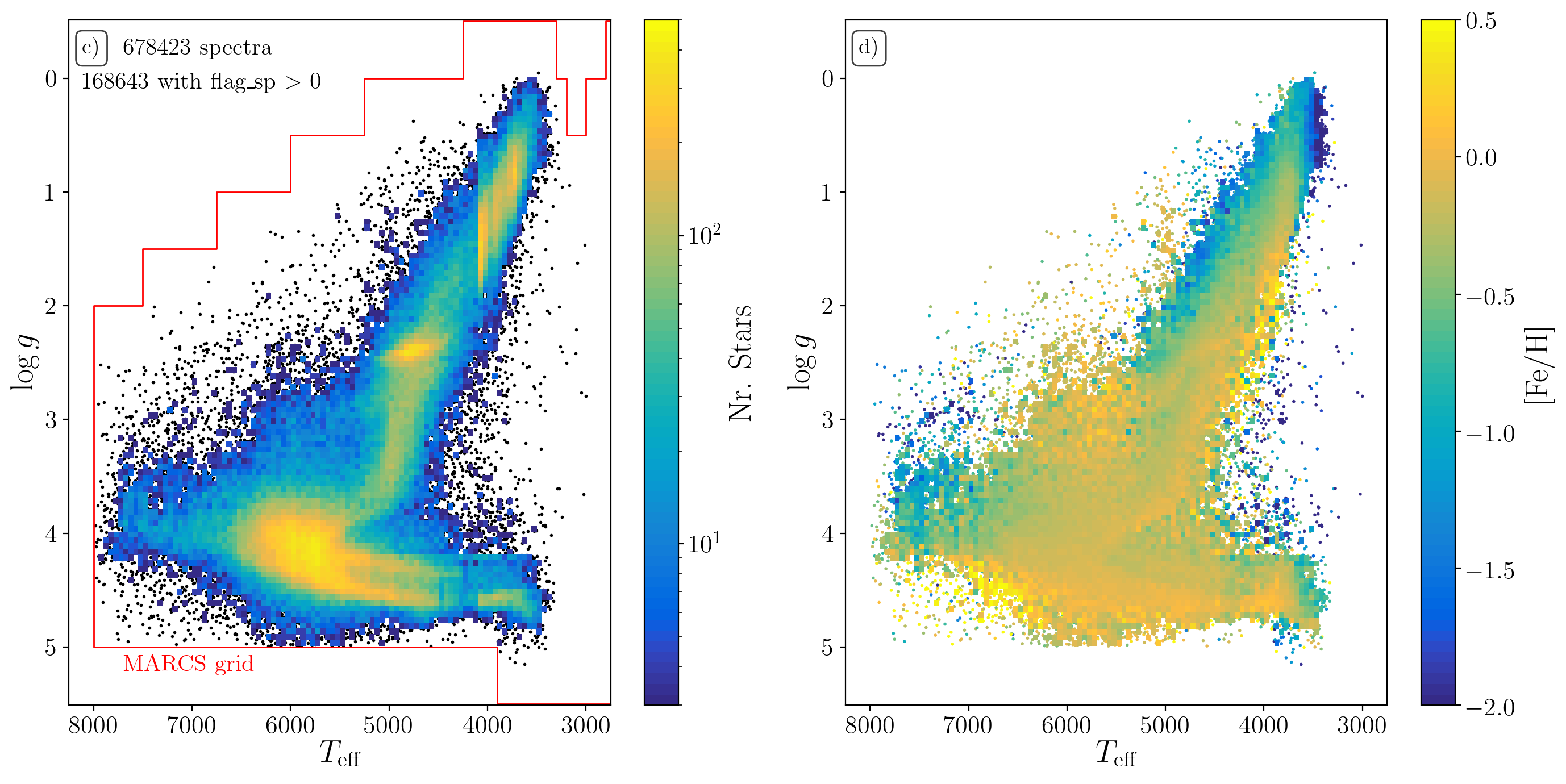}
\caption{
\textbf{Density and mean \feh distribution in Kiel diagrams ($T_\text{eff}$ vs. $\log g$) for the spectra of GALAH+DR3}.
\textbf{Panels a) and c)} show the density distribution of all unflagged and flagged spectra respectively (note that cool stars are typically flagged as unreliable due to issues with \feh as outlined in Sec.~\ref{sec:caveats}.
\textbf{Panels b) and d)} show the same distribution as panels a) and c), but here coloured by mean \feh values for each bin.
The total number of spectra and those with or without flags are annotated in the upper left corner of the left panels. In the same panels, red lines indicate the grid limit of the {\sc marcs} atmosphere grid, which marks the limits of the synthesis computations.}
\label{fig:hrd_galah_dr3}
\end{figure*}

Ultimately, we have decided to adopt the zero points for the lines and elements that are stated in Table~\ref{tab:solar_reference_values2}. For convenience we also list the values estimated by \cite{Asplund2009} to allow the identification of lines with issues such as possibly erroneous $\log gf$ values. We also list the results for the aforementioned sky flat of GALAH+~DR3 as well as the average abundances for stars in the Solar circle with near-Solar iron abundances\footnote{We select these stars via $-0.1 < \texttt{fe\_h} < 0.1$, $\texttt{r\_est} < 500$, $\texttt{snr\_c2\_iraf} > 40$, $\texttt{flag\_cannon} = 0$, $4500 < \texttt{teff} < 6500$, and for abundances of element X additionally $\texttt{flag\_X\_fe} = 0$.}. We expect that this data release will be used in combination with APOGEE~DR16 to explore the Galaxy, so we also list the values from APOGEE~DR16 for the asteroid 4 Vesta as well as the differences of the overlapping observations\footnote{We have used all x-matches, including repeats, via 2MASS IDs and then further restricted the overlap sample to stars with $\texttt{flag\_sp} = 0$, $\texttt{snr\_c2\_iraf} > 100$, $\texttt{ASPCAPFLAG} = 0$, $\texttt{SNR} > 100$, and for abundances additionally to reasonable, finite measurements ($\mathrm{[X/Fe]} > -5\,\mathrm{dex}$) of unflagged elements with $\texttt{flag\_X\_fe} = 0$ and $\texttt{X\_FE\_FLAG} = 0$.} of GALAH+~DR3 and APOGEE~DR16.

We want to stress that more work is needed to further scrutinise the line selection and abundance zero points. Due to time and computation restrictions during the implementation of the new non-LTE grids, we have only been able to run these elements combined, rather than line-by-line. However, we have found that the line-by-line analysis of element abundances is important for several elements (e.g. Al, Ca, and Ba), which are estimated from several different HERMES bands, and thus suffer from unreliable wavelength solutions in either band, and has to be done in future releases to improve the accuracy and precision of abundance measurements further.

\paragraph*{Abundances of Arcturus and other GBS stars.}

For the validation of our abundance accuracy, we also turn to the best studied giant star, Arcturus, and the \Gaia FGK benchmark stars.
For Arcturus, we use the seminal study by \citet{Ramirez2011}, which is also used by APOGEE as reference. We list the values for Arcturus in Table~\ref{tab:arcturus_reference_values}, which in general show good agreement between our measurements and those of \citet{Ramirez2011}, both performed in the optical.

For the GBS, we use the compilation by \cite{Jofre2018a} to compare average [X/Fe] differences (see Table~\ref{tab:gbs_delta_xfe}). All values suggest good agreement in light of the total median GBS uncertainties of $0.11-0.16\,\mathrm{dex}$ for each of the elements.

\begin{table}
\centering
 \caption{Average differences of GALAH+~DR3 abundances with respect to the compilation by \citet{Jofre2018a}.}
\label{tab:gbs_delta_xfe}
 \begin{tabular}{cc}
  \hline \hline
 DR3-GBS	&	16/50/84th percentile	\\
\hline
$\Delta\mathrm{[Mg/Fe]}$ & $0.03_{-0.07}^{+0.05}\,\mathrm{dex}$\\
$\Delta\mathrm{[Si/Fe]}$ & $0.03_{-0.03}^{+0.05}\,\mathrm{dex}$\\
$\Delta\mathrm{[Ca/Fe]}$ & $0.00_{-0.17}^{+0.20}\,\mathrm{dex}$\\
$\Delta\mathrm{[Ti/Fe]}$ & $0.02_{-0.11}^{+0.06}\,\mathrm{dex}$\\
$\Delta\mathrm{[Sc/Fe]}$ & $0.06_{-0.03}^{+0.08}\,\mathrm{dex}$\\
$\Delta\mathrm{[V/Fe]}$ & $0.00_{-0.10}^{+0.18}\,\mathrm{dex}$\\
$\Delta\mathrm{[Cr/Fe]}$ & $-0.03_{-0.04}^{+0.09}\,\mathrm{dex}$\\
$\Delta\mathrm{[Mn/Fe]}$ & $0.03_{-0.14}^{+0.17}\,\mathrm{dex}$\\
$\Delta\mathrm{[Co/Fe]}$ & $0.07_{-0.12}^{+0.04}\,\mathrm{dex}$\\
$\Delta\mathrm{[Ni/Fe]}$ & $0.04_{-0.03}^{+0.13}\,\mathrm{dex}$\\
  \hline
\end{tabular}
\end{table}

\paragraph*{Abundances of the Solar twins.}

We compare the abundances of Solar twins in the Solar neighbourhood in Fig.~\ref{fig:solar_twin_comparison} with the results from the studies performed by \citet{Spina2016} and \citet{Bedell2018}. We follow the definition of these studies and select high-quality Solar twin abundances with the selection of $\Delta T_\text{eff} < 100\,\mathrm{K}$, $\Delta \log g < 0.1\,\mathrm{dex}$, and $\Delta \mathrm{[Fe/H]} < 0.1\,\mathrm{dex}$ with respect to the Solar listed in Tab.~\ref{tab:solar_reference_values1}.
For such stars, these and other studies \citep[e.g.][]{Nissen2015} have found tight correlations with abundances and stellar ages, that is, chemical clocks. Given that these studies have been performed with significantly higher $S/N$ and resolution, they are useful indicators to assess our abundance zero points, if we assume that firstly the age-abundance relations they found apply to our selection (typically further away than their sample) and secondly our age estimates agree on average with theirs. For the comparisons in this section, we do not use the stellar ages estimated as part of the VAC, but the ones calculated on-the-fly by the spectroscopic analysis pipeline. For the comparison in Fig.~\ref{fig:solar_twin_comparison}, we shift the age scale by the difference of our Solar age and the $1.26\,\mathrm{Gyr}$ lower one reported by \cite{Bonanno2002}. We plot the age-abundances distribution of the Solar twins from GALAH+~DR3 in Fig.~\ref{fig:solar_twin_comparison} together with the fitted relations from \citet{Bedell2018} and state the mean difference between these curves and our data for each panel. We see good good agreement in the plots for O, Na, Na, Si, Ca, Sc, Ti, \ion{Ti}{II}, Mn, Zn, Y, and Ba, thus confirming our abundance zero points. We see some smaller differences for Mg, Cr, Ni, Cu{. For Ni and Cu, the difference is still well within the uncertainties. We are unsure of the origin of the larger differences for Mg and Cr; especially since our other zero point comparisons (Table~\ref{tab:solar_reference_values2}) show agreement to better than $0.05\,\mathrm{dex}$. Systematic offsets could, however, be introduced by the different line selections and information (for Mg and Cr for example, only one line is in common between each analysis). For C and V the data are inconclusive.

% Based on the outcome of GALAH_DR3/validation/comparisons/comparison_solar_twins/solar_twin_comparisons.ipynb
\begin{figure*}
\centering
\includegraphics[width=\textwidth]{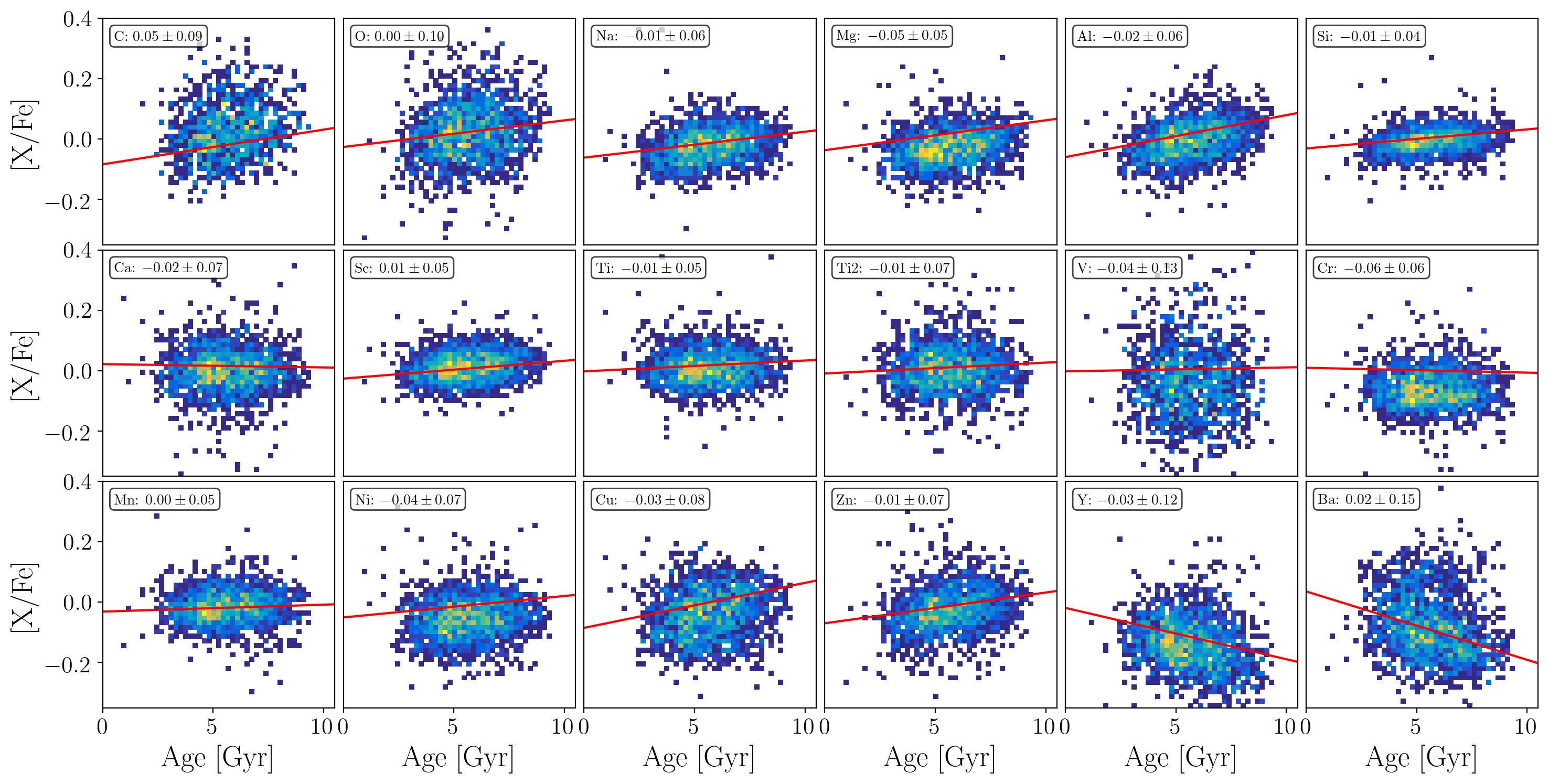}
\caption{
\textbf{Chemical abundances $\mathrm{[X/Fe]}$ of solar twin stars as a function of on-the-fly computed stellar age}
Ages are shifted by $-1.26\,\mathrm{Gyr}$. We note that this age is different from the one reported in the VAC on stellar ages. We over plot in red the functions calculated by \citet{Spina2016} and \citet{Bedell2018}.}
\label{fig:solar_twin_comparison}
\end{figure*}

\paragraph*{Abundances of the cluster stars.}

Similar to GALAH~DR2, we could assess the element abundances for selected clusters with numerous observed members. These are, however, not useful in a straight forward manner to estimate the accuracy and precision of our measurements, due to internal processes like atomic diffusion and dredge-up changing the observed photospheric abundances for different evolutionary stages for open clusters \citep[see e.g.][]{Gao2018, BertelliMotta2018, Souto2018, Souto2019} as well as the presence of multiple populations in several globular clusters, leading to the spread in metallicities \citep[see e.g.][]{Carretta2009b} and anti-correlations in several elements, like Na-O or Mg-Al \citep[see e.g.][]{Carretta2009, DOrazi2010}.

{Since we lack good calibrators for these abundances, we choose to include overviews of several abundances for open clusters, as we expect the evolutionary effect within these to be seen, but predictable. Significant differences in abundances above the expected evolutionary effects are therefore indicative of systematic trends within our analysis. A more detailed analysis of abundances and trends in open and globular clusters will be performed in the studies by \citet{Spina2020b} and D.~M.~Nataf et al. (in prep.) respectively, but we make the overview plots available in our online documentation. For the vast majority, our trends agree with the literature values, change{such as} the OCCAM survey \citep{Donor2020}, as shown in Fig.~\ref{fig:oc_abundances}, where we plot the abundances of Si, Cr, Cu, and Ba for a selection of open clusters.

% Created with GALAH_DR3/validation/comparisons/comparison_clusters/GALAH_DR3_Comparison_Clusters.ipynb
\begin{figure*}
\centering
\includegraphics[width=0.95\textwidth]{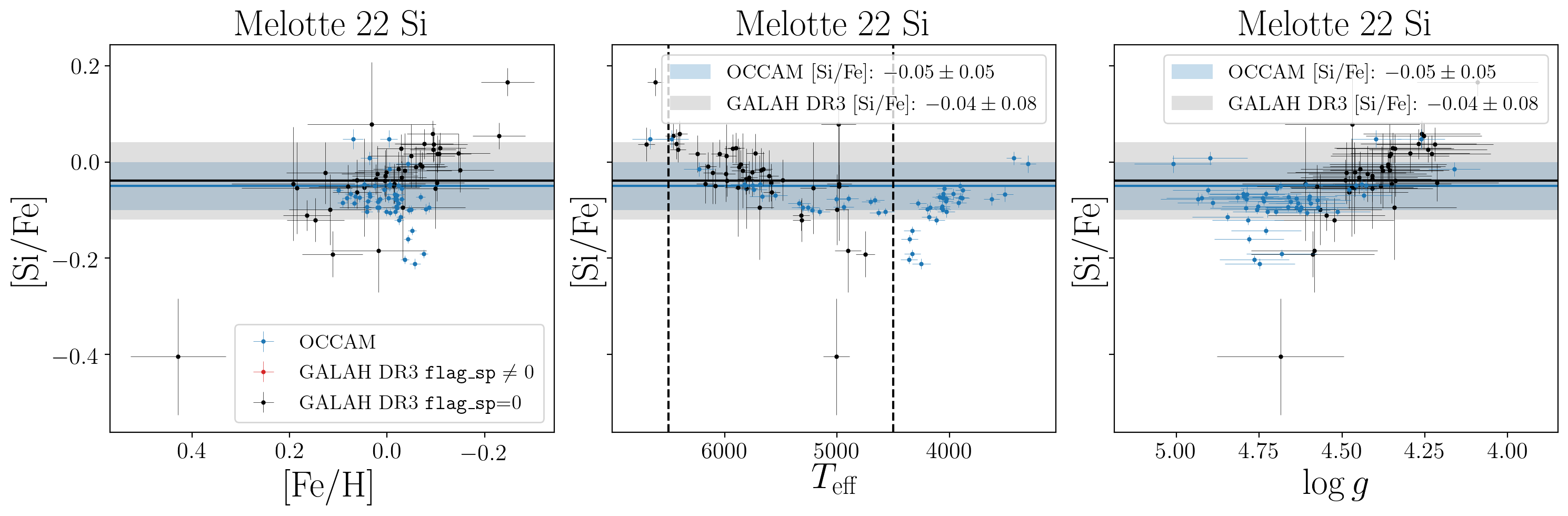}
\includegraphics[width=0.95\textwidth]{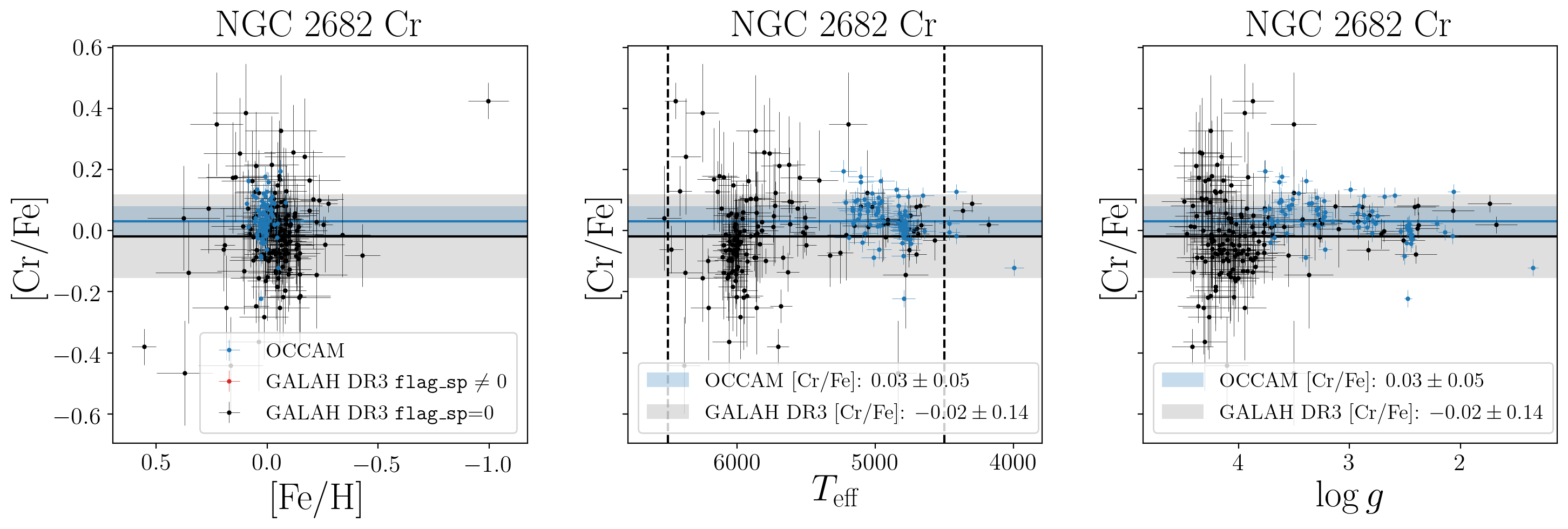}
\includegraphics[width=0.95\textwidth]{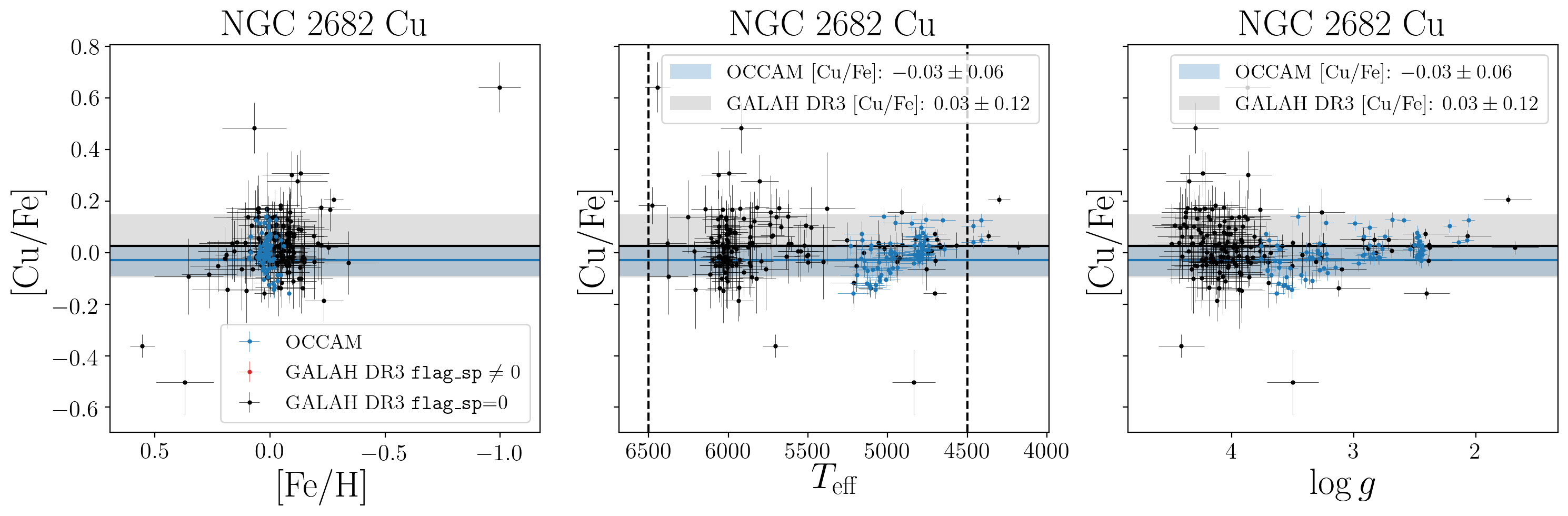}
\includegraphics[width=0.95\textwidth]{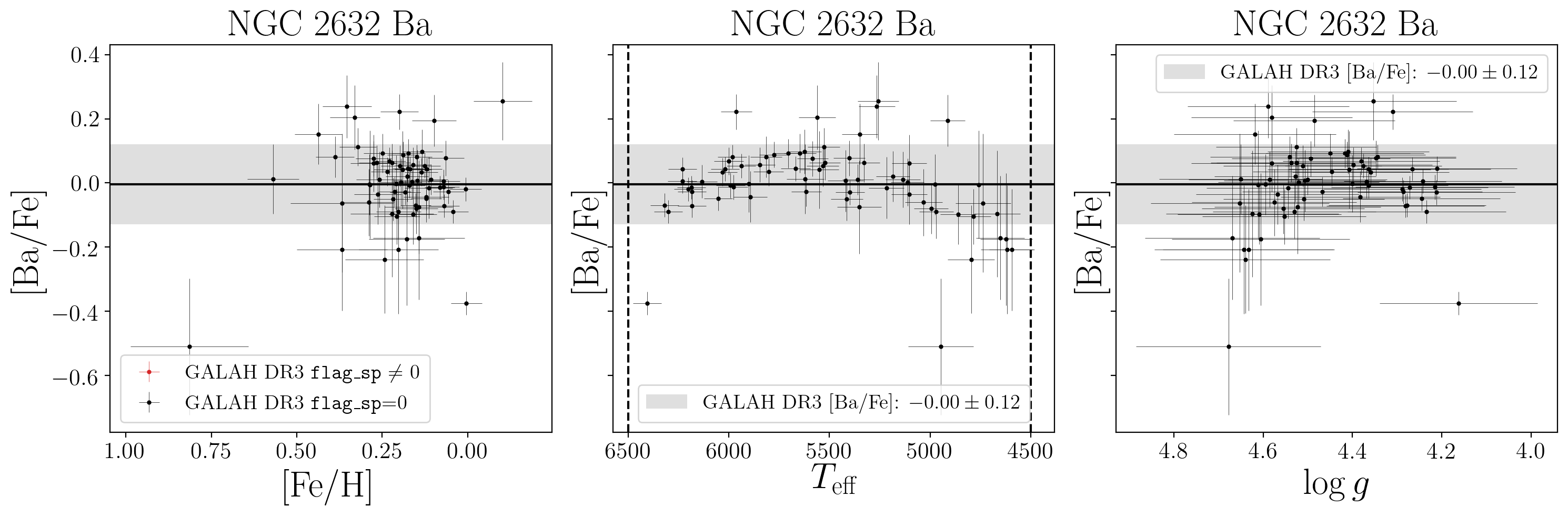}
\caption{
\textbf{Element abundances [X/Fe] as a function of stellar parameters \feh, \Teff, and \logg for a selection of elements X from the four clusters with information from both GALAH+~DR3 (unflagged in black, flagged in red) and the OCCAM survey \citep[unflagged data, blue][]{Donor2020} in Fig.~\ref{fig:oc_stellar_params}.} Horizontal bars indicate the mean abundances of the clusters from GALAH in grey (estimated from unflagged measurements of for stars with $4500 < T_\mathrm{eff} < 6500\,\mathrm{K}$) and the OCCAM survey (blue).}
\label{fig:oc_abundances}
\end{figure*}

{Some of the differences that we find between dwarfs and giants in two of the open clusters with many members, namely M~67 (NGC~2682) and Ruprecht~147, are however significantly higher than would be expected from such evolutionary effects. We list the 16/50/84th percentiles for both clusters separated into dwarfs ($T_\text{eff} \geq 5500\,\mathrm{K}$ or $\log g \geq 3.5\,\mathrm{dex}$) and giants ($T_\text{eff} < 5500\,\mathrm{K}$ and $\log g < 3.5\,\mathrm{dex}$) in Table~\ref{tab:m67_ab}. While for M~67 the agreement seems quite good for most elements in view of the average precision uncertainties, we find significant differences between dwarfs and giants for Al, Ni, Zn, Ba, and La for both clusters{. We also find significant differences in at least one cluster for O, Na, Si, \ion{Ti}{II}, V, Zr, Ce, Nd, and Sm. We will elaborate more on these findings in Sec~\ref{sec:pecularities_groups}, where we describe caveats of abundances for Solar and metal-rich giants, especially for the elements Al, \ion{Ti}{II}, Ni, and Ba, which seem to show in general elevated values for Solar and metal-rich giants.

\begin{table}
\centering
\caption{Comparison of GALAH DR3 element abundances of dwarfs and giants in M67 and Ruprecht 147. Values are only listed if more than 5 pairs with unflagged [X/Fe] were available.} \label{tab:m67_ab}
\begin{tabular}{ccccc}
\hline
 & \multicolumn{2}{c}{M67} & \multicolumn{2}{c}{Ruprecht 147} \\
Elem. & Dwarfs & Giants & Dwarfs & Giants \\
\hline
{[Fe/H]} & $-0.05_{-0.08}^{+0.11}$ & $-0.05_{-0.09}^{+0.06}$ & $0.04_{-0.07}^{+0.07}$ & $0.01_{-0.04}^{+0.08}$ \\
{[$\alpha$/Fe]} & $0.01_{-0.05}^{+0.07}$ & $0.01_{-0.05}^{+0.04}$ & $-0.01_{-0.03}^{+0.03}$ & $0.02_{-0.05}^{+0.02}$ \\
{[Li/Fe]} & $1.38_{-0.28}^{+0.21}$ & - & $1.26_{-0.27}^{+0.22}$ & - \\
{[C/Fe]} & $0.06_{-0.13}^{+0.11}$ & - & $-0.00_{-0.09}^{+0.12}$ & - \\
{[O/Fe]} & $0.05_{-0.15}^{+0.14}$ & $0.12_{-0.14}^{+0.14}$ & $-0.00_{-0.10}^{+0.15}$ & $0.19_{-0.17}^{+0.05}$ \\
{[Na/Fe]} & $0.06_{-0.11}^{+0.07}$ & $0.06_{-0.13}^{+0.12}$ & $0.09_{-0.05}^{+0.05}$ & $0.17_{-0.05}^{+0.05}$ \\
{[Mg/Fe]} & $-0.02_{-0.08}^{+0.10}$ & $-0.05_{-0.02}^{+0.12}$ & $-0.04_{-0.04}^{+0.04}$ & $-0.02_{-0.04}^{+0.03}$ \\
{[Al/Fe]} & $0.01_{-0.11}^{+0.10}$ & $0.17_{-0.07}^{+0.05}$ & $-0.00_{-0.03}^{+0.06}$ & $0.19_{-0.03}^{+0.01}$ \\
{[Si/Fe]} & $-0.00_{-0.06}^{+0.07}$ & $0.21_{-0.19}^{+0.18}$ & $-0.01_{-0.03}^{+0.03}$ & $0.07_{-0.08}^{+0.02}$ \\
{[K/Fe]} & $-0.01_{-0.11}^{+0.14}$ & $-0.00_{-0.14}^{+0.14}$ & $0.02_{-0.07}^{+0.04}$ & $-0.07_{-0.07}^{+0.04}$ \\
{[Ca/Fe]} & $0.05_{-0.08}^{+0.09}$ & $0.04_{-0.09}^{+0.10}$ & $0.08_{-0.06}^{+0.06}$ & $0.14_{-0.06}^{+0.01}$ \\
{[Sc/Fe]} & $0.04_{-0.07}^{+0.06}$ & $-0.02_{-0.06}^{+0.05}$ & $0.07_{-0.08}^{+0.05}$ & $-0.04_{-0.01}^{+0.02}$ \\
{[Ti/Fe]} & $0.04_{-0.08}^{+0.12}$ & $-0.01_{-0.04}^{+0.04}$ & $-0.01_{-0.04}^{+0.05}$ & $-0.00_{-0.03}^{+0.01}$ \\
{[Ti2/Fe]} & $0.02_{-0.10}^{+0.12}$ & $0.05_{-0.06}^{+0.06}$ & $-0.02_{-0.08}^{+0.07}$ & $0.15_{-0.03}^{+0.03}$ \\
{[V/Fe]} & $0.05_{-0.17}^{+0.24}$ & $0.17_{-0.11}^{+0.22}$ & $-0.06_{-0.10}^{+0.19}$ & - \\
{[Cr/Fe]} & $-0.03_{-0.10}^{+0.14}$ & $0.02_{-0.05}^{+0.06}$ & $-0.08_{-0.03}^{+0.08}$ & $0.00_{-0.02}^{+0.02}$ \\
{[Mn/Fe]} & $0.00_{-0.07}^{+0.09}$ & $0.05_{-0.08}^{+0.09}$ & $0.01_{-0.03}^{+0.04}$ & $0.06_{-0.07}^{+0.03}$ \\
{[Ni/Fe]} & $0.01_{-0.10}^{+0.12}$ & $0.09_{-0.07}^{+0.06}$ & $-0.08_{-0.04}^{+0.07}$ & $0.10_{-0.06}^{+0.02}$ \\
{[Cu/Fe]} & $0.02_{-0.08}^{+0.11}$ & $0.02_{-0.05}^{+0.04}$ & $0.03_{-0.08}^{+0.03}$ & $0.01_{-0.03}^{+0.04}$ \\
{[Zn/Fe]} & $0.06_{-0.11}^{+0.13}$ & $-0.11_{-0.12}^{+0.12}$ & $0.03_{-0.06}^{+0.06}$ & $-0.15_{-0.07}^{+0.03}$ \\
{[Rb/Fe]} & - & - & - & - \\
{[Sr/Fe]} & - & - & - & - \\
{[Y/Fe]} & $0.01_{-0.15}^{+0.14}$ & $0.01_{-0.20}^{+0.25}$ & $0.07_{-0.16}^{+0.06}$ & $0.09_{-0.04}^{+0.03}$ \\
{[Zr/Fe]} & $0.73_{-0.51}^{+0.66}$ & $-0.02_{-0.07}^{+0.08}$ & - & $-0.02_{-0.09}^{+0.04}$ \\
{[Mo/Fe]} & - & - & - & - \\
{[Ru/Fe]} & - & $0.17_{-0.09}^{+0.60}$ & - & - \\
{[Ba/Fe]} & $0.03_{-0.15}^{+0.10}$ & $0.17_{-0.15}^{+0.14}$ & $0.05_{-0.14}^{+0.10}$ & $0.35_{-0.07}^{+0.05}$ \\
{[La/Fe]} & $0.47_{-0.30}^{+0.37}$ & $-0.02_{-0.09}^{+0.13}$ & $0.22_{-0.07}^{+0.34}$ & $-0.08_{-0.02}^{+0.05}$ \\
{[Ce/Fe]} & $0.14_{-0.25}^{+0.44}$ & $-0.07_{-0.09}^{+0.13}$ & $-0.04_{-0.16}^{+0.10}$ & $-0.05_{-0.03}^{+0.05}$ \\
{[Nd/Fe]} & $0.46_{-0.17}^{+0.45}$ & $0.16_{-0.09}^{+0.14}$ & - & $0.10_{-0.05}^{+0.05}$ \\
{[Sm/Fe]} & $0.37_{-0.31}^{+0.62}$ & $-0.20_{-0.13}^{+0.57}$ & - & - \\
{[Eu/Fe]} & - & $-0.01_{-0.08}^{+0.09}$ & - & - \\
\hline
\end{tabular}
\end{table}

\paragraph*{Element abundances of wide binaries.}

We use wide binaries from GALAH, selected using the algorithms presented in \citet{ElBadry2018c}, for the validation of our elemental abundances, in the same way as described earlier for \feh. We plot the difference in element abundances of the two components for different nucleosynthesis channels in Fig.~\ref{fig:wide_binary_ab}. For this comparison, we limit ourselves to those stars with similar \feh (within $0.25\,\mathrm{dex}$) and similar \vrad (within $1\,\mathrm{km\,s^{-1}}${), and no raised stellar parameter flags. The average differences of these stars, which are believed to (on average) share very similar composition, are typically small as a function of \feh, \Teff, and \logg, as shown in the left, middle and right panels for the $\upalpha$-enhancement, O, Na, Si, Mn, and Ba, confirming that our analysis works rather well for stars with similar astrometric information.

For completeness, we list the average differences of all analysed elements in Tab.~\ref{tab:wide_binary_ab} together with their average quoted uncertainties from GALAH+~DR3. If our assumptions of equal chemical compositions among the binary stars are correct, one would expect no bias between their measured element abundances and a scatter corresponding to the measurements uncertainty. For many of these elements, we see an overall small average difference between the binary components among the elements for which we had enough abundance estimates of both components. Only for O, V, Zr, La, Ce, and Sm we see a disagreement above $0.05\,\mathrm{dex}$. For these elements, however, we note that the scatter is larger compared to the other elements suggesting less precise measurements. This is confirmed by the larger precision uncertainties, confirming that our precision estimates are reliable, although not always to scale. While the scatter of the binary differences and the average reported uncertainties agree roughly for most elements, they differ significantly for Li,  O, \ion{Ti}{II}, V, La, Ce, and Sm, suggesting that our uncertainties due to precision are underestimated for these particular elements or accuracy uncertainties play an important role, which is neglected in our analysis. We can also not exclude the possibility that there are actual differences in the abundances of the wide binaries, which is expected at least for Li for binary components at different \Teff.

% Created with GALAH_DR3/validation/comparisons/comparison_wide_binaries/GALAH_DR3_wide_binaries.ipynb
\begin{figure*}
\centering
\includegraphics[width=\textwidth]{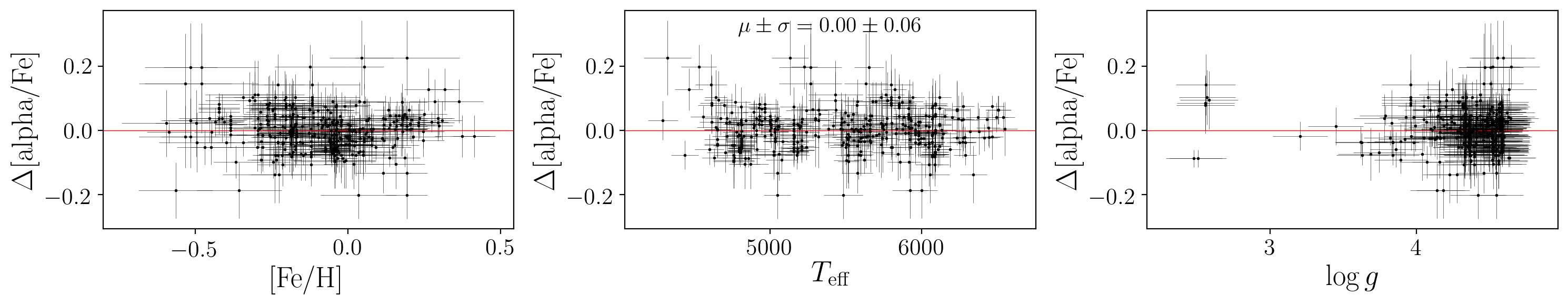}
\includegraphics[width=\textwidth]{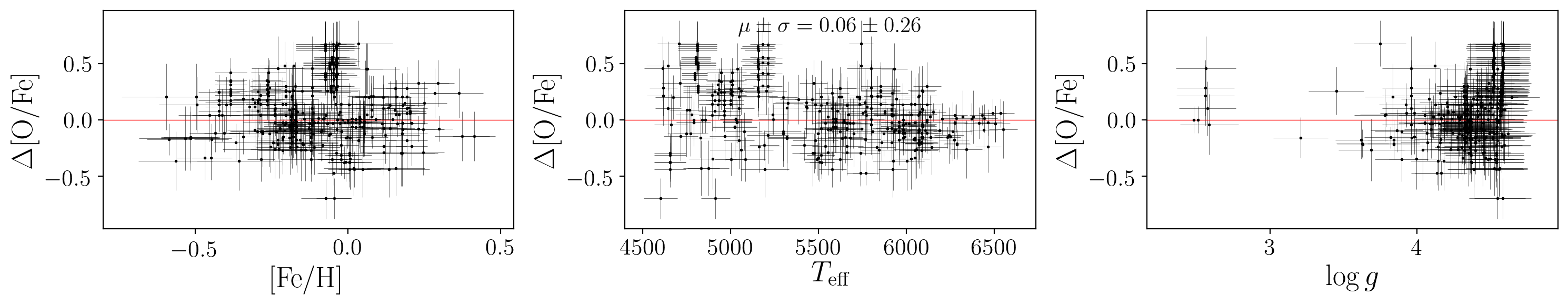}
\includegraphics[width=\textwidth]{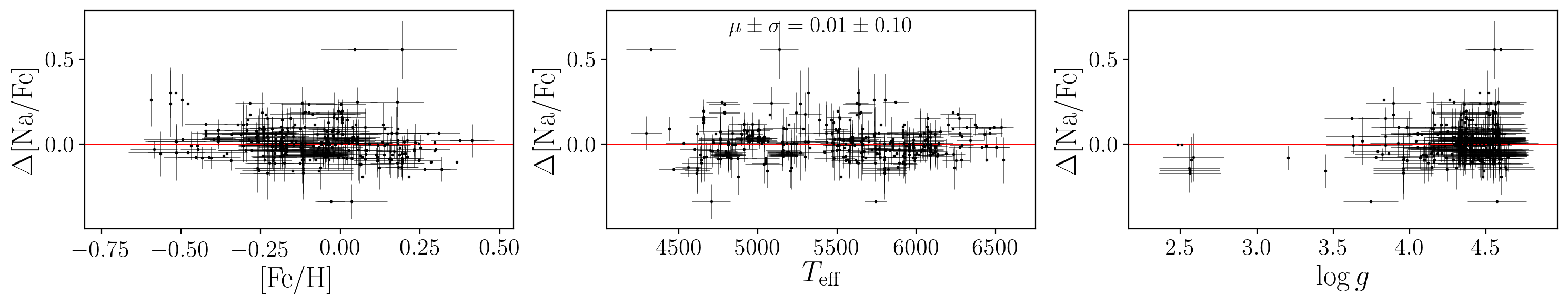}
\includegraphics[width=\textwidth]{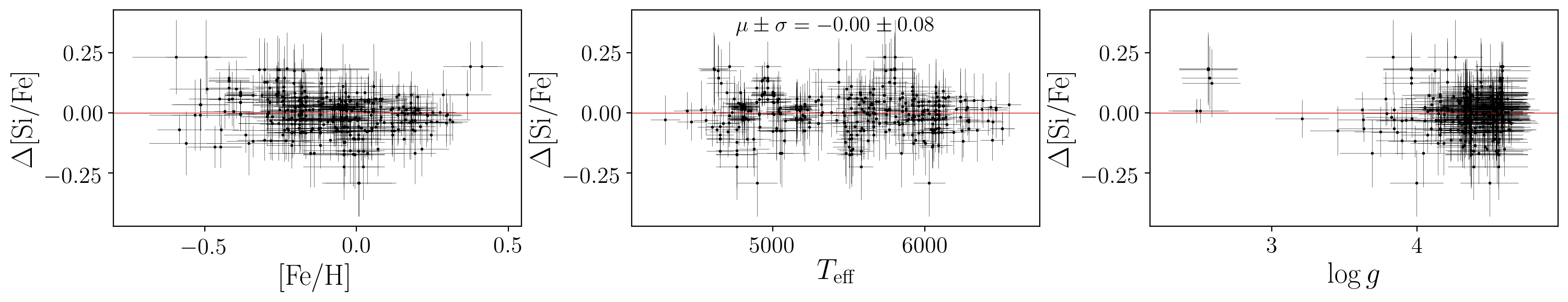}
\includegraphics[width=\textwidth]{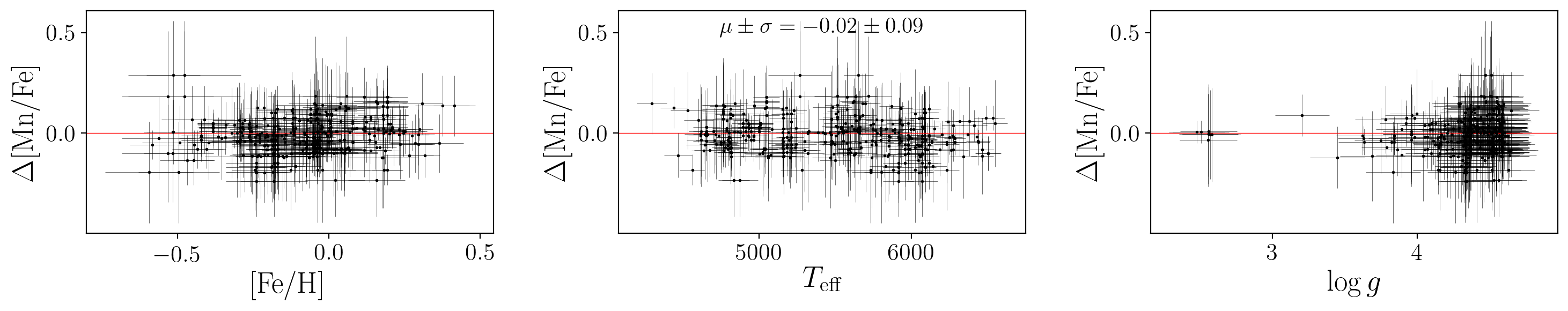}
\includegraphics[width=\textwidth]{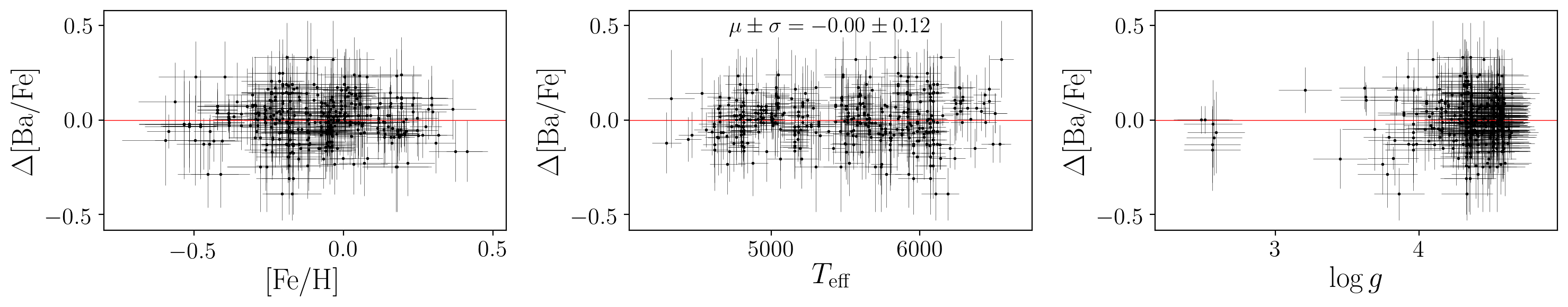}
\caption{
\textbf{Comparison of unflagged element abundances for wide binaries for the selected elements alpha, O, Na, Si, Mn, and Ba. Pairs were identified with the algorithm by \citet{ElBadry2018c}.}}
\label{fig:wide_binary_ab}
\end{figure*}

% Created with GALAH_DR3/validation/comparisons/comparison_wide_binaries/GALAH_DR3_wide_binaries.ipynb
\begin{table}
\centering
\caption{Comparison of element abundances of wide binaries in GALAH DR3. Values are only listed if more than 5 pairs with unflagged [X/Fe] were available.} \label{tab:wide_binary_ab}
\begin{tabular}{ccc}
\hline
Elem. & 16/50/84th perc. & Avg. \texttt{e\_X\_fe} \\
\hline
$\alpha$ & $-0.00_{-0.06}^{+0.06}$ & 0.04 \\
Li & $-0.10_{-0.04}^{+0.27}$ & 0.08 \\
C & $0.00_{-0.11}^{+0.09}$ & 0.10 \\
O & $-0.02_{-0.29}^{+0.20}$ & 0.13 \\
Na & $0.00_{-0.09}^{+0.07}$ & 0.06 \\
Mg & $0.03_{-0.11}^{+0.08}$ & 0.09 \\
Al & $-0.01_{-0.08}^{+0.08}$ & 0.07 \\
Si & $-0.00_{-0.06}^{+0.08}$ & 0.06 \\
K & $-0.03_{-0.12}^{+0.12}$ & 0.09 \\
Ca & $-0.01_{-0.09}^{+0.09}$ & 0.08 \\
Sc & $0.02_{-0.09}^{+0.07}$ & 0.06 \\
Ti & $0.00_{-0.08}^{+0.08}$ & 0.08 \\
Ti2 & $0.02_{-0.14}^{+0.13}$ & 0.08 \\
V & $-0.05_{-0.22}^{+0.23}$ & 0.11 \\
Cr & $0.01_{-0.08}^{+0.07}$ & 0.08 \\
Mn & $0.02_{-0.09}^{+0.08}$ & 0.09 \\
Ni & $-0.01_{-0.10}^{+0.10}$ & 0.08 \\
Cu & $0.01_{-0.08}^{+0.07}$ & 0.06 \\
Zn & $0.00_{-0.11}^{+0.13}$ & 0.11 \\
Rb & $-$ & - \\
Sr & $-$ & - \\
Y & $0.03_{-0.19}^{+0.20}$ & 0.15 \\
Zr & $0.10_{-0.15}^{+0.13}$ & 0.10 \\
Mo & $-$ & - \\
Ru & $-$ & - \\
Ba & $0.01_{-0.12}^{+0.10}$ & 0.09 \\
La & $-0.04_{-0.27}^{+0.33}$ & 0.11 \\
Ce & $-0.11_{-0.10}^{+0.22}$ & 0.11 \\
Nd & $-$ & - \\
Sm & $-0.19_{-0.16}^{+0.18}$ & 0.13 \\
Eu & $-$ & - \\
\hline
\end{tabular}
\end{table}

\subsection{Precision of element abundances} \label{sec:precision_ab}

We assess the precision of our element abundances by comparing the internal {\sc sme} covariance uncertainties with those from repeat observations of the same star in Fig.~\ref{fig:repeat_uncertainties_abundance_selection}. In contrast with the case for the stellar parameter estimation, we see that the covariance errors from the individual line measurements are typically in good agreement for almost all lines. The standard deviations of the measurements are also consistent irrespective of the fibre combination. 

We note, however, that our final estimates of the internal {\sc sme}-based uncertainties are lower than the those from the repeat observations, when a large number of lines is fitted in combination rather than line-by-line{, in particular Fe, which we discussed in Sec.~\ref{sec:precision_sp}. This suggests that either the {\sc sme}-internal method has problems in estimating realistic errors when many pixels are involved, or that our spectrum quality indicator (\texttt{snr\_c2\_iraf}) is not representative in those cases. With the exeption of Fe, we have however managed to estimate abundances of elements with many lines in a line-by-line manner, such that the fitting and repeat uncertainty show a similar behaviour. Unlike for the case of the stellar parameter estimation, we only report the final abundance uncertainty based on the maximum of the raw internal covariance error for the abundance fit and the $S/N$-scaled repeat observation uncertainty.

\begin{figure*}
\centering
\includegraphics[width=\textwidth]{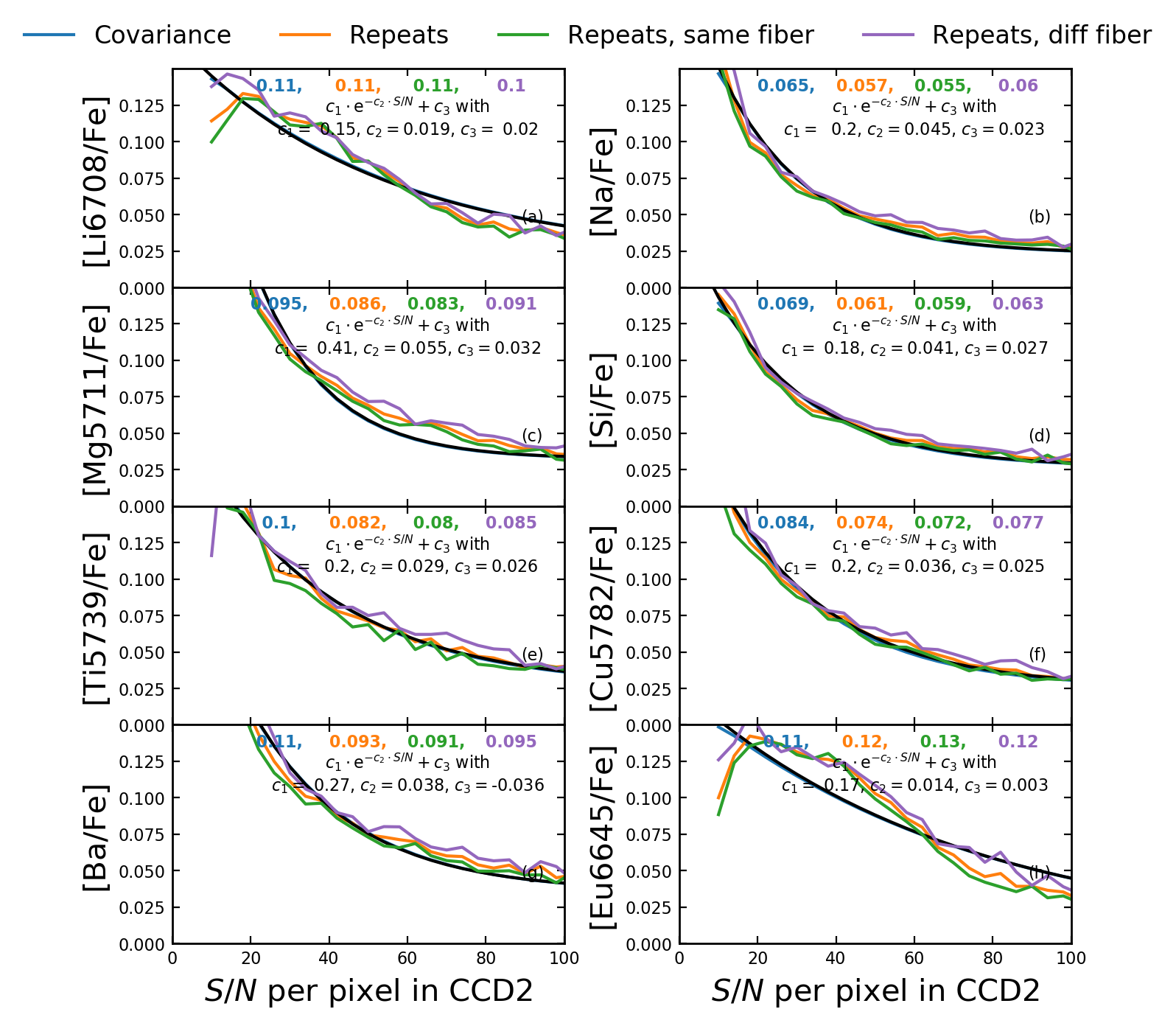}
\caption{
\textbf{Standard deviation of element abundances for eight different elements / lines in different bins of $S/N$ ($\texttt{snr\_c2\_iraf}$).} Shown are the mean (internal) covariance fit uncertainties from {\sc sme} (blue), as well as the standard deviations for repeat observations of the same star for all fibre-combinations (orange), same fibre combination (green) and different fibre combinations (purple). An exponential function (black) was fitted to all (orange) repeats was performed. Number in the top state the expected or mean uncertainty at $\texttt{snr\_c2\_iraf} = 40$.}
\label{fig:repeat_uncertainties_abundance_selection}
\end{figure*}

\subsection{Flagging of element abundances} \label{sec:flagging_ab}

{As for all of our previous GALAH releases, we want to stress that we discourage the use of flagged element abundances without consideration of the possible systematic trends that these probably less reliable measurements can introduce. However, setting flags for element abundance measurements is extremely difficult and will never be perfect. In this section we lay out which flags we have implemented. We also point to identified possible caveats, where no flags are raised, but caution should be applied.

{The major abundance flags are reported as \texttt{flag\_X\_fe} for each element X and are listed in Table~\ref{tab:flag_sp_galah_dr3}. For the final element abundances, we only report measurements that have passed all our quality checks with no flag raised or are an upper limit (with flag 1, see our description of upper limits in Sec.~\ref{sec:spec_details}). For ease of use we have raised a flag value 32 if the measurement is not reported, because it did not pass the quality checks.

{For each individual abundance measurement (\texttt{ind\_X1234\_fe} for line/combined measurements X1234), we have run a series of quality checks, which are summarised in the \texttt{ind\_flag\_X1234} as listed in Table~\ref{tab:flag_sp_galah_dr3}. These include a check of the $\chi^2$ values (value 2), where we raise a warning, if the $\chi^2$ value is above a very high $S/N$-depending threshold (estimated to be around 16 for $S/N = 10$, 20 for $S/N = 35$, and 160 for $S/N = 100$ from the general $\chi^2$-$S/N$ trend for the elements, compared to the typical median $\chi^2$ value of 0.5 to 1.0 at $S/N = 35$). A warning flag 4 for an imprecise or likely saturated measurement was raised, if the fitting uncertainty was above $0.3\,\mathrm{dex}$ or twice the expected uncertainty from repeat observations. We raise a flag value 8, if the wavelength solution around the Li line at $6708\,\text{\AA}$ is questionable, that is, when our Li fits with and without an additional \vrad fit just at this particular spectrum region result in a difference above $10\,\mathrm{km/s}$. We report individual measurements, but raise a flag 16, if we do not trust the stellar parameters ($\texttt{flag\_sp} \geq 128$). Flag value 32 is raised, if the measurement was not successful, that is if no stellar parameters were available or the line was too shallow or too blended.

{As we elaborate in the next section, we have identified several possible caveats, for which we have not implemented an algorithm to raise a flag, either because we could not find a way to automatically identify the caveats, or we are unsure if the measurements should be flagged. For the abundances, we therefore advice to carefully consider the caveats described in Secs.~\ref{sec:possible_systematics} and \ref{sec:pecularities_groups} in particular.

\section{Possible caveats: analysis shortcoming or physical correlation?} \label{sec:caveats}

In the previous sections we have laid out the methods by which we flag unphysical results and spectra with peculiarities for which our pipeline is likely to underperform. However, we cannot visually inspect all of the more than 30 million measurements that have been performed for this data release. Furthermore, we aim to not follow up all of the possible correlations in full detail, because many of these pose problems to understand the possible astrophysical nature of these trends (as shown later in Sec.~\ref{sec:rvcaveats} for atomic diffusion causing systematic differences of surface abundances in open cluster stars). Instead, we choose to leave such efforts for future scientific follow-up.

In this section, we address several possible caveats, for which we either have not yet found or implemented a solution, or believe that these results could indeed be of an astrophysical nature. We give examples for peculiar abundance patterns and show an example where the pattern (of Am/Fm stars) is truly representing the observed surface abundances when assuming ionisation equilibrium. In other cases, especially for the most metal-rich as well as coolest giant stars, we are aware that our pipeline is likely introducing systematic trends that may be ascribed firstly to our use of standard 1D hydrostatic model atmospheres (although the effect is partly mitigated by fudge parameters that can be tuned to mimic the effects of convection), secondly to our partly unreliable or incomplete molecular line data, and thirdly to the lack of true continuum points to use for the spectrum normalisation in these stars. In the latter case, the pseudo-continuum placement can correlate strongly with the stellar parameters (especially \feh) and lead to systematic trends in the reported measurements. We address possible influences on the reliability of surface gravities and finally also lay out considerations of the uncertainties and describe how these will be improved in future data releases.

\subsection{Abundance differences with {\sc sme} version}

{\sc sme} is an actively developed software with regular improvements and bug fixes as part of version publications. For GALAH DR3, we used {\sc sme} version 536. This particular version has recently been found to have a code inconsistency which may cause non-trivial changes in synthetic fluxes and abundances (A. Gavel and A. J. Korn, priv. comm. in Feb. 2021) with our used synthesis library \textit{linux.x86\_64.64g}. It is unlikely that this bug significantly influences our results thanks to a cancellation effect in our abundance zero-point calibration. We have performed several tests with the more recent {\sc sme} version 580 on a large sample of roughly  20\,000 stars, including reference stars such as the GBS, the stars with asteroseismic data, and open/globular cluster stars. For the majority of stars, we find that effects on abundances will be essentially null. This is because the bug appears to introduce a bias in abundances for all stars, which cancels in our calculation of solar-relative abundances. In practice, these effects are on the order of $0.05\,\mathrm{dex}$ for dwarfs and giants. Less than 10\% of the latter stars with very low surface gravity exhibit complex effects with biases as large as $0.15\,\mathrm{dex}$. We have unfortunately not found a predictable systematic trend with stellar parameters, that would have allowed us to correct all abundance measurements. We want to stress though that our accuracy and precision estimates point towards {\sc sme} 536 delivering reliable results after the zero point correction. In the future, we will implement newer versions of {\sc sme}.

\subsection{Stacked spectra} \label{sec:stacking}

In this data release, for the first time we also include stacked spectra of repeat observations. We select the higher $S/N$ observations for the main catalogue, but also report the individual spectra in the extended catalogue.
We identify observations of the same star via matching coordinates. However, we have found few cases, where calibration observations were performed already with configurations of stellar observations, such as bias frames, sky or dome flats. In few cases, the observations were not marked as calibration frames clearly, and have been used for the stacking of stellar spectra. We therefore caution the use of metal-poor stacked spectra, which can be identified in the 11th digit of the \texttt{sobject\_id} being 2 instead of 1, for example 160522002102FFF, where some of the science observations 0023-0025 have been stacked with dome flat observations 0016-0018, causing underestimated \feh, which cannot be picked up by the pipeline, because the analysis is reasonable for the wrongly stacked spectra. By comparison the differences of \feh, \vrad, and \Teff for stacked and unstacked spectra, we find less than 800 observations\footnote{More than 50 stacking issues per field were found for 1412310030, 1501010025, 1501030027, 1501030030, 1501080015, 1501120025, 1605220021, and 1710310021. 94\% of the spectra were already flagged. Further not all spectra of these fields wrongly stacked. We thus do not to flag all of them a priori.} being potentially wrongly stacked - 753 with $\texttt{flag\_sp} > 256$.

\subsection{Radial velocities in table version 1} \label{sec:rvcaveats}

In the main tables with the suffix v1 we report the radial velocities as estimated with the reduced spectra, which already applied barycentric corrections. Right before this data release, we have identified a wrong implementation of barycentric corrections was used and the reported radial velocities as part of the main tables were shifted incorrectly (within less than $0.4\,\mathrm{km\,s^{-1}}$).

{In our updated table versions with suffix v2, we have corrected these incorrect estimates and provide correct measurements from {\sc sme} in the VAC on \vrad with the columns \texttt{rv\_sme\_v2} together with the old, incorrect estimates under \texttt{rv\_sme\_v1}. Based on feedback from the scientific community, we have further exchanged the entries in the main catalogues for \texttt{rv\_galah} and \texttt{e\_rv\_galah} by even more reliable measurements of \texttt{rv\_obst} (when available) as outlined in Sec.~\ref{sec:rv_vac} and provide a flag \texttt{use\_rv\_flag} in the VAC on \vrad to outline where the used values stem from.

\subsection{Possible systematic trends} \label{sec:possible_systematics}

Below we present a list of possible systematic trends, as found up until the publication of GALAH+~DR3 during the validation. These do not appear in particular order.

\paragraph*{High abundances of V, Co, Rb, Sr, Zr, Mo, Ru, La, Nd, and Sm.}

In this data release, we try to push the boundary of what can be extracted from the observed spectra with the aim to deliver as many abundance measurements as possible. This does, however, not only push the limits of deciding what measurement is reliable, that is, significantly different from a continuum measurement, but leads to complicated cases where lines are blended, leading to possible wrong systematics. We therefore especially caution the use of elevated abundance measurements (especially above [X/Fe] of $0.3\,\mathrm{dex}$, as indicated in Fig.~\ref{fig:abundance_overview}) for  V, Co, Rb, Sr, Zr, Mo, Ru, La, Nd, and Sm, as we suspect that these are most likely affected by blending issues close to the detection limit. Only visual inspection could however confirm this, which is not possible for the vast number of measurements at hand and we therefore advise the user to inspect the published spectra before using these measurements blindly.

For V, we caution the use of measurements with \texttt{nr\_V\_fe} 2 and 3, that is, using \ion{V}{I} 4832.
For Co, we caution the use of measurements with \texttt{nr\_Co\_fe} = 2 or 8, that is, measurements purely based on lines \ion{Co}{I} 6490 and 7713. While we have not been able to narrow down the exact cause, we assume that measurements only based on these lines are caused by imperfect telluric corrections in CCD 3 for \ion{Co}{I} 6490 and spikes or imperfect telluric corrections in CCD4 for \ion{Co}{I} 7713. Such reduction issues introduce strong emission and absorption lines, which will either weaken the line beyond detectibility or strengthen it so that only high V or Co abundances can reproduce the observation. However, the resultant fits and their $\chi^2$ values will not appear suspicious and such reduction issues are therefore hard to identify and flag automatically.

\paragraph*{1D LTE / 1D non-LTE and microturbulence.}

Our spectrum synthesis is performed by assuming 1D LTE and 1D non-LTE. However, modelling stellar atmospheres with a 1-dimensional description is neglecting 3-dimensional, time-dependent effects, which can only partially be mitigated by fudge factors, like \vmic. While allowing this factor to be fitted as part of the analysis, our tests have shown that the abundance precision decreases. We have therefore implemented an empirical relation, estimated by \citet{Gao2018} for GALAH, over the whole parameter space, as outlined in Sec.~\ref{sec:spec_details}, as shown in panel a) of Fig.~\ref{fig:vmic_comparison}.

During the validation of element abundances, we have discovered several temperature-dependent trends. These occur in regions where our analysis approach is prone to systematic trends anyway, that is, the coolest/most line-rich ($<4500\,\mathrm{K}$) and hottest/most line-poor ($>6500\,\mathrm{K}$) regions. We cannot exclude that the found systematic trends can also be partially caused by over- or underestimated \vmic (in addition to a systematically incorrect normalisation for the most line-righ spectra). Comparisons with other \vmic-relations, see e.g. the relations by \citet{DutraFerreira2016} based on 3D atmosphere calculations (see panel b) suggest large deviations for certain stars, leading to a difference of up to $2\,\mathrm{km\,s^{-1}}$ (see panel c). The tests by \citet{Jofre2017} also showed that different stellar types are affected differently by inaccurate \vmic, with strongest implications for (more metal-rich) giant stars among the analysed sample of GBS.

While our long-term goal is to implement 3D non-LTE calculations, we believe that it is worth testing the implementation of $v_\text{mic}$ as a free parameter or the relations estimated by \citet{DutraFerreira2016} for certain parts of the parameter space, if the advantages outweigh the loss in abundance precision. Using $v_\text{mic}$ as a free parameter showed for example significant improvements of trends with $T_\text{eff}$ for the APOGEE survey \citep{Holtzman2018}.

\begin{figure*}
\centering
\includegraphics[width=0.99\textwidth]{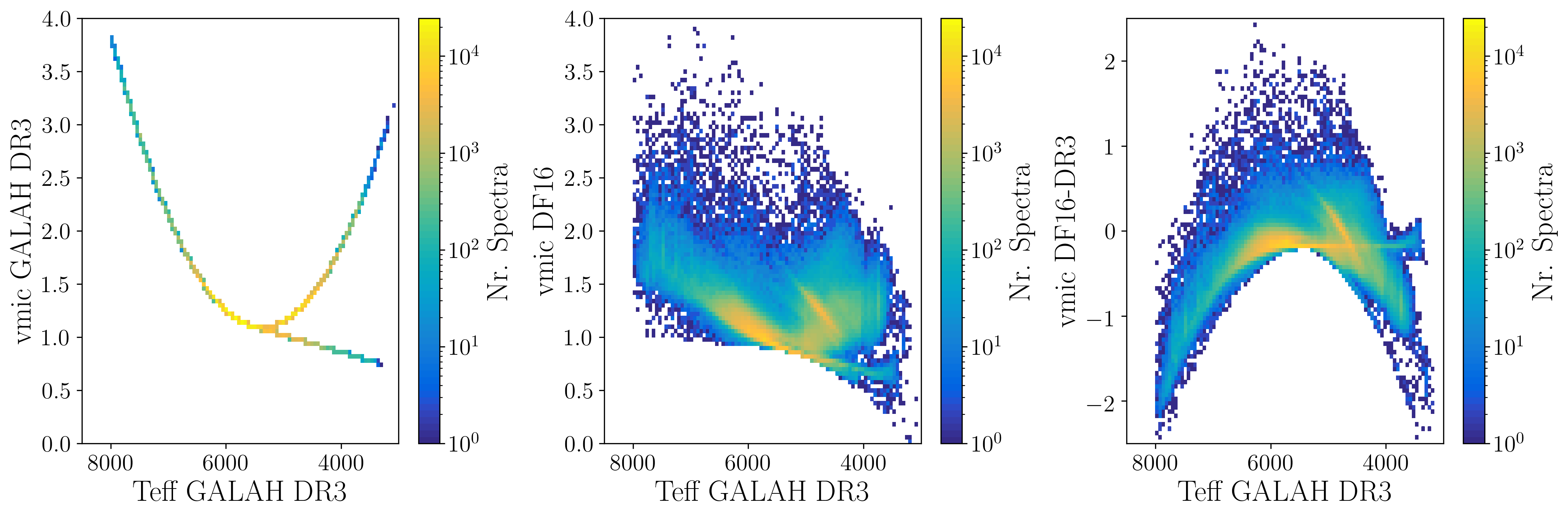}
\caption{
\textbf{Comparison of $v_\text{mic}$ calculated via relations used for GALAH+~DR3 and \citet{DutraFerreira2016}.}
Relations of $v_\text{mic}$ as a function of $T_\text{eff}$ calculated via relations used for GALAH+~DR3 and \citet{DutraFerreira2016} in the left and middle panels, respectively. A comparison of the two $v_\text{mic}$ values as a function of \Teff is plotted in the right panel and shows good agreement for the majority of stars, i.e. cool and warm dwarfs and stars around the RC. However, the two relations disagree strongly for the most luminous and hottest stars. For the latter, GALAH+~DR3 $v_\text{mic}$ values are up to two times higher. We note  that the relation by \citet{DutraFerreira2016} is based on 3D models with $4500 \leq T_\text{eff} \leq 6500\,\mathrm{K}$ and $2.5 \leq \log g \leq 4.5\,\mathrm{dex}$ for strictly Solar \feh.
}
\label{fig:vmic_comparison}
\end{figure*}

\paragraph*{Consistency of atmosphere composition for spectrum synthesis.}

For computational reasons, we estimate the abundances of all elements independently, and assume scaled-Solar patterns for most other elements during that optimisation. However, our approach might introduce systematic trend for elements which are often correlated (e.g. C and O), surrounded by lines that are deviating from the scaled-Solar pattern, or when the abundance pattern in general differs from the scaled-Solar pattern, thus leading to differences in the continuum and molecular lines strengths. If computationally possible, it would therefore be preferable to fit all elements partially \citep{Brewer2016} or fully self-consistent \citep{Ting2019}, which could also allow to estimate abundances not only via atomic lines, but also molecular features, which follow molecular equilibria \citep{Ting2018}.

\paragraph*{Metallicity/abundance trends.}

For numerous open and globular clusters we have found trends of \feh with temperature and/or evolutionary stage at the coolest and hottest ends of the \Teff range or in general for young clusters.

Stellar clusters are not the main focus of our survey, and many of the observations that were performed for them are outside of the typical GALAH magnitude, distance, and age range. Most of the open and globular clusters targeted by our observations are much more distant, which leads to less reliable distance estimates, with implications for our distance-dependent \logg estimates of their stars. Many of stars in the open clusters stars are typically younger than the GALAH targets, with astrophysical implications on additional features in their spectra.

\citet{Baratella2020} found that \vmic is overestimated and thus \feh is underestimated when using Fe lines in clusters, a trend that we also observe in some of our cluster observations. We therefore cannot a priori exclude wrong \vmic values as the influence of cluster abundance trends (see comments on \vmic above).

We note, however, that for open clusters, differences in \feh as well as other abundances have been found to be of astrophysical nature, e.g. atomic diffusion \citep[e.g.][]{Souto2018, Gao2018, BertelliMotta2018, Souto2019, Liu2019, Semenova2020} or stellar activity \citep[e.g.][]{Spina2020}. Furthermore, astrophysical abundance trends, like anti-correlations of light elements \citep[see][and references therein]{Bastian2018}, have also been found in globular clusters and are partially hard to disentangle from other abundance trends, e.g. those introduced by our analysis pipeline. We will follow this up for globular clusters with a dedicated study (D.~M.~Nataf et al., in prep.).

\paragraph*{Binarity.}

A central assumption of our observations is that each fibre observes only one star. We try to ensure this by only selecting point sources from 2MASS with a sufficient separation from other bright neighbours. Our selection does however not exclude stars that are not extended within 2MASS, for example spectroscopic binaries.

Our means to identify (spectroscopic) binaries are, however, limited, because as part of GALAH we usually only take three spectra within typically 1 hour per star, and can only resolve spectroscopic binaries if the lines of both components are resolved with the given broadening induced by our instrument and stellar rotation. Although we try to identify and flag stars as part of our validation (see Sec.~\ref{sec:flagging_sp}), we expect that we are not able to identify a significant fraction of stars as binaries. \citet{PriceWhelan2020} find 19\,635 high confidence close-binaries among 232,495 APOGEE sources (8\%), and \citet{ElBadry2018b} find that for 2645 of 20\,142 analysed main sequence targets (13\%), more than one star contributes significantly to the spectrum. Based on the results of \citet{PriceWhelan2020} we would expect at least 10\% of the stars above $>6000\,\mathrm{K}$ (23\% of GALAH+~DR3) and more than 40\% of stars with $>7000\,\mathrm{K}$ (3\% of GALAH+~DR3) to be binaries.

The implications of not identifying a star as a binary can be manifold. Firstly the binarity changes the astrometric solution, which is not always identified via \Gaia warnings or quality values like the RUWE value. This can falsify the estimated distance of objects. Secondly, the photometry of a binary system can deviate significantly from that of the primary component, depending on the flux contribution of the secondary. Thirdly, the flux contribution within the spectrum lead to inaccurate fits when assuming a single star as quantified by \citet{ElBadry2018, ElBadry2018b}, which leads to inaccurate stellar parameters as well as element abundances. For binaries, the measured \vrad also only reflects (at best) the value at the time of observation and is thus not indicative of its Galactic orbit. We note that we have not made use of the assessment of \vrad changes among our 51\,539 spectra with dedicated repeat observations (typically on different nights).

\paragraph*{Stars with uncertain/unreliable astrometry.}

As part of our spectroscopic analysis we rely on the quality of astrometric measurements, to infer reliable absolute photometry and then \logg. While we flag stars with high RUWE values above 1.4 \citep{Lindegren2018, Lindegren2018b}, we caution the user to not blindly use all measurements, especially those of stars with uncertain astrometry.

We have used more elaborate distance estimates from \citet{BailerJones2018} which infer more trustworthy distances based on a Galactic prior for stars with parallax uncertainties beyond $20\%$. Especially for very distant stars, like some of our observations of LMC stars, this Galactic prior leads to an underestimated distance and thus likely overestimated \logg (see Eqs.~\ref{eq:logg} and \ref{eq:lbol}).

In general, we note that for stars with more constrained distance estimates, like open clusters \citep{CantatGaudin2020}, globular clusters \citep{Baumgardt2019} and stars of the LMC \citep{deGrijs2014}, a reanalysis would be leading to more reliable stellar parameters and abundances, when using these distances instead of the ones solely estimated from \Gaia parallaxes.

\paragraph*{Influences of isochrone choice.}

For computational reasons we have limited the isochrones used for the on-the-fly mass estimation to a grid of $0.5..(0.5)..13.5\,\mathrm{Gyr}$. We note that for the youngest stars this might not be a good choice, as we see some noding in the on-the-fly mass and age estimates, especially for hot stars and secondary RC stars. In the future we would like to make use of a better set of isochrones in terms of sampling (more ages on a logarithmic scale), which will hopefully also include different alpha-enhancements and will take into account atomic diffusion as well as stellar rotation. For a better quantification of the uncertainties, for example when using (Markov Chain) Monte Carlo sampling, it would also be useful to be able to sample ages above the age of the universe.

\paragraph*{High extinction.}

86\% of the stars of this data release have estimated $E(B-V) < 0.2\,\mathrm{mag}$ from \citet{Schlegel1998} and 95\% below $0.2\,\mathrm{mag}$. Similarly, 90\% and 98\% of the stars have estimated $A_{K_S} < 0.1\,\mathrm{mag}$ and $0.2\,\mathrm{mag}$, respectively. If a star has a high and uncertain extinction, this can influence the bolometric luminosity that we estimate and thus introduce biases in the surface gravity and thus all subsequent analyses. Our pipeline especially is only optimised for $E(B-V) < 0.48\,\mathrm{mag}${, the limit that includes 95\% of the initial input catalog. We therefore caution that trends found among stars with high extinction, and where $A_{K_S}$ estimated via the RJCE method and $E(B-V)$ differ significantly should be treated with caution..

Potassium is estimated from the \ion{K}{I} 7699 resonance line. This line is also a good tracer of interstellar potassium which leads to contamination of the stellar line in highly extinct regions. In the future we aim to estimate the extinction for example via diffuse interstellar bands and possibly use correlation of extinction and line strength of interstellar potassium \citep{Munari1997} to correct the spectra and measurements of stellar [K/Fe]. For this DR, we however caution the user to check the extinction of stars when using [K/Fe], as we measure this abundance without any corrections causing a rather hard to predict systematics (depending on the velocities of star and ISM) of [K/Fe].

\paragraph*{Upper limits.}

While we report upper limits for advanced users, we strongly recommend that all users take great care in using these measurements. For all elements, but especially for neutron-capture elements, these estimates are pushing the limits of what we can be extracted from the data and are by definition only an upper limit, not a measurement. We therefore strongly recommend to check upper limit estimates against the data and inspect spectra when possible.

\subsection{Peculiarities for certain groups of stars} \label{sec:pecularities_groups}

For some groups of stars, we have found peculiar trends of abundances, for which we cannot exclude astrophysical reasons rather than influences of our analysis and suggest follow-up studies to disentangle those.

\paragraph*{Solar and metal-rich giants, especially red clump stars.}

An ongoing disagreement concerns the stellar parameters of metal-rich giants, and especially metal-rich red clump stars. Already in GALAH~DR2 our analysis has yielded unreasonable stellar parameters (in the case of DR2 the estimated \logg were deviating significantly by up to $0.7\,\mathrm{dex}$ from those expected from astro-/photometry, while \Teff and \feh agreed with other literature estimates/expectations).

\begin{figure*}
\centering
\includegraphics[width=0.9\textwidth]{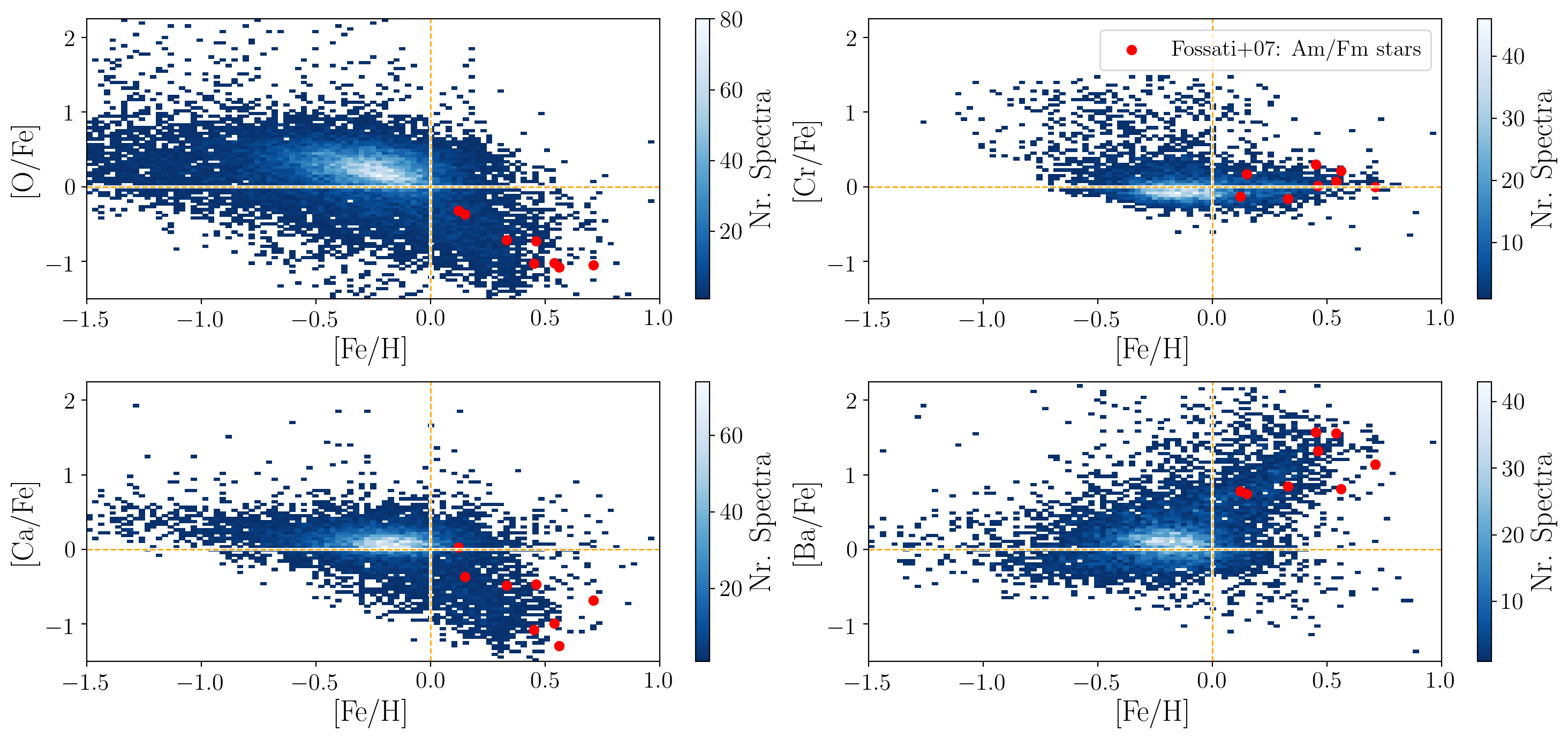}
\caption{
\textbf{Element abundances [O/Fe], [Ca/Fe], [Cr/Fe], and [Ba/Fe] for GALAH stars with $T_\text{eff} > 6500\,\mathrm{K}$.} Overplotted are studies of Am/Fm stars by \citet{Fossati2007} in red.}
\label{fig:GALAH_DR3_AmFmStars}
\end{figure*}

For DR3, the use of astrometry and photometry allows us to get more accurate \logg. For the metal-rich ($\mathrm{[Fe/H]} > 0$) giants and RC stars, however, we notice that the estimated iron abundances show a significant trend of underestimated \feh with increasing metallicity{, when comparing with GALAH DR2. This is an indicator that our synthetic spectra are inaccurate for this specific type of stars or spectra. As discussed above, \citet{Jofre2017} showed that for giant stars, an over-/underestimated \vmic can change the measured abundances of some lines significantly, by up to $1.5\,\mathrm{dex}$. The reasons for underestimated \feh are however more diverse and also include missing/unreliable molecular line data, the underestimation of blending and incorrect continuum normalisation. We believe that we can exclude incorrect estimates of \logg estimates, e.g. as a result of poor mass-estimates from missing isochrone models in the super-Solar \feh regime, because photometric and spectroscopic positions in the CMD and Kiel diagrams agree well.

We find systematically higher abundances of Na, Al, Sc, \ion{Ti}{II}, Ni, and Ba among metal-rich RC stars when compared to RGB stars\footnote{These can be identified as unexpected extensions of high [X/Fe] elevated above the majority of stars in Fig.~\ref{fig:abundance_overview}, especially when selecting only high-$S/N$ spectra of giants.} with increasing disagreement from 0 at Solar \feh to $\Delta \mathrm{[X/Fe]} > 0.4\,\mathrm{dex}$ above $\mathrm{[Fe/H]} > 0.2\,\mathrm{dex}$ for these elements. However, another neutron capture element Y is not as affected.

When using the K2 sample with asteroseismic classifications of evolutionary stages within this DR (Stello, priv. communication), we find a significant difference of around $0.3\,\mathrm{dex}$ between RC and RGB stars. The reasons for this might be manifold and could for example suggest non-scaled-Solar abundance patterns for C and N among the RC stars, as shown by \citet{Tautvaisiene2013}. The follow-up of these spectroscopic shortcomings are beyond the scope of this paper, but should also assess line saturation and discuss the implications of different formation depths of atomic lines \citep[see e.g.][]{Gurtovenko2015}, which could possibly explain the different effect for different lines within the GALAH range as well.

\paragraph*{Abundance patterns of Am/Fm stars.}

While following up peculiar abundance trends of the hottest stars, we identified a group of stars with high [Ba/Fe] among the stars with $T_\text{eff} > 6500\,\mathrm{K}$, coinciding with those identified by \citet{Fossati2007,Fossati2008} for a handful of stars, see the agreement of their measurements with peculiar pattern of some of the hot GALAH stars in Fig.~\ref{fig:GALAH_DR3_AmFmStars}. Similar to \citet{Xiang2020} who identified tens of thousands of these Am/Fm stars we measure typically higher [Ba/Fe] than for the Sun, but lower alpha-enhancement than in the Sun for these typically young stars, when assuming ionisation equilibrium.

\paragraph*{Young star parameters.}

We stress that our stellar parameters for the youngest stars  (below $0.5\,\mathrm{Gyr}$) are likely unreliable. This is caused by our analysis setup with an isochrone grid selection favouring older stars, tying \vmic to an empirical relation and estimating stellar parameters mainly from iron lines \citep{Baratella2020}, but also neglecting stellar rotation, possible stellar activity and magnetic fields \citep{Spina2020} which can alter the shape of stellar lines quite drastically.

\paragraph*{Unexpected over-/underdensities.}

Below we list several unexpected over- and underdensities, which are likely introduced by our analysis, that is, not the RC area or the red giant bump. While using the recent versions of {\sc sme}, we have identified several overdensities in the parameter space, coinciding with grid points of the chosen atmosphere grids. We especially warn the user of these overdensities at $3500\,\mathrm{K}$ as well as $4750..(250)..8000\,\mathrm{K}$. We further have found an under-density around of stars with temperatures below $4750\,\mathrm{K}$, which coincide with regions a different atmosphere grid spacing. Comparisons with the IRFM temperatures show however that the temperatures of these stars are not drastically different and we have therefore decided to not flag them. We have further identified an overdensity at $4650\,\mathrm{K}$ and \logg of $4.7\,\mathrm{dex}$, which we can ascribe to an issue in the isochrone interpolation due to sparsely available isochrone points.

\subsection{Uncertainties}

For this data release, we include more accuracy and precision estimates than for GALAH~DR2. However, for several stellar parameters and abundances, the means of accuracy estimation are limited, because there are no benchmark values available. We therefore want to caution the user that the accuracy uncertainties might be underestimated and also not complete in terms of their parameter dependence.

For hot stars we have identified a systematic trend causing increasingly underestimated \Teff for hotter stars above $6000\,\mathrm{K}$. The comparison with the GBS shows agreement of our and the literature values within the uncertainties, but our absolute accuracy value for \Teff is likely underestimating the uncertainty for the hottest stars.

We have not been able to find enough benchmark values to test the accuracy of \feh as a function of stellar parameters and therefore only employ an absolute value for the \feh accuracy. More benchmark measurements, especially with similar conditions to the survey setup (instead of nearby bright stars as validators for distant faint stars), for all stellar parameters would be useful.

For GALAH+~DR3, our precision estimates are based on the repeat uncertainties and internal fitting uncertainties from {\sc sme}, which for some parameters have been rescaled to match in overall shape. As we continue to develop our pipeline, and obtain more repeat observations in the future, we will be able to also expand the precision estimation not only as a function of an average $S/N$, but $S/N$ in particular line regions as well as \Teff, \logg, and \feh, similar to the APOGEE survey \citep{Joensson2020}.

\begin{landscape}
\begin{figure}
\includegraphics[width=\columnwidth]{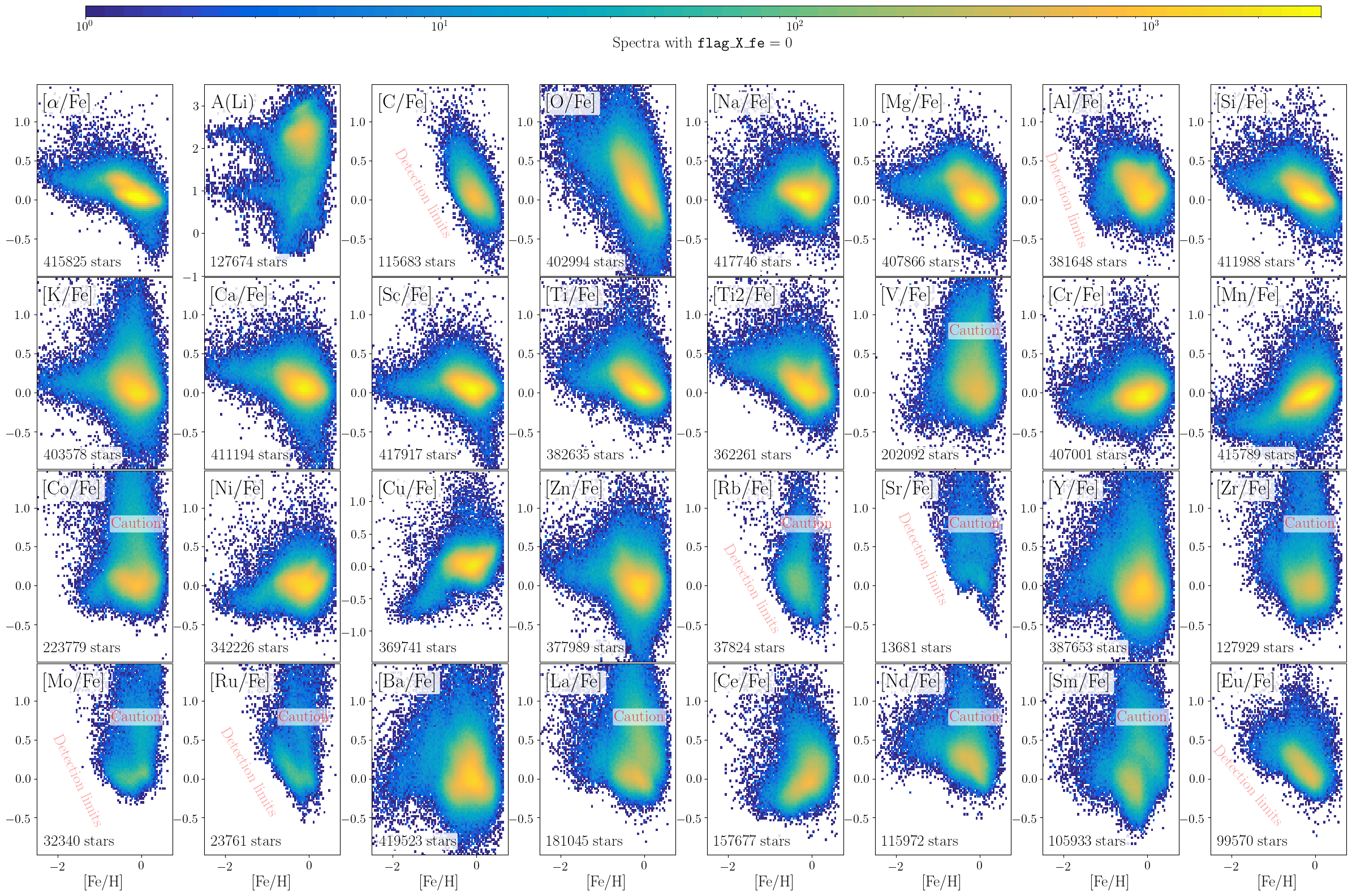}
\caption{
\textbf{Distribution of the element abundances included in GALAH Data Release 3 over the iron abundance \feh.} Shown are relative abundances [X/Fe] for stars with $\texttt{X\_FE\_FLAG} = 0$ (with exception of Li, for which we show the absolute abundance). Colours indicate the stellar density, truncated at a maximum of 6000 per density bin.} 
\label{fig:abundance_overview}
\end{figure}
\end{landscape}

\section{Catalogues included in this release}  \label{sec:catalogues}

\subsection{Catalogue versions before and after \Gaia eDR3} \label{sec:version2}

{We have published two versions of our data catalogues, version 1 before \Gaia eDR3 and version 2 after \Gaia eDR3. We recommend the use of version 2, as it includes the following updates and fixes:
\begin{itemize}
    \item We have solved the bug concerning barycentric values of \texttt{rv\_galah} (see Sec.~\ref{sec:rvcaveats}). In version 2, we now report our recommended values of \vrad based on GALAH measurements, including a correct {\sc sme}-estimate of \vrad, that is \texttt{rv\_sme\_v2}.
    \item We have crossmatched our targets with the \Gaia eDR3, parallax zero point corrections by \citet{Lindegren2020b} and (photo-)geometric distance estimates from \citet{BailerJones2020}. We provide all this information in a new VAC (see Sec.~\ref{sec:vac_gaia}).
    \item We have propagated the information from \Gaia eDR3 to estimate better added values for the VACs on stellar ages as well as dynamics.
\end{itemize}
%________________________________________________________________
\subsection{Main catalogues} \label{sec:main_catalogue}

The main catalogues can be downloaded from the DataCentral\footnote{\url{https://cloud.datacentral.org.au/teamdata/GALAH/public/}.} and accessed via TAP\footnote{\url{https://datacentral.org.au/vo/tap}.}. 

We provide two main catalogues. The first one \texttt{allstar} includes one entry per star and is a cleaned version of the extended \texttt{allspec}, with each entry representing the highest $S/N$ measurement for each star and only the combined, final abundance estimates that are unflagged or upper limits.

The \texttt{allspec} catalogue includes an entry for each spectrum (multiple entries for some stars). It extends the \texttt{allstar} catalogue by also having the raw stellar parameter and abundance measurements, that is, raw measurements without zero point or bias corrections and uncalibrated fitting uncertainties (\texttt{cov\_*}) for each stellar parameter and individual line abundances (\texttt{ind\_*}), with more extensive flags.

{The flags for both the main stellar parameters (\texttt{flag\_sp}) and the final and raw abundance measurements are listed in Table~\ref{tab:flag_sp_galah_dr3} and explained in Secs.~\ref{sec:flagging_sp} and \ref{sec:flagging_ab}, respectively.

For illustration we plot the distribution of all element abundances in Fig.~\ref{fig:abundance_overview}.

{After the publication of \Gaia EDR3, we have added \texttt{dr3\_source\_id} to the main catalogues and updated the reported \texttt{rv\_galah} and \texttt{e\_rv\_galah} as well as \texttt{rv\_gaia} and \texttt{e\_rv\_gaia}, as described in Sec.~\ref{sec:rv_vac}. The new versions of the catalogues can be found via their suffix v2.

We list the table schema of the catalogue in Table~\ref{tab:main_catalog_schema}, but they can also be found in the FITS header or online at \url{https://datacentral.org.au/services/schema/}. It includes the following categories for the \texttt{allstar} and \texttt{allspec} catalogues:
\begin{enumerate}
\item Stellar paramaters (see Fig.~\ref{fig:hrd_galah_dr3})
\item Stellar parameter flags (both warning and flags)
\item Final uncertainties for each parameter
\item Combined alpha-abundance (for unflagged/upper limit measurements, see Sec.~\ref{sec:spec_details})
\item Combined element abundances (for unflagged/upper limit measurements, see Sec.~\ref{sec:spec_details}) and bitmask of the line selection
\item Most important products of the reduction pipeline
\item Crossmatches with \Gaia~DR2, \citet{BailerJones2018}, \Gaia~DR2 RUWE, 2MASS, \end{enumerate}
For the \texttt{allspec} catalogue, it also includes
\begin{enumerate}
\item Individual element abundances (including flagged measurements)
\item Uncalibrated fitting uncertainties
\item More products of the reduction pipeline
WISE $W2$
\end{enumerate}

%________________________________________________________________
\subsection{Value-Added-Catalogues (VACs)} \label{sec:value_added_catalogues}

This data release of GALAH is accompanied by four Value-Added-Catalogues, one for a cross-match with \Gaia~EDR3, one for stellar ages and masses, one with kinematic as well as dynamic information for each star/spectrum, one for more elaborate radial velocity estimates, and a fourth one with additional estimates for double-lined spectroscopic binaries. After the publication of \Gaia EDR3, we have updated all our catalogues and their latest versions have the suffix v2. Subsequently, we outline the important information used to calculate the additional values for each VAC and note the changes from v1 to v2.

\subsubsection{Crossmatch with \Gaia EDR3} \label{sec:vac_gaia}

{As of December 2020, we provide the crossmatch of our allspec-catalogue with \Gaia EDR3 \citep{Brown2020, Fabricius2020}, with improved astrometric \citep{Lindegren2020a} and photometric \citep{Riello2020} information. We performed this match via the \texttt{dr2\_source\_id}, specifically with the smallest angular distance\footnote{In the future, we suggest to perform this crossmatch via GALAH's 2MASS ID and the yet-to-come match of \Gaia EDR3 and 2MASS identifiers. We also note that for 57 sources, a sorting via Gaia G magnitude would have yielded different identifiers.} within the internal crossmatch catalogue of \Gaia DR2 and \Gaia EDR3 \citep{Torra2020}. For each star, we further list the parallax zero point as queried from \citet{Lindegren2020b} to correct the parallaxes of the sources with 5 or 6-parameter astrometric solutions from \Gaia EDR3 \citep{Lindegren2020a}. We further provide the photogeometric and geometric distance estimated by \citet{BailerJones2020} which were matched via the \texttt{dr3\_source\_id}. The table schema of the VAC is included in the FITS file but can also be found at \url{https://datacentral.org.au/services/schema/}.

\subsubsection{Stellar age and mass estimates} \label{sec:vac_age}
To estimate stellar properties like age, mass, and distance we use the Bayesian Stellar Parameter Estimation code {\sc BSTEP} \citep{Sharma2018}. {\sc BSTEP} provides a Bayesian estimate of intrinsic stellar parameters from observed parameters by making use of stellar isochrones. 
For details of the adopted priors see \citet{Sharma2018}, in short, a flat prior on age and metallicity was used and for density distribution of stars a combination of an exponential stellar disc and a diffuse stellar halo was used.
For results presented in this paper, we use the PARSEC release v1.2S + COLIBRI stellar isochrones \citep{Marigo2017}. We use the following observables, \Teff, \logg, \feh, \alphafe, 2MASS $J$ and $Ks$  photometry, and parallax from \Gaia. The effective observed metallicity was estimated using the formula 
\begin{eqnarray}
\log\left(\frac{Z}{Z_{\odot}}\right) & = & {\rm [Fe/H]}
+ \log(10^{[\alpha/{\rm Fe}]} 0.694+ 0.306).
\end{eqnarray}
by \citet{Salaris2006}, with $Z_{\odot}=0.0152$ in accordance with the isochrones used. 
This was compared with the surface metallicity reported by the isochrones, which takes the evolutionary changes in surface metallicity $Z$ into account. 
The code provides an estimate of age, actual mass, initial mass, initial metallicity, surface metallicity, radius, distance, extinction $E(B-V)$, luminosity, surface gravity, temperature and the probability of being a red clump star. 
For each estimated parameter we report a mean value and standard deviation together with the 16th, 50th, and 84th percentiles.
The isochrone grid consisted of 16\,768\,422 grid points.
An $81 \times 121$ grid spanning $-2<\log Z/Z_{\odot}<0.5$ and $6.6<\log {\rm age/Gyr}<10.12$ was used. The mass dimension of the grid was resampled by interpolation, such that $\Delta \log T_{\rm eff}< 0.004$ and $\Delta \log g< 0.01$. 
For each parameter we report a mean value and standard deviation based on 16 and 84 percentiles. 
{After the publication of \Gaia EDR3, we have recalculated all entries in this VAC (provided in the table with suffix v2) based on new \Gaia EDR3 parallaxes and their uncertainties, corrected by the zero point shifts by \citet{Lindegren2020b}.
The table schema of the VAC is included in the FITS file but can also be found at \url{https://datacentral.org.au/services/schema/}.

\subsubsection{Kinematic and dynamic information} \label{sec:vac_dynamics}

We provide a Value-Added-Catalogue with kinematic and dynamic information, that builds upon the 5D astrometric information by \Gaia~DR2 and radial velocities preferably from GALAH+~DR3 (97.3\%) of all spectra and otherwise from \Gaia~DR2 (0.5\% of all spectra). Where possible, we use distance that take into account uncertainties, preferably those estimated via isochrone matching as part of the BSTEP grid-based modelling ({95.7\% of all spectra, see description in Sec.~\ref{sec:vac_age}), otherwise we use the prior-informed values (4.1\% of all spectra) by \citet{BailerJones2018}.

For the calculation of orbit information we use version 1.6 of the python package {\sc galpy} \citep{Bovy2015}. More specifically, we use its {\sc orbit} module for coordinate/velocity transformation as well as orbit energy computation. To estimate actions, eccentricity, maximum orbit Galactocentric height, and apocentre/pericentre radii, we use {\sc galpy}'s {\sc actionAngleStaeckel} approximations via the Staeckel fudge  \citep{Binney2012} with a focus of 0.45 and the method by \citet{Mackereth2018}.

For our calculations we use the best fitting axisymmetric potential by \citet{McMillan2017} with a Solar radius of $8.21\,\mathrm{kpc}$, consistent with the latest measurement by \citet{Abuter2019} of $8.178 \pm 0.013_\text{stat.} \pm 0.022_\text{sys.}\,\mathrm{kpc}$, and circular velocity at this radius of $233.1\,\mathrm{km\,s^{-1}}$. We use the total motion of the Sun in the V-direction of $248.27\,\mathrm{km\,s^{-1}}$ by evaluation the proper motion measurements from \citet{Reid2004} at our chosen Solar radius. We further place the Sun $25\,\mathrm{pc}$ above the plane \citep{Juric2008} and use the peculiar Solar velocities $U_\odot = 11.1_{-0.75}^{+0.69}\,\mathrm{km\,s^{-1}}$ and $W_\odot = 7.25_{-0.36}^{+0.37}\,\mathrm{km\,s^{-1}}$ by \citet{Schoenrich2010}, but $V_\odot = 15.17\,\mathrm{km\,s^{-1}}$. This value is higher than the $12.24_{-0.47}^{+0.47}\,\mathrm{km\,s^{-1}}$ from \citet{Schoenrich2010}, but given the ongoing debate \citep[see e.g. the discussion in][]{Sharma2014} of this value, we choose our value for internal consistency between the chosen total and peculiar motions of the Sun in our reference frame with a given circular velocity.

For the Sun, this leads to actions of $J_R = 7.7\,\mathrm{kpc\,km\,s^{-1}}$, $J_\phi = L_Z = 2038.3\,\mathrm{kpc\,km\,s^{-1}}$, and $J_Z = 0.4\,\mathrm{kpc\,km\,s^{-1}}$ on an orbit with eccentricity 0.073, a pericentre radius of $8.15\,\mathrm{kpc}$, apocentre radius of $9.43\,\mathrm{kpc}$ and a total energy of $E_{n,\odot} = -1.53\cdot 10^5\,\mathrm{km^2\,s^{-2}}$.

We provide columns for the heliocentric cartesian coordinate ($X$, $Y$, $Z$) and velocity frames ($U$, $V$, $W$) as well as the Galactocentric cylindrical coordinate ($R$, $\phi$, $z$) and velocity frames ($v_R$, tangential speed in azimuthal direction $v_T = R \cdot \frac{\mathrm{d}\phi}{\mathrm{d}t}$, $v_z$) together with the actions ($J_R$, $J_\phi = L_Z$, $J_Z$), eccentricity ($e$), maximum Galactocentric orbit height ($z_\text{max}$), pericentre and apocentre radii ($R_\text{peri}$, $R_\text{ap}$), as well as orbit energies for the best value input. We further realise 10,000 Monte Carlo samplings per star/spectrum by sampling \Gaia astrometry within the uncertainties\footnote{In version 2 we sample values using the \Gaia covariances and replace the distance and radial velocity sample with our own independent sample. In version 1 we neglected the covariances, but also provide a script to take them into account.}. For the distance sampling in version 2, we sample with 2 Gaussians (half the sample with a standard deviation based on the 16th percentile below the median and the other half with 84th percentile above the median) linearly in the distance modulus. For version 1, we assumed Gaussian uncertainties when using the BSTEP distances or sample from a 2-sided Gaussian in distance based on the bold assumption that the distributions left and right of the mode stated by \citet{BailerJones2018} are Gaussian and we can thus describe them via the stated lower and higher percentiles\footnote{We want to stress however, that given the excellent parallax quality for the vast majority of our sample (see Fig.~\ref{fig:plx_snr_quality}), these choices are only affecting less than 5\% of the observed stars with parallax uncertainties above 20\%, for which we caution the user to carefully assess the quality of both the astrometry as well as our distance and thus kinematic/dynamic estimates.}. We then provide the 5th, 50th, and 95th percentile for the user for each orbit parameter. For the 269 spectra in version 2 and 101 spectra in version 1, where the dynamic calculations yield unbound orbits, we only report the kinematic properties.

{After the publication of \Gaia EDR3, we have recalculated all entries in this VAC based on updated distances as well as \vrad and provide them in a table with the suffix v2.
We indicate, which distance is used, with the flag \texttt{use\_dist\_flag}, with the values 0 (649453 spectra for which we use the distances from version 2 of the VAC on ages described in Sec.~\ref{sec:vac_age}), 1 (27639 spectra) when using the photogeometric distances from \citet{BailerJones2020}, 2 (41 spectra) when using the geometric distances from \citet{BailerJones2020}, and 4 when using distances by inverting parallaxes (0 spectra). All methods take into account the zero point shifts by \citet{Lindegren2020b} and the flags are based on the availability of each distance method (preferring BSTEP over photogeometric over geometric over inverted-parallax distances), where flag \texttt{use\_dist\_flag} = 8 indicated that no distance was available (1290 spectra).
We use updated RV from version 2 of our RV VAC by selecting the RV based on the \texttt{use\_rv\_flag} described in Sec.~\ref{sec:rv_vac}. The table schema of the VAC is included in the FITS file but can also be found at \url{https://datacentral.org.au/services/schema/}.

\begin{figure*}
\centering
\includegraphics[width=\textwidth]{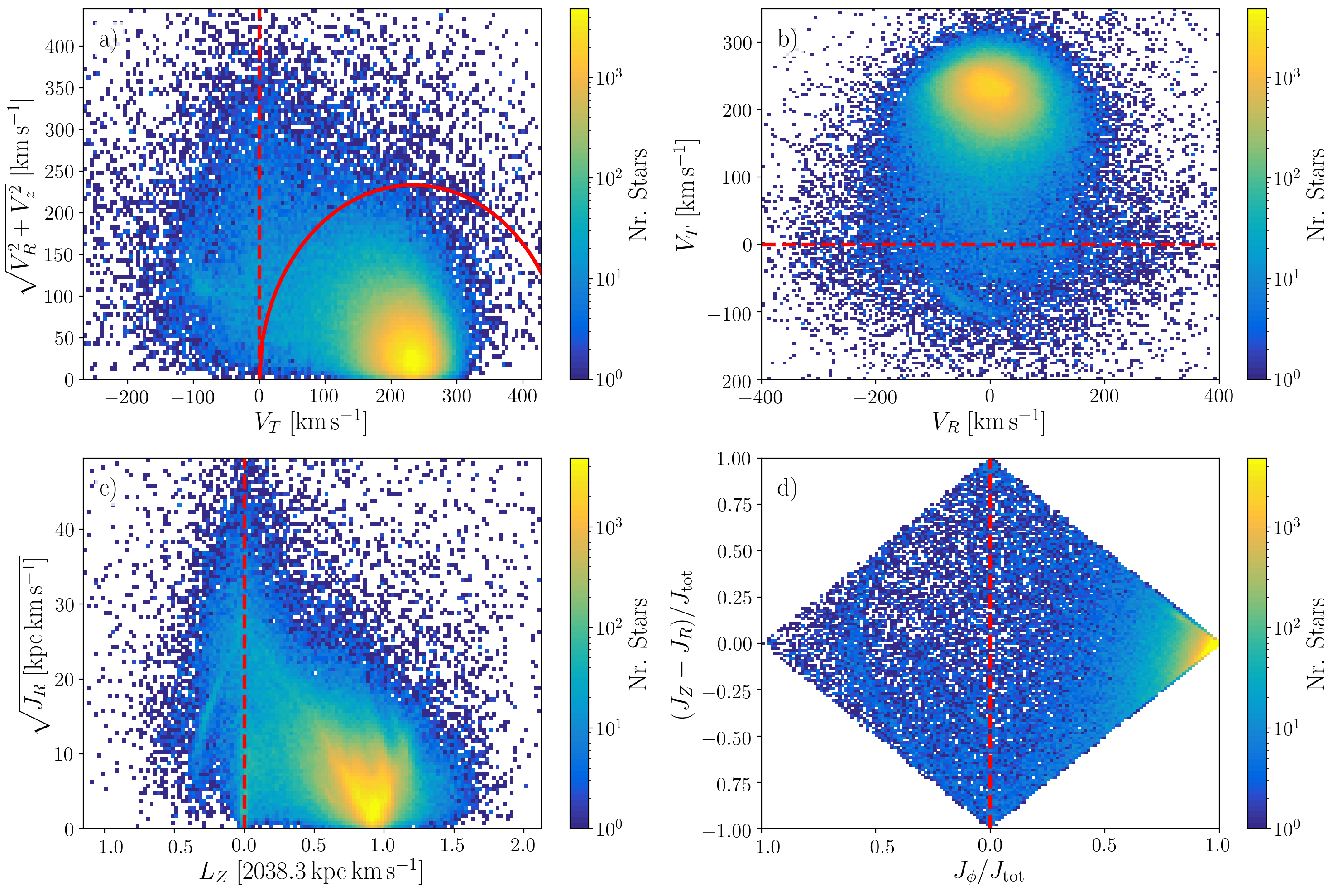}
  \caption{
  \textbf{Coverage of stellar kinematics (space velocities) and dynamics (actions) for the stars observed as part of GALAH.}
  \textbf{Panel a)} shows a Galactocentric version of the Toomre diagram \citep[compare to e.g.][]{Bonaca2017, Feuillet2020}, 
  \textbf{panel b)} the Galactic space velocities \citep[compare to e.g.][]{Belokurov2018, Feuillet2020},
  \textbf{panel c)} two actions \citep[compare to e.g.][]{Trick2019, Feuillet2020}, and 
  \textbf{panel d)} the distribution of actions \citep[compare to e.g.][]{Vasiliev2019}.
  The vast majority of stars in GALAH+~DR3 has both azimuthal / transversal Galactocentric velocities and angular momenta very similar to the Sun ($V_\odot = 248.27\,\mathrm{km\,s^{-1}}$, $J_\phi = L_Z = 2038.3\,\mathrm{kpc\,km\,s^{-1}}$).
  Red dashed lines in each panel indicate an angular momentum or azimuthal velocity of $0\,\mathrm{kpc\,km\,s^{-1}}$ or $0\,\mathrm{kkm\,s^{-1}}$ respectively. The Red line in panel a) indicates a total velocity of $233.1\,\mathrm{km\,s^{-1}}$.
  We note that the overdensity at low $V_T{\sim}-85\,\mathrm{km\,s^{-1}}$ in panels a) and b) as well as the streak at $-0.25 L_{Z,\odot}$ in panel c) and $-0.6 J_\phi/J_\text{tot}$ in panel d) coincide with the location of the distant targeted star of the globular cluster $\omega$\,Cen with mean \Gaia parallax uncertainties of 46\%.
  }
  \label{fig:DR3_dynamics_overview}
\end{figure*}

The space velocities $(V_R, V_T, V_z)$ in the Galactocentric frame are shown in a Toomre diagram in Fig.~\ref{fig:DR3_dynamics_overview}a. Most of the stars observed as part of GALAH+~DR3 have disc-like kinematics similar to the local standard of rest, but an extension of stars with lower rotational velocity than the disc ($V \ll 0\,\mathrm{km\,s^{-1}}$) are shown and indicate that several stars with halo-like kinematic properties are part of GALAH+~DR3.

\begin{figure*}
\centering
\includegraphics[width=\textwidth]{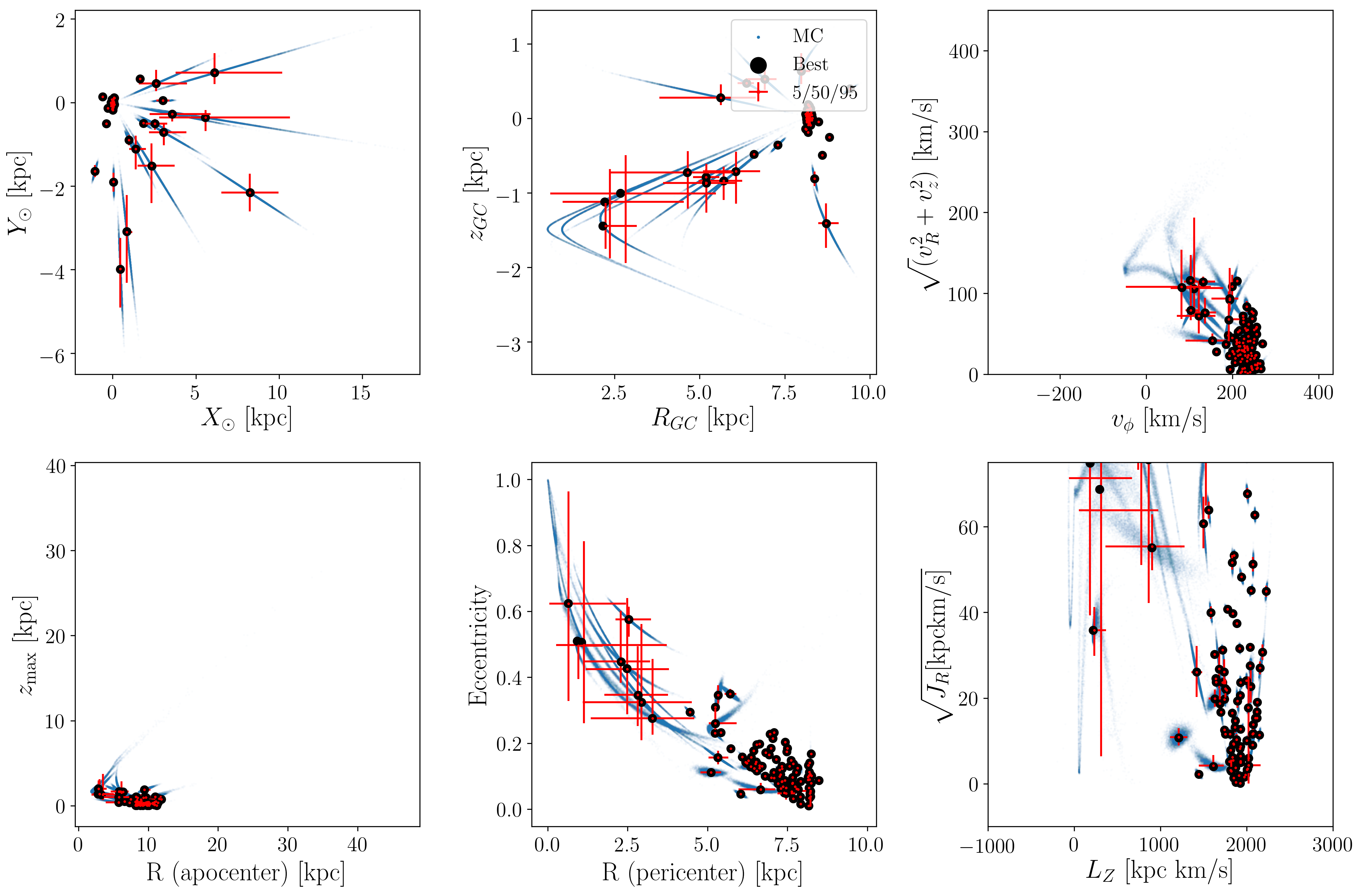}
  \caption{
  \textbf{Overview of phase space and dynamic stellar properties for randomly chosen stars from GALAH+~DR3, including their sampling within the measurement uncertainties.} The black points indicate the values calculated from the best 6D information. Blue points indicate 1000 samples from the 6D information per star within the uncertainties. Red error bars indicate the distribution between 50th percentile (middle of the cross) and the 5th and 95th percentile, respectively.}
  \label{fig:MC_output}
\end{figure*}

For the computed phase space and dynamic properties, we report a variety of statistical values. In addition to the best-value, that is computed by using the best values as input, we also sample the distribution for each property within the uncertainties via Monte Carlo sampling with size 1000 and report the 5th, 50th, and 95th percentiles of these distributions. An example of the sampling of parameters for 100 randomly selected stars is shown in Fig.~\ref{fig:MC_output}.  We also provide the code to perform this sampling with different sampling choices. Whereas we currently sample the properties by assuming their input parameters are uncorrelated, we also provide the code to sample with the \Gaia correlation matrices. The latter are currently not applying a distance prior and are thus problematic for large distances. However, we stress, that the vast majority of the stars from GALAH+~DR3 have very precise parallax measurements, for which the sampling choice is negligible (see Fig.~\ref{fig:plx_snr_quality}).

The distribution of heliocentric coordinates ($X,Y$) and Galactocentric cylindrical coordinates ($R,z$) is shown in Fig.~\ref{fig:dr3_apogee_lamost}a and d. The vast majority of targets are distributed within $4\,\mathrm{kpc}$ from the Sun and covers a large fraction of the disc. Because of the target selection of the GALAH main program ($\vert b \vert > 10\,\mathrm{deg}$), relatively few stars are observed close to the Galactic plane. We remind, however, that GALAH+~DR3 includes also observations from TESS-HERMES, K2-HERMES, and several smaller projects that targeted the Galactic bulge and clusters. The distribution in Fig.~\ref{fig:dr3_apogee_lamost}a is hence also including observations with $\vert b \vert < 10\,\mathrm{deg}$ especially towards the Galactic centre at ($R,z$) = 0. A combination of distance uncertainties and special targeting of clusters and K2/TESS fields is causing unrealistic streaks in both Fig.~\ref{fig:dr3_apogee_lamost}a and d.

\subsubsection{Radial velocities}\label{sec:rv_vac}

As outlined in Sec.~\ref{sec:rv}, we provide a Value-Added-Catalogue for radial velocities. In version 2 of this VAC, we list all values of measurements from {\sc sme} (now with suffix \texttt{sme}, rather than suffix galah as in version 1), measurements from \citet{Zwitter2020} with suffix \texttt{obst} based on an improved algorithm based on the work by \citet{Zwitter2018}. The latter estimates were performed with a grid of 718 template spectra (created from observed HERMES spectra{ in selected \Teff, \logg, \feh, and \alphafe bins). These make use of the whole spectrum rather than just a specific wavelength regions used for the stellar parameter estimation with the main pipeline (providing \vrad estimates under \texttt{rv\_sme\_v2}).  We furthermore provide \vrad estimates which correct for gravitational redshift (\texttt{rv\_obst}, which we recommend to use). The catalogue also includes corrections for incorrect barycentric velocity shifts as outlined in Sec.~\ref{sec:rvcaveats} and \vrad from \Gaia DR2, as reported in \Gaia EDR3.
{Based on feedback from the scientific community, we report our recommendation for \vrad estimated from GALAH in the columns \texttt{rv\_galah} and \texttt{rv\_galah}, together with a flag \texttt{use\_rv\_flag} to indicate their origin (0 for 563260 spectra using \texttt{rv\_obst}, 1 for 86173 spectra using \texttt{rv\_sme\_v2} and 2 for 14190 spectra, when we the latter column is empty, but we recommend to use the available \texttt{dr2\_radial\_velocity} value from \Gaia DR2 or 4 for 14800 spectra for which no RV is available in either survey.
The table schema of the VAC is included in the FITS file but can also be found at \url{https://datacentral.org.au/services/schema/}.

\subsubsection{Double-lined spectroscopic binary stars} \label{sec:vac_binaries}

Binary stellar systems represent a significant fraction of stars in our Galaxy. Therefore, their effect on observations, as well as their impact on the Galactic environment, have to be properly taken into account when studying Galactic structure and evolution. To this end, we present a sample of 12\,760 binary systems for which the properties of their stellar components were derived in a separate analysis from that described in Sec.~\ref{sec:analysis}. In order to compute individual parameters for both stars ($T_{\rm eff[1,2]}$, $\log g_{[1,2]}$, $V_{r[1,2]}$, $v_{\rm mic[1,2]}$, $v_{\rm broad[1,2]}$, $R_{[1,2]}$), together with a common metallicity and extinction for the binary system (\feh, $E(B-V)$), we combine information from GALAH spectra, \Gaia DR2 parallax, and data from several photometric surveys (APASS; \citealt{APASS}, \textit{Gaia} DR2; \citealt{Brown2018}, 2MASS; \citealt{Skrutskie2006}, WISE; \citealt{Cutri2013}) into a joint Bayesian scheme. The details of the analysis are described in \citet{Traven2020}, and the catalogue of derived parameters is available at CDS\footnote{\url{http://cdsarc.u-strasbg.fr/viz-bin/cat/J/A+A/638/A145}}. 

The binary stars presented in this VAC were detected in a sample of 587\,153 spectra from the second GALAH internal data release. We investigated direct products of the reduction pipeline before implementation of some improvements described in Sec.~\ref{sec:reductions}. Detection of binarity was performed using a t-SNE classification and a cross-correlation analysis \citep{Merle2017, Traven2017} of GALAH spectra. The final sample of this catalogue consists of systems with mostly dwarf components, a significant fraction of evolved stars, and also several dozen members of the giant branch.  The statistical distributions of derived stellar properties can be further used for population studies (G.~Traven et al., in prep.), and show trends which are expected for a population of close binary stars ($a < 10\,\mathrm{au}$) with mass ratios $0.5 \leq q \leq 1$. Our results also indicate that the derived metallicity of binary stars is statistically lower than that of single dwarf stars observed in the same magnitude-limited sample of the GALAH survey. Among other reasons, this might point to an anti-correlation between the binary fraction and metallicity of close binary stars, as recently explored by e.g. \cite{Moe2019}, \cite{Bate2019}, and \cite{PriceWhelan2020}.

%________________________________________________________________
\section{GALAH+~DR3 in context} \label{sec:galah_in_context}

\begin{figure*}
\centering
\includegraphics[width=\textwidth]{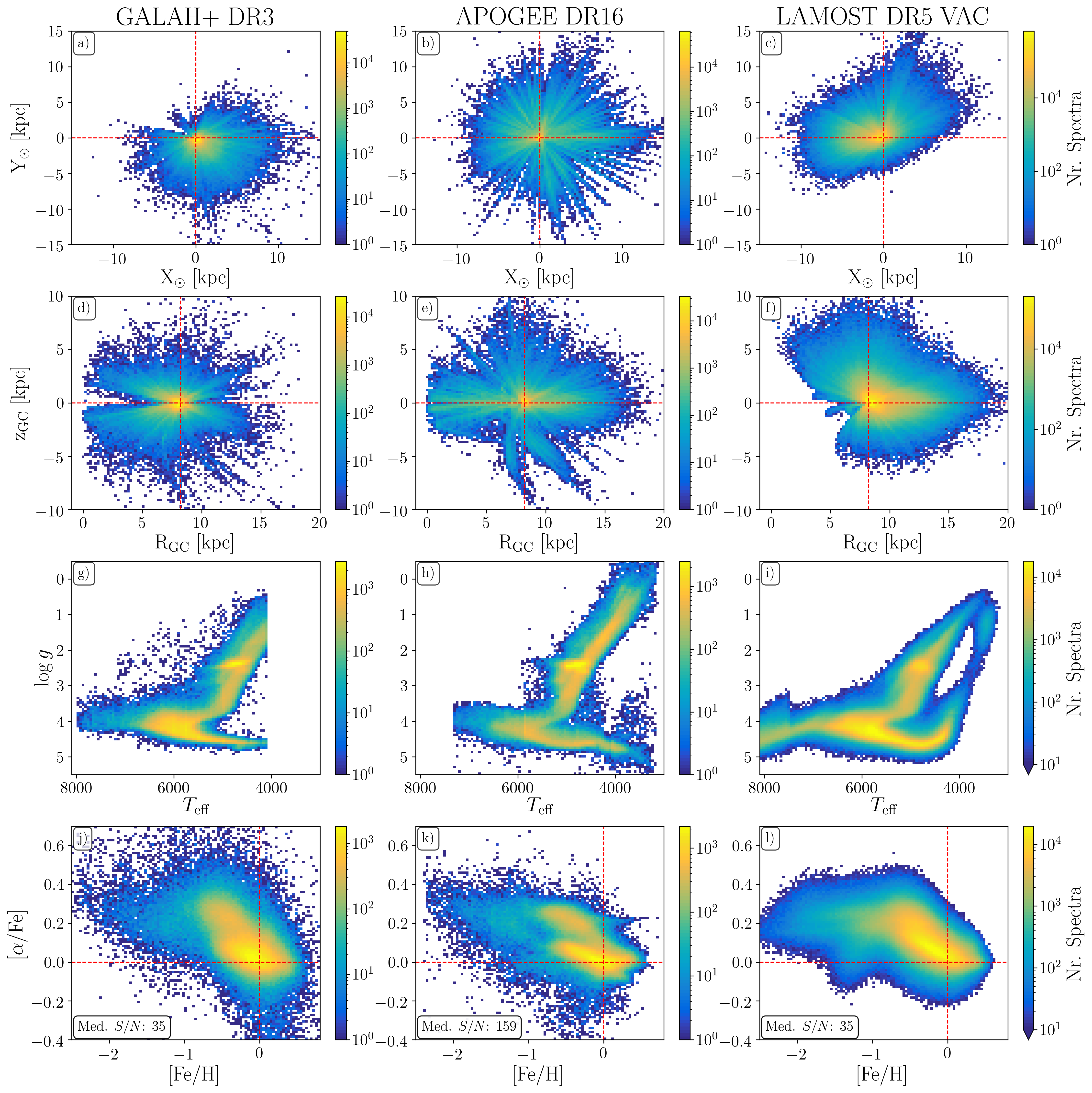}
  \caption{
  \textbf{Comparison of GALAH+~DR3 (left panels) with APOGEE~DR16 (middle panels) and LAMOST DR5 VAC (right panels).} The surveys trace different Galactic regions as shown in the heliocentric cartesian frame (a-c) as well as the Galactocentric cylindrical frame (d-f) across different stellar types (seen in the overview of the \Teff-\logg coverage in panels g-i) across different stellar populations (shown in the Tinsley-Wallerstein diagrams, \feh vs. \alphafe, in panels j-l). Numbers in the bottom left of panels j-l indicate the median SNR for CCD2 of GALAH, SNR for APOGEE, and SNR G for LAMOST for the shown stars, which are indicative of the precision that can be reached by the spectrum analysis. We note that the colour bars of all panels are have different scales.}
  \label{fig:dr3_apogee_lamost}
\end{figure*}

The GALAH collaboration releases millions of abundance measurements for 678\,423 spectra of 588\,571 stars. In this section, we put this achievement into perspective. This release provides, to the best of our knowledge, the largest number of element abundances from high-resolution ($R{\sim}28\,000$) spectra published so far for a well-selected sample of stars with the promise of most precise dynamic and age information. This number is, however, still rather small compared to the roughly 1.5 billion stars observed by \Gaia \citep{Brown2016}, which aims to observe about 1\% of all Milky Way stars, and also limited mainly to stars in the Solar vicinity within $4\,\mathrm{kpc}$. However thanks to our information on stellar orbits, we can learn infer results beyond the volume of the present day stellar positions in the Galaxy \citep{May1986}.

To be able to perform Galactic archaeology on a truly galactic scale, it is therefore vital to be able to use the measurements of other large scale stellar surveys. In Sec.~\ref{sec:global_ga} we compare some key properties, like spatial coverage as well as observed stellar types and the major abundance tracers \feh and \alphafe from GALAH+~DR3 with those from two other ongoing surveys, namely APOGEE \citep[DR16][]{SDSSDR16, Joensson2020} and LAMOST \citep[DR5][]{Deng2012, Zhao2012, Xiang2020}. Both of these surveys provide an millions of measurements for multiple elements, but we note that there are several other surveys which also provide abundance measurements (but for typically fewer elements and/or stars), like the Geneva-Copenhagen-Survey \citep{Nordstroem2004, Casagrande2011}, SEGUE \citep{Yanny2009}, RAVE \citep{Steinmetz2020a, Steinmetz2020b}, and \Gaia-ESO \citep{Gilmore2012}.
In Sec.~\ref{sec:gce}, we highlight the potential to further our understanding of Galactic and stellar chemical evolution with Li, one of the 30 elements measured by GALAH. In Sec.~\ref{sec:cde}, we then showcase some of the specific advantages of GALAH for the exploration of the chemodynamic evolution of the Milky Way.

\begin{figure*}
\centering
\includegraphics[width=\textwidth]{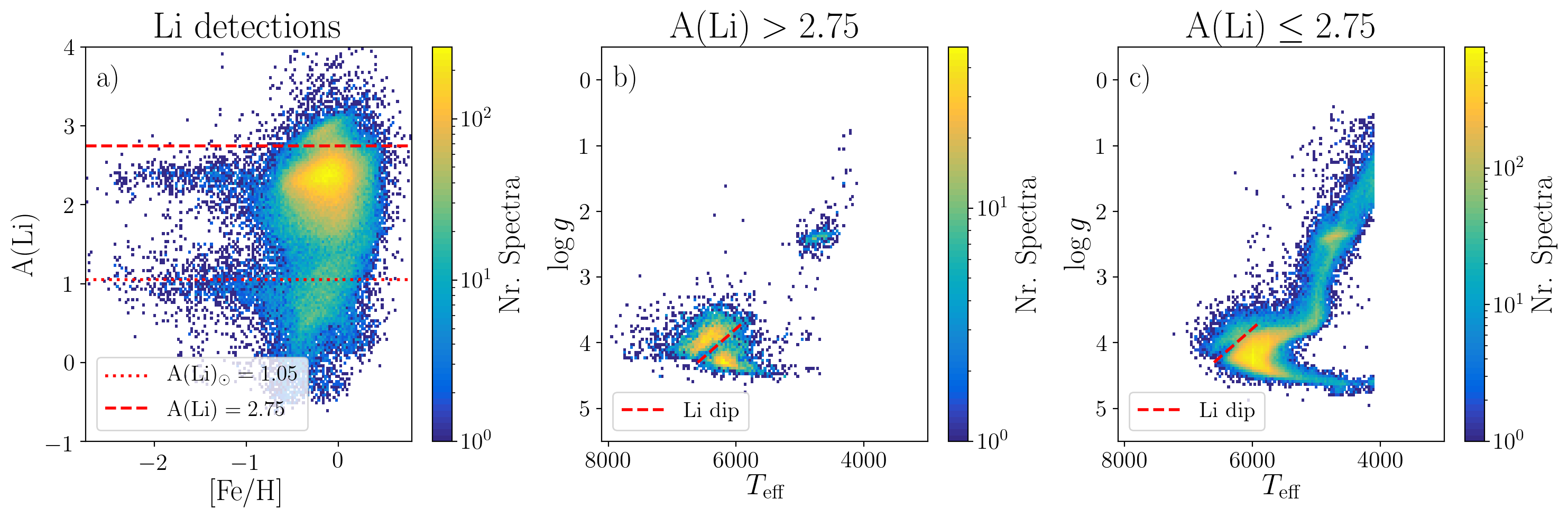}
  \caption{
  \textbf{Overview of Lithium in the abundance and parameter plane.}
\textbf{Panel a)} shows the distribution of A(Li) as a function of \feh for all sucessful measurements (no flags). The abundances cover a range of almost $5\,\mathrm{dex}$, with the majority of stars between 2 and $3\,\mathrm{dex}$. We also show dotted and dashed lines of Solar and primordial A(Li), respectively. Several stars extend towards low \feh with A(Li) on the Spite plateau ($\mathrm{A(Li)}{\sim}2.3\,\mathrm{dex}$).
\textbf{Panels b) and c)} show the distribution of stars in the Kiel diagram with A(Li) above or below the primordial ($\mathrm{A(Li)} = 2.75\,\mathrm{dex}$). Li-rich stars are either warm and cool dwarfs (with th exception of stars at the Li dip, indicated as dashed red line) or Li-rich giants, mainly in the red clump stage. Detected Li-poorer stars, however, cover the whole parameter range.
}
\label{fig:DR3_Li_overview}
\end{figure*}

\subsection{Galactic archaeology on a global scale}  \label{sec:global_ga}

To understand how we can use the available surveys on a global scale, two key points need to be considered. Firstly, if the surveys are complementary and secondly if the surveys are on the same scale.

{While a detailed comparison of overlapping stars between the surveys would be very helpful to assess systematic trends, it should be performed in a dedicated study and include comparisons of stellar parameters as well as abundances for different parameter selections. Some basic comparisons can be found in our open-source repository, but a detailed study is beyond the scope of this paper.

{In this section, we are rather aiming to give explanations why different surveys may not depict the same trends on first glance (based on different selection functions), while highlighting the potential of combining the different surveys.

GALAH+~DR3 includes 678\,423 combined spectra of 588\,571 stars, obtained at high-resolution (28\,000) in 4 narrow optical bands (covering 1000\,\AA).
APOGEE~DR16 includes 473\,307 combined spectra of 437\,445 stars, obtained at high-resolution (22\,500) in the H-band (15\,000-17\,000\,\AA).
LAMOST DR5 VAC includes 8\,162\,566 combined spectra of 6\,091\,116 stars, obtained as low-resolution (1,800) in the full optical range (4000-9000\,\AA). It is important to note that this VAC is estimated by data-driven models trained on GALAH DR2 and APOGEE DR14.

The overlap of GALAH+~DR3 and APOGEE~DR16 is 15\,047 stars, that is 3\% of the each survey. The overlap of GALAH+~DR3 and LAMOST DR5 is 47\,118 stars, that is 8\% and 1\% of the respective survey. The overlap of APOGEE~DR16 and LAMOST DR5 is 111\,626 stars, that is 26\% and 2\% of the respective surveys.

These numbers show that the surveys are very complementary in the stars that they target, but also have a non-negligible overlap between them. Even more important, this overlap allows us to test if these surveys are on the same scale and even to cross-calibrate them to bring them on the same scale \citep[see e.g.][]{Ho2017, Casey2017, Wheeler2020, Xiang2020, Nandakumar2020}.

For the subsequent comparison we limit the samples to those stars with \texttt{flag\_sp} = 0 , \texttt{flag\_fe} = 0, and \texttt{flag\_alpha\_fe} = 0 for GALAH, \texttt{ASPCAPFLAG} = 0 for APOGEE, and \texttt{FLAG\_SINGLESTAR} = 0, \texttt{QFLAG\_CHI2} = ``good'' as well as SNR ratios for at least 30 for either G, R, or I for LAMOST's DR5 VAC.

Because the three surveys operate on different sites, they are typically observing different regions of the sky. This can be seen in Fig.~\ref{fig:dr3_apogee_lamost}a-f, where we show the spatial distribution of stars in heliocentric cartesian coordinates ($X$ vs. $Y$ in Fig.~\ref{fig:dr3_apogee_lamost}a-c) and Galactocentric cylindrical coordinates ($R$ vs. $z$ in Fig.~\ref{fig:dr3_apogee_lamost}d-f). While GALAH observes stars of the Southern hemisphere, LAMOST targets mainly the Northern hemisphere (compare Fig.~\ref{fig:dr3_apogee_lamost}a and c), and APOGEE observes both hemispheres. When looking at the Galactic spatial distribution, we see the selection function of GALAH, especially $\vert b \vert > 10\,\mathrm{deg}$ introducing a lack of stars in the plane (panel d), whereas APOGEE is mainly targeting the plane (small $z$ in panel e) and LAMOST (panel f) targets all regions except the inner Galaxy.

Fig.~\ref{fig:dr3_apogee_lamost}f-h depicts the distribution of \Teff and \logg for the surveys, which now all deliver results for all different stellar types and evolutionary stages (for example APOGEE, which mainly focussed on the observation and analysis of giants in previous releases, now also delivers dwarf parameters with DR16).

The elemental abundances obtained by these surveys are data of particular interest for Galactic archaeology. A detailed comparison of those between the surveys is beyond the scope of this paper. Typically, different surveys operate at different resolutions and reach different $S/N$ in different wavelength regions, thus selecting different lines for their analyses. Different lines again, can form at different optical depths and may be blended differently; all possible factors for possibly different abundance measurements \citep{Jofre2019}.

For $\upalpha$-element abundances another important consideration is how these are defined and computed. For GALAH+~DR3, we provide individual element abundances for Mg, Si, Ca, and Ti, but also compute a combined \alphafe value from error-weighted combinations of well selected individual lines from these elements (see Sec.~\ref{sec:spec_details}), resulting in the distribution shown in Fig.~\ref{fig:dr3_apogee_lamost}j. Because of the differences in yields between these different $\upalpha$-elements, the enhancement pattern of different $\upalpha$-process elements looks slightly different to each other, and a combined $\upalpha$-enhancement label is only a compromise to reach a higher precision. For DR3, this compromise is dominated by Si and Ti, followed by Mg as the most precisely measured elements, with rather little contribution from Ca. For APOGEE~DR16, on the other hand, [$\upalpha$/M], which we convert to \alphafe in Fig.~\ref{fig:dr3_apogee_lamost}k is computed using all lines in the APOGEE wavelength range and adjusting all of the [X/Fe] at the same time by the same amount. For LAMOST DR5 we show the VAC estimates trained on GALAH~DR2 by \citet{Xiang2020}. When comparing these distributions, it is important to keep in mind the quality of data that was used for the analysis. The median $S/N$ for GALAH and LAMOST is 35, which is 4.5 times lower than the median $S/N$ of 159 achieved by APOGEE. We therefore expect that the scatter for GALAH and LAMOST is larger, as can be seen in Fig.~\ref{fig:dr3_apogee_lamost}j-l. Furthermore it is important to keep in mind that these distributions trace different regions of the sky, different distributions of stellar types, and thus likely also different distributions of stellar populations. Especially for APOGEE, we expect a larger ratio of stars from the bulge and high-$\upalpha$ disc, which will change the colourmap distribution.

When comparing with APOGEE~DR16 abundances quantitatively (see Table~\ref{tab:solar_reference_values2}), we find an excellent agreement for most abundance zero points, that is sky flats and vesta, including $0.00 \pm 0.01\,\mathrm{dex}$ for \feh and $-0.01\pm0.05\,\mathrm{dex}$ for \alphafe. The difference for all stars with unflagged abundances between APOGEE~DR16 and GALAH shows a slightly lower \feh for GALAH ($-0.05 \pm 0.14\,\mathrm{dex}$) and slightly higher \alphafe ($0.02 \pm 0.07\,\mathrm{dex}$). For a comparison of the other elements we refer to Table~\ref{tab:solar_reference_values2}.

\subsection{Galactic and stellar chemical evolution} \label{sec:gce}

In this section we briefly aim to show the potential of GALAH+~DR3 for the exploration of Galactic and stellar chemical evolution, while leaving the true exploration to the scientific community. One would ideally like to take all abundance measurements into account for such an endeavour, including GALAH's main goal of the chemical tagging experiment, but here we aim to show how much potential the exploration of a single element has to offer.

In Fig.~\ref{fig:DR3_Li_overview}, we plot the distribution of lithium in different projections. Fig.~\ref{fig:DR3_Li_overview}a shows the absolute abundance A(Li), as a function of \feh. We indicate two important values, the theoretical prediction of $\mathrm{A(Li)} = 2.75$ from the Big Bang Nucleosynthesis \citep{Pitrou2018} and the photospheric abundance of the Sun $\mathrm{A(Li)_\odot} = 1.05$ \citep{Asplund2009}. First of all, it is important to notice that we only plot the stars with unflagged Li measurements (\texttt{flag\_sp} = 0 and \texttt{flag\_li\_fe} = 0). These are 127\,674 measurements or 18.8\% of all GALAH+~DR3 stars. 

In this projection, several substructures are noticeable. While the mean abundance of all stars is $\mathrm{A(Li)} = 2.2_{-0.7}^{+0.3}\,\mathrm{dex}$, we actually see a large spread of A(Li) between $-0.5$ and $4.0\,\mathrm{dex}$ across for $\mathrm{[Fe/H]} > -1\,\mathrm{dex}$. Among many others, \citet{Ramirez2012} and \citet{Bensby2018} explored this pattern in their studies extensively by analysing its correlation with stellar parameters, stellar populations as well as age and temperature and found a strong correlation for example between temperature and A(Li). When we only plot the stars with the largest A(Li), especially above the theoretical primordial value of 2.75, we find that only specific groups of stars exhibit these abundances, see Fig.~\ref{fig:DR3_Li_overview}b, namely hot dwarf stars and few lithium-rich giants. Because of the dredge-up, we would not expect such high amounts of Li in giants, which questions our understanding of stellar physics and evolution. With the new more reliable data from GALAH+~DR3, more scenarios of Li production during binary interaction or the He-flash in giant stars \citep[see e.g.][]{Casey2019, Kumar2020} can be tested more reliably, indicating that lithium-rich giant stars require multiple formation channels \citep{Martell2020}.

Thanks to the hundreds of thousands of Li measurements, we are also able to study phenomena which previously have mainly been analysed in cluster stars, such as the occurrence of the Li dip \citep{Boesgaard1986}, a region among the warm dwarf stars, for which deep mixing induced by rotation and meridional circulation causes
strong Li depletion. The first analysis of this region with GALAH+~DR3 by \citet{Gao2020} has identified a significant offset between the warm and cool side of this Li dip of $0.4-0.5\,\mathrm{dex}$. Down to metallicities of $\mathrm{[Fe/H]}{\sim}-1\,\mathrm{dex}$ this offset appears metallicity-independent which sheds new light on the famous disagreement between predicted Li abundance and the one measured in cool, old, metal-poor on the Spite plateau of $\mathrm{A(Li)}{\sim}2.3\,\mathrm{dex}$ \citep{Spite1982}. In particular, \citet{Gao2020} speculate that the most metal-poor stars on the warm side of the dip may have experienced insignificant Li depletion as well as insignificant Galactic Li enrichment, naturally explaining why their abundances closely reflect those predicted by standard Big Bang nucleosynthesis.

Several of these metal-poor stars ($\mathrm{[Fe/H]} < -1\,\mathrm{dex}$) have actually been identified as stars of the accreted \Gaia-Enceladus-Sausage. With GALAH data, both \citet{Molaro2020} and \citet{Simpson2020} show that the distribution of A(Li) from the accreted stars, like the GES agrees with different populations of the Milky Way, an important confirmation that the Cosmological Lithium problem is not a consequence of formation environment \citep[see also][]{Nissen2012,Cescutti2020}.

\begin{figure*}
\centering
\includegraphics[width=\textwidth]{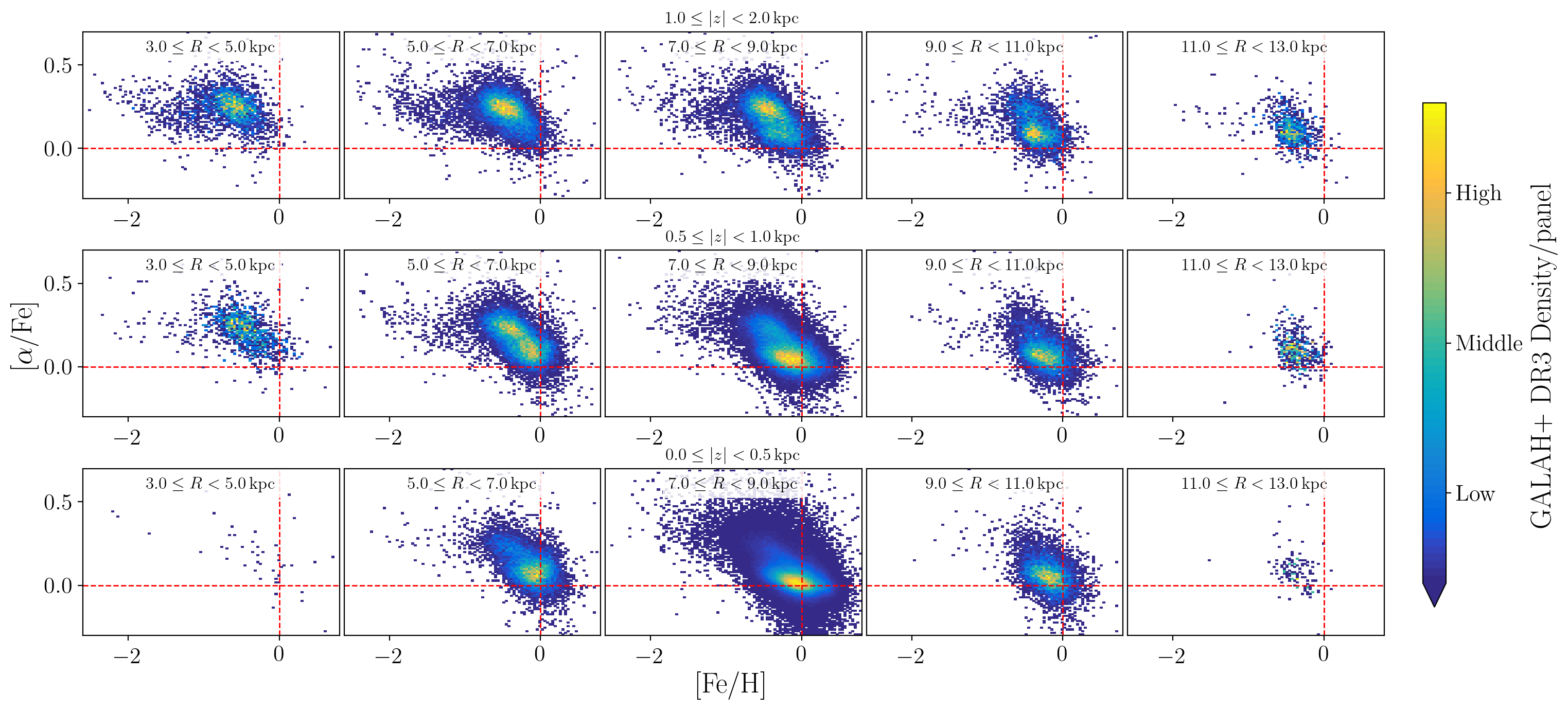}
  \caption{
  \textbf{Spatial coverage of GALAH shown with Tinsley-Wallerstein diagrams, \feh vs. \alphafe, for different regions ($R$,$z$) of the Galaxy.}
  With 81.2\% of stars within $2\,\mathrm{kpc}$, the majority of stars are located in the bottom middle panels. However, we see a evidence of a similar structure as was observed by \citet{Hayden2015} with APOGEE data, that is, firstly a gradient of \feh for the low-$\upalpha$ disc with decrasing \feh towards larger radii, and secondly a coordinate independent chemical composition of the majority of high-$\upalpha$ disc.}
  \label{fig:RZ_alpha}
\end{figure*}

\subsection{Chemodynamical evolution}  \label{sec:cde}

To assess the potential of GALAH+~DR3 in terms of exploring the chemodynamic evolution of the Milky Way, we show the distribution of the data in plots that have been used in seminal studies for Galactic exploration.

Similar to \citet{Hayden2015}, we plot the distribution of $\upalpha$-enhancement versus iron abundance for stars of GALAH+~DR3 in different spatial bins, that is different bins in Galactic radius (from inner Galaxy on the left to outer Galaxy on the right) as well as Galactic height (from the Galactic plane in the bottom to more than $1\,\mathrm{kpc}$ above or below the plane in the top) in Fig.~\ref{fig:RZ_alpha}. The Solar vicinity, which is located in the bottom centre of this figure hosts by far the most stars of GALAH+~DR3 and consists mainly of low-$\upalpha$ (thin) disc stars. When looking at larger Galactic heights, stars of the high-$\upalpha$ (thick) disc become dominant, in good agreement with the results by \citet{Hayden2015} based on APOGEE data. However, we find a less pronounced abundance gradient of the low-$\upalpha$ disc with Galactic radius, which we believe is attributed to less reliable iron abundances of distant, metal-rich giants (mainly expected in the inner Galaxy), as outlined in Sec.~\ref{sec:pecularities_groups}. In agreement with \citet{Hayden2015}, we see a clear separation of the overdensities of low- and high-$\upalpha$ disc stars and a spatial invariance of the position of the peak/distribution of the high-$\upalpha$ disc stars. With the improved distances thanks to the \Gaia mission, we are able to also explore the most distant bins (beyond $2\,\mathrm{kpc}$) of this spatial distribution and find several stars with \feh below $-1\,\mathrm{dex}$ at larger Galactic heights ($\vert z \vert > 0.5\,\mathrm{kpc}$), coinciding with the chemical composition of the metal-weak extension of the high-$\upalpha$ disc as well as most halo stars, including the recently identified \Gaia-Enceladus-Sausage stars (see further explanations below).

\begin{figure*}
\centering
\includegraphics[width=\textwidth]{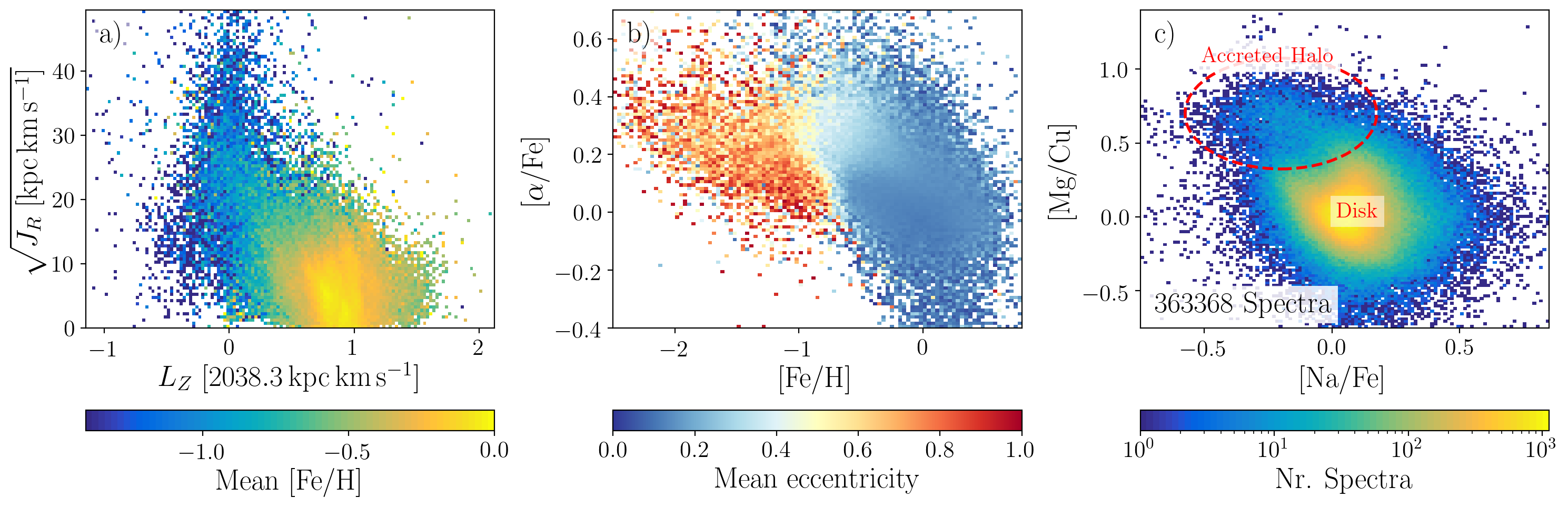}
  \caption{
\textbf{The potential of chemodynamic studies, shown with an overview of actions and abundances.}
\textbf{Panel a)} shows the action distribution from Fig.~\ref{fig:DR3_dynamics_overview}c, but here coloured by mean \feh per bin \citep[see e.g.][]{Trick2019, Feuillet2020}.
\textbf{Panel b)} shows the Tinsley-Wallerstein diagram, \alphafe over \feh from Fig.~\ref{fig:dr3_apogee_lamost}j, but here coloured by mean eccentricity per bin \citep[see][]{Mackereth2019}.
\textbf{Panel c)} shows another abundance plane, but the elements Na, Fe, Mg, and Cu tracing different element/nucleosynthesis groups with the majority of GALAH stars around (0,0), where the majority of disc stars is located, and a distinct overdensity of accreted halo stars in the upper left \citep[see ][]{Das2020}.
}
  \label{fig:DR3_action_xfe}
\end{figure*}

With the provided VAC on stellar ages, we are also able to assess the data set by this important property. Many recent studies \citep[e.g.][]{Haywood2013, Bensby2014, Minchev2017, Hayden2017, Haywood2019} have shown the potential of including ages when assessing the Milky Way populations. \citet{Buder2019} have further shown that age and chemistry combined (as more conserved properties than kinematics/dynamics) can help to dissect the disc populations. Among the 415653 stars in GALAH+DR3 with unflagged stellar parameters, \feh, and \alphafe as well as ages, we find 1.8\% with $\mathrm{[Fe/H]} \leq -1$. When assigning the other stars to young ($\leq 8\,\mathrm{Gyr}$) and old ($> 8\,\mathrm{Gyr}$) as well as low-$\upalpha$ ($\mathrm{[\upalpha/Fe]} \leq 0.2$) and high-$\upalpha$ ($\mathrm{[\upalpha/Fe]} > 0.2$) groups, we find 62.5\% young low-$\upalpha$ stars, 8.8\% young high-$\upalpha$ stars \citep[compare to 5.8\% found within APOKASC by][]{Martig2015}, and 26.9\% old stars (21.5\% low-$\upalpha$ and 5.4\% high-$\upalpha$).

The vast majority of GALAH targets, especially the 62.5\% young low-$\upalpha$ stars, are expected to move on orbits very similar to the Sun. In Fig.~\ref{fig:DR3_dynamics_overview} this is confirmed in all panels of kinematic and dynamic properties, where most stars are located close to the Sun ($V_\odot = 248.27\,\mathrm{km\,s^{-1}}$, $J_\phi = L_Z = 2038.3\,\mathrm{kpc\,km\,s^{-1}}$) and exhibit only small radial and vertical velocities / actions.

Although halo stars are not the main target of GALAH, roughly 1\% of all GALAH targets are expected to belong to the chemical or kinematic halo \citep{DeSilva2015}. While the definition of halo stars is contentious, we at least aim to assess their rough number by looking at different kinematic and dynamic properties. For this, we look at the distribution of azimuthal / transversal velocity $V_T$ with respect to the combined radial and vertical velocity $\sqrt{V_R^2 + V_z^2}$ in Fig.~\ref{fig:DR3_dynamics_overview}.

The majority of stars move on almost circular orbits at Solar radius ($V_T{\sim}v_\text{circ} = 233.1\,\mathrm{km\,s^{-1}}$). Half of all GALAH stars differ by less than $57\,\mathrm{km\,s^{-1}}$ from this total velocity. Only 8.2\,\%, 4.4\,\%, and 2.4\,\% are more than 140, 180, and $233.1\,\mathrm{km\,s^{-1}}$ from this total velocity. In the literature, the latter two values have been used to assign stars to the kinematic halo, and while such distinct cuts are debatable, their numbers are significantly higher than the initially estimated 1\% \citep{DeSilva2015}, partially due to the additional surveys like K2-HERMES contributing to GALAH+Dr3. These stars do not move coherently with the (local) disc, but are on kinematically hotter orbits. This suggests that they are for example halo stars or belong to the bulge. 1.2\,\% of the stars even move on retrograde orbits.

Similar to \citet{Belokurov2018} we can identify a -like overdensity of the \Gaia-Enceladus-Sausage (GES) \citep[see e.g.][and references therein]{Helmi2020} in Fig.~\ref{fig:DR3_dynamics_overview}b along an extended range of $-400 < V_R < 400\,\mathrm{km\,s^{-1}}$ along small Galactocentric azimuthal velocities, that is, following closely the dashed red line indicating $V_T{\sim}0\,\mathrm{km\,s^{-1}}$. While the stars stick out in this projection, the shown properties are not conserved and it is therefore advisable to also inspect the conserved properties of actions.

The distribution of stars in action space is shown in a view of vertical angular momentum (normalised to the Solar value) and radial action in Fig.~\ref{fig:DR3_dynamics_overview}c. Most of the stars in this diagram show a similar vertical angular momentum radial action as the Sun ($L_Z = 2038.3\,\mathrm{kpc\,km\,s^{-1}}$, $J_R = 7.7\,\mathrm{kpc\,km\,s^{-1}}$). Similar to the analyses by \citet{Trick2019}, a much richer substructure can be seen when compared to Fig.~\ref{fig:DR3_dynamics_overview}b. The overdensity of stars around $L_Z{\sim}0 \,\mathrm{kpc\,km\,s^{-1}}$ with higher radial actions is typical for stars of the Galactic halo, especially those of the GES \citep[see e.g.][and references therein]{Helmi2020}. When looking at the distribution of stellar actions relative to their total actions in Fig.~\ref{fig:DR3_dynamics_overview}d, it again becomes evident that most stars are on near-circular orbits ($L_Z{\sim}L_\text{tot}$). However, this projection also allows the identification of accreted stars and stars of streams \citep[see e.g.][]{Vasiliev2019, Myeong2019, Monty2020}. Stars of the GES are to be found in the lower corner of this plot, and stars of the Sequoia \citep{Myeong2019} in the left corner.

It should be noted, that the globular cluster $\omega$\,Cen, targeted by GALAH and thus part of this data release, sticks out in all panels of Fig.~\ref{fig:DR3_dynamics_overview} either as overdensity at low $V_T{\sim}-85\,\mathrm{km\,s^{-1}}$ in Fig.~\ref{fig:DR3_dynamics_overview}a and b or as streak at $-0.25 L_{Z,\odot}$ in Fig.~\ref{fig:DR3_dynamics_overview}c and $-0.6 J_\phi/J_\text{tot}$ in Fig.~\ref{fig:DR3_dynamics_overview}d. Although beyond the scope of this paper, our release provides new and diverse data to follow up the connection of this globular cluster and possible remnant of a tidally disrupted dwarf galaxy \citep[e.g.][]{Bekki2003} in combination with other stars with similar chemodynamic properties \citep{Myeong2018b} and assess if the streaks are only mainly caused by the high parallax uncertainty of 46\% for stars in $\omega$\,Cen or might coincide with a true extension.

When combining dynamic information (such as actions and eccentricities) with chemistry (like \feh and $\upalpha$-enhancement) in chemodynamic projections, we can see the potential of GALAH+~DR3 in action. Until the recent years, analyses of the Milky Way had usually been performed either from a spectroscopic/chemical or dynamical point of view. Thanks to the advent of \Gaia and stellar spectroscopic surveys, we can now bring together both disciplines.

In Fig.~\ref{fig:DR3_action_xfe}a, we plot the distributions in action bins coloured by their mean \feh. Similar to Fig.~\ref{fig:DR3_dynamics_overview}c, we see a right substructure, which strongly suggests a correlation of resonances with certain iron abundances. Furthermore, we see a gradient of iron abundance with lower angular momenta dropping from $\mathrm{[Fe/H]} = -0.13_{-0.22}^{+0.20}\,\mathrm{dex}$ at $1.00_{-0.05}^{+0.05}L_{Z,\odot}$ to $\mathrm{[Fe/H]} = -0.99_{-0.47}^{+0.39}\,\mathrm{dex}$ at $0.00_{-0.05}^{+0.05}L_{Z,\odot}$. In Fig.~\ref{fig:DR3_action_xfe}b we plot a chemical overview, coloured by the dynamic property of eccentricity, as performed previously \citep[e.g.][]{Schuster2012, Mackereth2019}. Here we see that the low-$\upalpha$ disc stars are typically on rather circular orbits (with eccentrics well below 0.5), whereas high-$\upalpha$ disc stars exhibit higher eccentricities around mean values of 0.5. The most striking feature in this projection is the stars with low \feh, which almost exclusively show eccentricities above 0.5 (stars with \feh below $-1.0\,\mathrm{dex}$ move on orbits with typical eccentricities of $e = 0.70_{-0.39}^{+0.23}$). This is strong evidence that these stars, with chemical composition that are very distinct from the stellar disc and bulge, and orbits very different from the disc are accreted \citep[see discussions in][]{Belokurov2018, Helmi2018, Mackereth2019, Helmi2020}.

That these stars are not only different in their dynamics, can be seen in a chemical projection in Fig.~\ref{fig:DR3_action_xfe}c, where we follow up the distinct chemical signatures of accreted halo stars as found by \citet{Nissen2010} and \citet{Nissen2011} in the projections similar to those proposed by \citet{Hawkins2015} and \citet{Das2020}. When assessing different nucleosynthesis channels via different elements, that is Al or Na, $\upalpha$ like Mg, and Cu or Mn, the accreted halo stars clearly stick out as a distinct overdensity because of their different chemical enrichment history compared to the majority of the Milky Way disc stars. A follow-up of these findings will be presented in the chemodynamical study of accreted halo stars by S.~Buder et al. (in prep.).

\section{Conclusions and outlook} \label{sec:conclusions}

With this third data release of the Galactic Archaeology with HERMES (GALAH) survey, we are providing the most complete set of information in terms of chemical composition, dynamics, and stellar ages to the public. The new data provides abundances for up to 30 elements and with the additional astrometric information provided by the \Gaia satellite, we are able to estimate very precise orbits for almost all stars. This data is extremely valuable for different disciplines of astrophysics and will bring together observers with theorists.

\subsection{The data and their usage}

In this manuscript, we describe the methodology behind the newly released data. This release incorporates data from GALAH's partner surveys, namely the K2/HERMES and TESS-HERMES surveys, yielding a total sample of 678\,423 spectra for 588\,571 stars. Regarding the use of our data, we conclude:
\begin{itemize}
    \item Use version 2 of our data catalogues. They include fixes to previously reported erroneous values for the radial velocities \texttt{rv\_galah} and a new VAC for the crossmatch with \Gaia eDR3 (see Sec.~\ref{sec:version2} and references therein for more details) updates to the VACs based on the new \Gaia data.
    \item Select the measurement method of \vrad based on the science cases from our VAC on \vrad. For comparison with other surveys, we recommend preferably \texttt{rv\_nogr\_obst} or otherwise \texttt{rv\_sme\_v2}. For dynamical studies, we recommend the use of \texttt{rv\_obst} (including gravitational redshift corrections) if available and report the best measurement for such dynamical studies as \texttt{rv\_galah} (see Sec.~\ref{sec:rv_vac} for details).
    \item Use only unflagged measurements of our main stellar parameter quality flag ($\texttt{flag\_sp} = 0$).
    \item Use only unflagged abundance measurements for each element X ($\texttt{flag\_X\_fe} = 0$). Even in this case, be careful of suspicious abundances and trends, which may not have been caught by our flagging algorithms. Systematic trends may be introduced through inherent limitations of the analysis pipeline. Please consult Sec.~\ref{sec:caveats} on caveats, including high abundances of V, Co, Rb, Sr, Zr, Mo, Ru, La, Nd, and Sm.
    \item We report both individual as well as stacked visit spectra. The latter result in higher $S/N$ and are generally the preferred values reported per star in the \texttt{allstar} catalogues. For studies of individual stars we recommend, however, to also compare all results of individual spectra of the stars from the \texttt{allspec} catalogues to confirm that the automatic stacking process was successful (see Sec.~\ref{sec:stacking} for details).
\end{itemize} 

\subsection{Scientific avenues for the use of GALAH DR3}

Since the advent of galactic archaeology \citep[][and references therein]{FreemanBlandHawthorn2002}, many large stellar surveys attempt to establish a narrative for the Galaxy by comparing vast amounts of stellar data (ages, kinematics, chemistry) to cosmological N-body $+$ hydrodynamic simulations \citep[e.g.][]{Kobayashi2011,ElBadry2018, Buck2019}. These comparisons assume the present-day Milky Way to be an axisymmetric system in dynamical equilibrium where measurables can be expressed as a function of Galactocentric radius, $R$ \citep{Sharma2011}. 

{Astronomers have long suspected there is much to learn from examining dynamical perturbations and their dependence on the stellar properties \citep{Minchev2009,Widrow2012}. We refer this field as Galactic seismology and identify it as a subset of Galactic archaeology. \citet[][their Appendix A]{BlandHawthorn2020} provide a brief history of Galactic seismology dating back to its first use in 1985. Indeed, in the \Gaia second data release (DR2) just two years ago \citep{Antoja2018}, a remarkable signature of incomplete phase-mixing was uncovered. If we consider a Galactic cylindrical coordinate frame defined by $(R,\phi,z)$, with velocity components $(V_R,V_\phi,V_z)$, the \Gaia team discovered a ``phase spiral'' in the $z-V_z$ plane. The vertical ($z$) oscillation frequency is anharmonic so this signal arises from a corrugated wave propagating across the Galactic disc. GALAH has been used to study this phenomenon in terms of stellar ages, actions and abundances \citep{BlandHawthorn2019,Laporte2019}, with further analyses already under way.

The exact formation of the halo and disc remains enigmatic, but the progress of cosmological simulations is now allowing us to by comparing properties like the chemical bimodality of the Milky Way's stellar disc those of simulated galaxies \citep[e.g.][and references therein]{Buck2020, Vincenzo2020}.

The large amount of stellar data provided by stellar spectroscopic surveys is bringing together expertise of previously independent research. Based on our data, the exoplanet community improves our understanding of exoplanets through their host stars with improved stellar parameters \citep[e.g.][]{Clark2020}, more realistic input for planet formation simulations \citep[e.g.][]{Bitsch2020} and will be able to explore exoplanet host stars in a chemo-kinematic or -dynamic context \citep[see e.g.][]{Carrillo2020}.

With the publication of the reduced spectra, we are going another step towards an open data community. Using the spectra will allow other scientists to not only verify our results, but also apply their analyses techniques for parts of the parameter space, for which our own pipeline is not optimised, e.g. the analysis of very hot stars, emission line stars, or very cool stars, among others. Furthermore does the publication of the spectra allow scientists to apply machine learning or clustering algorithms onto the data \citep[see e.g.][]{PriceJones2019}.

{Starting from such studies of spatially and dynamically bound groups of stars or solar twins, we are just at the beginning of understanding the correlations for field stars between abundances and orbits \citep[see e.g.][]{Coronado2020} as well as the abundances and ages for field stars \citep[see e.g.][]{Morel2020, Hayden2020, Sharma2020} as well as their limitations \citep[see e.g.][]{Feltzing2017, Ness2018}.

\subsection{GALAH DR4 and a sharper focus with GALAH Phase 2}

We have learned several lessons in the analysis for this data release, which will help us to improve our analysis in the future{, that is for the fourth data release of GALAH and thereafer. We have found several interesting trends, of which some are likely astrophysical, while others are not. We will follow these up in the future to hopefully minimise the unphysical trends. Several of these are likely to be addresses by improvements in the reduction of spectra with improved telluric corrections and improved stacking routines.
While an in-depth comparison of the data-driven vs. model-driven approaches is still to be conducted, first results from our work indicates that a quadratic model reaches its limitations when used to describe a very high-dimensional space, covering the stellar parameters along A-M type stars, as well as 30 element abundances. With the introduction of more higher-order models or flexible models and methods, for example neural networks or Gaussian process regression in stellar spectroscopy \citep{Ting2019, Wang2020}, we believe that such limitations can be overcome and will allow to further overcome the significant computational costs of on-the-fly spectrum synthesis. A major limitation of all spectroscopic analyses remains with the immensely uncertain oscillator strengths used to create synthetic spectra, with significantly more effort needed.
In the future, we aim to not only use improved synthetic model grids, based on 3D non-LTE computations, but also implement these more sophisticated interpolation routines combined with an Bayesian framework. The latter would allow us to include non-spectroscopic information like information from \Gaia eDR3 and DR3 in a probabilistic way and help us assess the uncertainties of our estimates more reliably.

One of the most limiting bottlenecks of Galactic archaeology are the still significant uncertainties of stellar ages which can be estimated to no better than 10\% \citep{Soderblom2010}, but are typically significantly higher. With the start of GALAH Phase 2, for which we adjust our target selection to observe more main-sequence turn-off stars to get more reliable age estimates, we also adjusted our observing strategy with longer exposure time to achieve higher spectral quality (and thus higher accuracy and precision). These adjustments will help us to more efficiently collect high-dimensional data of stars in our Solar vicinity and provide the community with a promising data set of chemistry, dynamics, and reliable ages.

%________________________________________________________________
\section*{Acknowledgements}

Based on data acquired through the Australian Astronomical Observatory, under programs: A/2013B/13 (The GALAH pilot survey); A/2014A/25, A/2015A/19, A2017A/18 (The GALAH survey phase 1), A2018 A/18 (Open clusters with HERMES), A2019A/1 (Hierarchical star formation in Ori OB1),  A2019A/15 (The GALAH survey phase 2), A/2015B/19, A/2016A/22, A/2016B/10, A/2017B/16, A/2018B/15 (The HERMES-TESS program), and A/2015A/3, A/2015B/1, A/2015B/19, A/2016A/22, A/2016B/12, A/2017A/14, (The HERMES K2-follow-up program). We acknowledge the traditional owners of the land on which the AAT stands, the Gamilaraay people, and pay our respects to elders past, present, and emerging.

This work has made use of data from the European Space Agency (ESA) mission \Gaia (\url{http://www.cosmos.esa.int/gaia}), processed by the \Gaia Data Processing and Analysis Consortium (DPAC, \url{http://www.cosmos.esa.int/web/gaia/dpac/consortium}). Funding for the DPAC has been provided by national institutions, in particular the institutions participating in the \Gaia Multilateral Agreement. 
This publication makes use of data products from the Two Micron All Sky Survey, which is a joint project of the University of Massachusetts and the Infrared Processing and Analysis Center/California Institute of Technology, funded by the National Aeronautics and Space Administration and the National Science Foundation.
This work was supported by the Australian Research Council Centre of Excellence for All Sky Astrophysics in 3 Dimensions (ASTRO 3D), through project number CE170100013 and the Swedish strategic research programme eSSENCE.
This work was supported by computational resources provided by the Australian Government through the National Computational Infrastructure (NCI) under the National Computational Merit Allocation Scheme (project y89).

The following software and programming languages made this research possible: \textsc{IRAF} \citep{Tody1986,Tody1993}, \textsc{configure} \citep{Miszalski2006}, \textsc{topcat} \citep[version 4.4;][]{Taylor2005}; Python (version 3.7) and its packages {\textsc{astropy}} \citep[version 2.0;][]{Robitaille2013,PriceWhelan2018}, {\textsc{scipy}} \citep{scipy}, {\textsc{matplotlib}} \citep{matplotlib}, {\textsc{pandas}} \citep[version 0.20.2;][]{McKinney2011}, {\textsc{NumPy}} \citep{numpy}, {\textsc{IPython}} \citep{ipython}, and  \textsc{galpy} \citep[version 1.3;][]{Bovy2015}. This research has made use of the VizieR catalogue access tool, CDS, Strasbourg, France. The original description of the VizieR service was published in A\&AS 143, 23. This research mad use of the TOPCAT tool, described in \citet{Taylor2005}. This publication makes use of data products from the Two Micron All Sky Survey, which is a joint project of the University of Massachusetts and the Infrared Processing and Analysis Center/California Institute of Technology, funded by the National Aeronautics and Space Administration and the National Science Foundation.

SB and KL acknowledge funds from the Alexander von Humboldt Foundation in the framework of the Sofja Kovalevskaja Award endowed by the Federal Ministry of Education and Research. SB acknowledges travel support from Universities Australia and Deutsche Akademische Austauschdienst. 
JK and TZ acknowledge financial support of the Slovenian Research Agency (research core funding No. P1-0188) and the European Space Agency (PRODEX Experiment Arrangement No. C4000127986).
AMA acknowledges support from the Swedish Research Council (VR 2016-03765), and the project grants ``The New Milky Way'' (KAW 2013.0052)
and ``Probing charge- and mass- transfer reactions on the atomic level'' (KAW 2018.0028) from the Knut and Alice Wallenberg Foundation.
KL acknowledges funds from the European Research Council (ERC) under the European Union's Horizon 2020 research and innovation programme (Grant agreement No. 852977)
KCF acknowledges support from the the Australian Research Council under award number DP160103747.
LS acknowledges financial support from the Australian Research Council (discovery Project 170100521) and from the Australian Research Council Centre of Excellence for All Sky Astrophysics in 3 Dimensions (ASTRO 3D), through project number CE170100013.
CK acknowledges funding from the UK Science and Technology Facility Council (STFC) through grant ST/ R000905/1, and the Stromlo Distinguished Visitorship at the ANU.
DMN acknowledges support from NASA under award Number 80NSSC19K0589, and support from the  Allan C. And Dorothy H. Davis Fellowship. YST is supported by the NASA Hubble Fellowship grant HST-HF2-51425 awarded by the Space Telescope Science Institute.
GT was supported by the project grant ``The New Milky Way'' from the Knut and Alice Wallenberg foundation and by the grant 2016-03412 from the Swedish Research Council. GT acknowledges financial support of the Slovenian Research Agency (research core funding No. P1-0188 and project N1-0040).
M{\v Z} acknowledges funding from the Australian Research Council (grant DP170102233).

We thank Chris Onken for crossmatching GALAH and SkyMapper DR3 data \citep[see][for more information]{Onken2019}.
We thank Kareem El-Badry for providing a vetted version of wide binaries from \Gaia~DR2 data and GALAH+~DR3 radial velocities.
{We thank Alvin Gavel and Andreas J. Korn for bringing the bug in SME 536 to our attention.
{We thank Scott Trager and many other colleagues who started a discussion about acknowledging the pioneering work by Beatrice M. Tinsley.
We thank the SDSS collaboration and APOGEE team for their efforts in documentation, which greatly benefited our efforts in documentation and validation.

\section*{Data Availability}

The data underlying this article are available in the Data Central at \url{https://cloud.datacentral.org.au/teamdata/GALAH/public/GALAH_DR3/} and can be accessed with the unique identifier \texttt{galah\_dr3} for this release and \texttt{sobject\_id} for each spectrum. For more information (including the single object viewer options and bulk downloads) we refer the reader to the Data Central documentation at \url{https://docs.datacentral.org.au/galah/dr3/overview/}.

\bibliographystyle{mnras}
\bibliography{bib} % if your bibtex file is called example.bib

%%%%%%%%%%%%%%%%%%%%%%%%%%%%%%%%%%%%%%%%%%%%%%%%%%

\noindent \rule{8.5cm}{1pt}
\noindent
% List of institutions
$^{1}$Research School of Astronomy \& Astrophysics, Australian National University, ACT 2611, Australia\\
$^{2}$Center of Excellence for Astrophysics in Three Dimensions (ASTRO-3D), Australia\\
$^{3}$Max Planck Institute for Astronomy (MPIA), Koenigstuhl 17, 69117 Heidelberg, Germany\\
$^{4}$Sydney Institute for Astronomy, School of Physics, A28, The University of Sydney, NSW 2006, Australia\\
$^{5}$Faculty of Mathematics and Physics, University of Ljubljana, Jadranska 19, 1000 Ljubljana, Slovenia\\
$^{6}$Theoretical Astrophysics, Department of Physics and Astronomy, Uppsala University, Box 516, SE-751 20 Uppsala, Sweden\\
$^{7}$Department of Astronomy, Stockholm University, AlbaNova, Roslagstullbacken 21, SE-10691 Stockholm, Sweden\\
$^{8}$School of Physics, UNSW, Sydney, NSW 2052, Australia\\
$^{9}$Max Planck Institute for Astrophysics, Karl-Schwarzschild-Str. 1, D-85741 Garching, Germany\\
$^{10}$Monash Centre for Astrophysics, Monash University, Australia \\
$^{11}$School of Physics and Astronomy, Monash University, Australia\\
$^{12}$Australian Astronomical Optics, Faculty of Science and Engineering, Macquarie University, Macquarie Park, NSW 2113, Australia\\
$^{13}$Department of Physics and Astronomy, Macquarie University, Sydney, NSW 2109, Australia \\
$^{14}$Istituto Nazionale di Astrofisica, Osservatorio Astronomico di Padova, vicolo dell'Osservatorio 5, 35122, Padova, Italy \\
$^{15}$Stellar Astrophysics Centre, Department of Physics \& Astronomy, Aarhus University, Ny Munkegade 120, DK-8000 Aarhus C, Denmark \\
$^{16}$Macquarie University Research Centre for Astronomy, Astrophysics \& Astrophotonics, Sydney, NSW 2109, Australia\\
$^{17}$Leibniz-Institut f{\"u}r Astrophysik Potsdam (AIP), An der Sternwarte 16, D-14482 Potsdam, Germany\\
$^{18}$Centre for Astrophysics, University of Southern Queensland, West Street, Toowoomba, QLD 4350, Australia\\
$^{19}$International Space Science Institute--Beijing, 1 Nanertiao, Zhongguancun, Hai Dian District, Beijing, 100190, China\\
$^{20}$Lund Observatory, Department of Astronomy and Theoretical Physics, Box 43, SE-221\,00 Lund, Sweden\\
$^{21}$ICRAR, The University of Western Australia, 35 Stirling Highway, Crawley, WA 6009, Australia\\
$^{22}$Kapteyn Astronomical Institute, University of Groningen, Landleven 12, 9747 AD Groningen, The Netherlands\\
$^{23}$Centre for Astrophysics Research, Department of Physics, Astronomy and Mathematics, University of Hertfordshire, Hatfield, AL10 9AB, UK\\
$^{24}$Centre for Astrophysics and Supercomputing, Swinburne University of Technology, Hawthorn, Victoria, 3122, Australia\\
$^{25}$Center for Astrophysical Sciences and Department of Physics \& Astronomy, The Johns Hopkins University, Baltimore, MD 21218\\
$^{26}$Department of Astronomy, Columbia University, Pupin Physics Laboratories, New York, NY 10027, USA \\
$^{27}$Center for Computational Astrophysics, Flatiron Institute, 162 Fifth Avenue, New York, NY 10010, USA\\
$^{28}$Istituto Nazionale di Astrofisica, Osservatorio Astronomico di Padova, vicolo dell'Osservatorio 5, 35122, Padova, Italy \\
$^{29}$Centre for Integrated Sustainability Analysis, School of Physics, The University of Sydney, NSW 2006, Australia \\
$^{30}$Institute for Advanced Study, Princeton, NJ 08540, USA \\
$^{31}$Department of Astrophysical Sciences, Princeton University, Princeton, NJ 08544, USA \\
$^{32}$Observatories of the Carnegie Institution of Washington, 813 Santa Barbara Street, Pasadena, CA 91101, USA \\
%%%%%%%%%%%%%%%%% APPENDICES %%%%%%%%%%%%%%%%%%%%%

\appendix

\section{Linelist, reference values, and table schema of the main catalogue}\label{sec:appendix}

Here we append all additional information used for the analysis.

\begin{table*}
\caption{Selected lines for the elemental abundance analysis. The full table is available online as supplementary material.}
  \label{tab:linelist}
\centering
\begin{tabular}{l l l l l p{2.8cm} l l}
\hline
Elem. & Ion & Wavelength [\AA] & LEP [eV] & $\log(gf)$ & Reference & Line mask [\AA] & Segment mask [\AA] \\                                                                                            
\hline    
Li & 1 &6707.7635 & 0.00000 & -0.00200000 & 1998PhRvA..57.1652Y & 6707.3000-6708.3000 & 6705.76-6709.76 \\                                                                                              
Li & 1 &6707.9145 & 0.00000 & -0.303000 & 1998PhRvA..57.1652Y & 6707.3000-6708.3000 & 6705.76-6709.76 \\                                                                                                
Li & 1 &6707.9215 & 0.00000 & -0.00200000 & 1998PhRvA..57.1652Y & 6707.3000-6708.3000 & 6705.76-6709.76 \\                                                                                              
Li & 1 &6708.0725 & 0.00000 & -0.303000 & 1998PhRvA..57.1652Y & 6707.3000-6708.3000 & 6705.76-6709.76 \\                                                                                                
C  & 1 &6587.6100 & 8.53700 & -1.02100 & 1993A\&AS...99..179H & 6587.2610-6587.9860 & 6585.61-6589.61 \\                                                                                                
O  & 1 &7771.9440 & 9.14600 & 0.369000 & NIST & 7771.3590-7772.5090 & 7769.50-7777.50 \\                                                                                                                
O  & 1 &7774.1660 & 9.14600 & 0.223000 & NIST & 7773.5220-7774.7820 & 7769.50-7777.50 \\                                                                                                                
O  & 1 &7775.3880 & 9.14600 & 0.00200000 & NIST & 7774.9120-7775.9620 & 7769.50-7777.50 \\                                                                                                              
Na & 1 &5682.6333 & 2.10200 & -0.706000 & GESMCHF & 5682.5170-5682.9970 & 5680.63-5691.20 \\                                                                                                            
Na & 1 &5688.2050 & 2.10400 & -0.404000 & GESMCHF & 5687.9170-5688.3920 & 5680.63-5691.20 \\                                                                                                            
Mg & 1 &5711.0880 & 4.34600 & -1.72400 & 1990JQSRT..43..207C & 5710.7570-5711.4280 & 5710.00-5713.09 \\                                                                                                 
Al & 1 &6696.0230 & 3.14300 & -1.56900 & 2008JPCRD..37..709K & 6695.7780-6696.1730 & 6695.00-6699.87 \\                                                                                                 
Al & 1 &6698.6730 & 3.14300 & -1.87000 & 2008JPCRD..37..709K & 6698.3920-6698.8950 & 6695.00-6699.87 \\                                                                                                 
Al & 1 &7835.3090 & 4.02200 & -0.689000 & 2008JPCRD..37..709K & 7834.8840-7835.5720 & 7834.00-7837.50 \\                                                                                                
Al & 1 &7836.1340 & 4.02200 & -0.534000 & 2008JPCRD..37..709K & 7835.8130-7836.4310 & 7834.00-7837.50 \\                                                                                                
Al & 1 &7836.1340 & 4.02200 & -1.83400 & 2008JPCRD..37..709K & 7835.8130-7836.4310 & 7834.00-7837.50 \\                                                                  

\dots & \dots & \dots & \dots & \dots & \dots & \dots & \dots \\
\hline \\
\end{tabular}
\newline
\footnotetext{p{2.8cm}}{References: 1982ApJ...260..395C: \cite{1982ApJ...260..395C},  1983MNRAS.204..883B|1989A\&A...208..157G: \cite{1983MNRAS.204..883B,1989A&A...208..157G},  1990JQSRT..43..207C: \cite{1990JQSRT..43..207C}, 
1992A\&A...255..457D: \cite{1992A&A...255..457D},  1993A\&AS...99..179H: \cite{1993A&AS...99..179H},  1993PhyS...48..297N: \cite{1993PhyS...48..297N},  1998PhRvA..57.1652Y: \cite{1998PhRvA..57.1652Y}, 
1999ApJS..122..557N: \cite{1999ApJS..122..557N},  2008JPCRD..37..709K: \cite{2008JPCRD..37..709K},  2009A\&A...497..611M: \cite{2009A&A...497..611M},  2009A\&A...497..611M:solar-gf: \cite{2009A&A...497..611M}, 
2014ApJS..211...20W: \cite{2014ApJS..211...20W},  2014ApJS..215...20L: \cite{2014ApJS..215...20L},  2014ApJS..215...23D: \cite{2014ApJS..215...23D},  2014MNRAS.441.3127R: \cite{2014MNRAS.441.3127R}, 
2015ApJS..220...13L: \cite{2015ApJS..220...13L},  2015ApJS..220...13L\_1982ApJ...260..395C: \cite{2015ApJS..220...13L,1982ApJ...260..395C},  2017MNRAS.471..532P: \cite{2017MNRAS.471..532P}, 
2017PhRvA..95e2507T: \cite{2017PhRvA..95e2507T},  BGHL: \cite{BGHL},  BIPS: \cite{BIPS},  BK: \cite{BK},  BK+BWL: \cite{BK,BWL},  BK+GESB82d+BWL: \cite{BK,GESB82d,BWL},  BKK: \cite{BKK}, 
BKK+GESB82c+BWL: \cite{BKK,GESB82c,BWL},  BLNP: \cite{BLNP},  BWL: \cite{BWL},  BWL+2014MNRAS.441.3127R: \cite{BWL,2014MNRAS.441.3127R},  BWL+GESHRL14: \cite{BWL,GESHRL14},  CB: \cite{CB},  DLSSC: \cite{DLSSC}, 
FMW: \cite{FMW},  GARZ|BL: \cite{GARZ,BL},  GESB82c+BWL: \cite{GESB82c,BWL},  GESB86: \cite{GESB86},  GESB86+BWL: \cite{GESB86,BWL},  GESMCHF: \cite{GESMCHF},  Grevesse2015: \cite{Grevesse2015},  HLSC: \cite{HLSC}, 
K06: \cite{K06},  K07: \cite{K07},  K08: \cite{K08},  K09: \cite{K09},  K10: \cite{K10},  K13: \cite{K13},  K14: \cite{K14},  KL-astro: astrophysical,  KR|1989ZPhyD..11..287C: \cite{KR,1989ZPhyD..11..287C}, 
LBS: \cite{LBS},  LD: \cite{LD},  LD-HS: \cite{LD-HS},  LGWSC: \cite{LGWSC},  LSCI: \cite{LSCI},  LWHS: \cite{LWHS},  MA-astro: astrophysical,  MC: \cite{MC},  MFW: \cite{MFW},  MRW: \cite{MRW},  NIST: \cite{NIST10}, 
NWL: \cite{NWL},  PQWB: \cite{PQWB},  RU: \cite{RU},  S: \cite{S},  SLS: \cite{SLS},  SR: \cite{SR},  VGH: \cite{VGH},  WLSC: \cite{WLSC},  WSL: \cite{WSL}.
}
\end{table*}

\begin{table*}
\caption{Reference values for Sun from GALAH DR3 (this work), \citet{Asplund2009}, and APOGEE DR16 VESTA \citep{SDSSDR16}. [M/H] is the pseudo-iron abundance sme.feh for GALAH DR3 and \textsc{m\_h} from APOGEE DR16. For APOGEE DR16 we use the a quadratic sum of $v_\text{macro}$ and $v \sin i$ as $v_\text{broad}$ value. We use values from the \hyperref{https://data.sdss.org/sas/dr16/apogee/spectro/aspcap/r12/l33/allStar-r12-l33.html}{SDSS website}, computed via [X/Fe] = [X/M] - [Fe/M] for the Vesta abundances of O, Na, V, and Ce.}\label{tab:solar_reference_values2}
\centering
\begin{tabular}{cc|cc|cccc}
\hline
Element &  $\lambda$ & $\mathrm{A(X_\odot)}$ & $\mathrm{A(X_\odot)}$ & [X/Fe] & [X/Fe] & [X/Fe] & [X/Fe] \\
        &   & GALAH DR3 & \citet{Asplund2009} & GALAH DR3 & GALAH DR3 & APOGEE DR16  & APOGEE DR16 \\
        & [\AA] & Zero point & Photosphere & Skyflat & Solar Circle & VESTA  & Overlap \\
\hline
Fe & combined & 7.38 & $7.50 \pm 0.04$ & $0.00 \pm 0.04$ & $-0.00\pm0.06$ & $-0.00 \pm 0.01$ & $-0.05 \pm 0.14$ \\
alpha & combined & - & - & $-0.00 \pm 0.02$ & $0.01\pm0.05$ & $-0.01 \pm 0.01$ & $0.02 \pm 0.07$ \\
Li & 6708 & 1.05 & $1.05 \pm 0.10$ & - & $1.16\pm0.49$ & - & - \\
C & 6588 & 8.45 & $8.43 \pm 0.05$ & - & $0.02\pm0.10$ & $0.02 \pm 0.02$ & $-0.02 \pm 0.12$ \\
O & combined & 8.77 & $8.69 \pm 0.05$ & $-0.12 \pm 0.05$ & $0.05\pm0.14$ & $0.05 \pm -$ & $0.14 \pm 0.22$ \\
Na & combined & 6.06 & $6.24 \pm 0.04$ & $0.01 \pm 0.02$ & $-0.00\pm0.10$ & $-0.01 \pm -$ & $0.09 \pm 0.20$ \\
Mg & 5711 & 7.60 & $7.60 \pm 0.04$ & $0.00 \pm 0.03$ & $0.01\pm0.09$ & $-0.00 \pm 0.01$ & $0.00 \pm 0.10$ \\
Al & combined & 6.41 & $6.45 \pm 0.03$ & $-0.00 \pm 0.02$ & $0.03\pm0.10$ & $0.01 \pm 0.02$ & $0.10 \pm 0.15$ \\
Si & combined & 7.47 & $7.51 \pm 0.03$ & $-0.02 \pm 0.03$ & $0.00\pm0.06$ & $-0.00 \pm 0.01$ & $0.03 \pm 0.11$ \\
K & 7699 & 5.07 & $5.03 \pm 0.09$ & $-0.09 \pm 0.04$ & $0.02\pm0.15$ & $-0.07 \pm 0.03$ & $0.03 \pm 0.23$ \\
Ca & combined & 6.18 & $6.34 \pm 0.04$ & $0.00 \pm 0.03$ & $0.03\pm0.08$ & $-0.01 \pm 0.02$ & $0.07 \pm 0.12$ \\
Sc & combined & 3.16 & $3.15 \pm 0.04$ & $-0.00 \pm 0.02$ & $0.02\pm0.08$ & - & - \\
\hline
\end{tabular}
\end{table*}
\begin{table*}
\caption{Continuation of Table~\ref{tab:solar_reference_values2}}
\centering
\begin{tabular}{cc|cc|cccc}
\hline
Element &  $\lambda$ & $\mathrm{A(X_\odot)}$ & $\mathrm{A(X_\odot)}$ & [X/Fe] & [X/Fe] & [X/Fe] & [X/Fe] \\
        &   & GALAH DR3 & \citet{Asplund2009} & GALAH DR3 & GALAH DR3 & APOGEE DR16  & APOGEE DR16 \\
        & [\AA] & Zero point & Photosphere & Skyflat & Solar Circle & VESTA  & Overlap \\
\hline
Ti & combined & - & $4.95 \pm 0.05$ & $-0.01 \pm 0.03$ & $0.02\pm0.07$ & $-0.02 \pm 0.05$ & $0.03 \pm 0.13$ \\
Ti & 4758 & 4.70 & $4.95 \pm 0.05$ & $-0.01 \pm 0.03$ & $0.02\pm0.07$ & - & - \\
Ti & 4759 & 4.72 & $4.95 \pm 0.05$ & $-0.01 \pm 0.03$ & $0.02\pm0.07$ & - & - \\
Ti & 4782 & 5.04 & $4.95 \pm 0.05$ & $-0.01 \pm 0.03$ & $0.02\pm0.07$ & - & - \\
Ti & 4802 & 5.05 & $4.95 \pm 0.05$ & $-0.01 \pm 0.03$ & $0.02\pm0.07$ & - & - \\
Ti & 4820 & 4.80 & $4.95 \pm 0.05$ & $-0.01 \pm 0.03$ & $0.02\pm0.07$ & - & - \\
Ti & 5739 & 4.82 & $4.95 \pm 0.05$ & $-0.01 \pm 0.03$ & $0.02\pm0.07$ & - & - \\
Ti2 & combined & - & $4.95 \pm 0.05$ & $-0.00 \pm 0.03$ & $-0.01\pm0.08$ & $0.11 \pm 0.09$ & $-0.01 \pm 0.22$ \\
Ti2 & 4720 & 5.12 & $4.95 \pm 0.05$ & $-0.00 \pm 0.03$ & $-0.01\pm0.08$ & - & - \\
Ti2 & 4765 & 4.85 & $4.95 \pm 0.05$ & $-0.00 \pm 0.03$ & $-0.01\pm0.08$ & - & - \\
Ti2 & 4799 & 4.85 & $4.95 \pm 0.05$ & $-0.00 \pm 0.03$ & $-0.01\pm0.08$ & - & - \\
Ti2 & 4866 & 5.12 & $4.95 \pm 0.05$ & $-0.00 \pm 0.03$ & $-0.01\pm0.08$ & - & - \\
V & combined & - & $3.93 \pm 0.08$ & $-0.01 \pm 0.02$ & $0.01\pm0.19$ & $-0.02 \pm -$ & $0.26 \pm 0.36$ \\
V & 4797 & 3.99 & $3.93 \pm 0.08$ & $-0.01 \pm 0.02$ & $0.01\pm0.19$ & - & - \\
V & 4832 & 3.99 & $3.93 \pm 0.08$ & $-0.01 \pm 0.02$ & $0.01\pm0.19$ & - & - \\
Cr & combined & 5.63 & $5.64 \pm 0.04$ & $-0.00 \pm 0.03$ & $-0.05\pm0.06$ & $0.04 \pm 0.06$ & $0.02 \pm 0.16$ \\
Mn & combined & 5.33 & $5.43 \pm 0.04$ & $0.00 \pm 0.03$ & $-0.01\pm0.06$ & $0.05 \pm 0.02$ & $-0.01 \pm 0.09$ \\
Co & combined & - & $4.99 \pm 0.07$ & - & $0.09\pm0.27$ & $0.29 \pm 0.14$ & $0.03 \pm 0.28$ \\
Co & 5647 & 5.00 & $4.99 \pm 0.07$ & - & $0.09\pm0.27$ & - & - \\
Co & 6490 & 4.85 & $4.99 \pm 0.07$ & - & $0.09\pm0.27$ & - & - \\
Co & 6632 & 4.93 & $4.99 \pm 0.07$ & - & $0.09\pm0.27$ & - & - \\
Co & 7713 & 5.06 & $4.99 \pm 0.07$ & - & $0.09\pm0.27$ & - & - \\
Ni & combined & - & $6.22 \pm 0.04$ & $0.01 \pm 0.02$ & $-0.05\pm0.08$ & $0.02 \pm 0.02$ & $0.01 \pm 0.10$ \\
Ni & 5847 & 6.23 & $6.22 \pm 0.04$ & $0.01 \pm 0.02$ & $-0.05\pm0.08$ & - & - \\
Ni & 6586 & 6.23 & $6.22 \pm 0.04$ & $0.01 \pm 0.02$ & $-0.05\pm0.08$ & - & - \\
Cu & combined & - & $4.19 \pm 0.04$ & $0.01 \pm 0.02$ & $-0.01\pm0.10$ & $-0.05 \pm 0.07$ & $0.02 \pm 0.25$ \\
Cu & 5700 & 3.74 & $4.19 \pm 0.04$ & $0.01 \pm 0.02$ & $-0.01\pm0.10$ & - & - \\
Cu & 5782 & 4.06 & $4.19 \pm 0.04$ & $0.01 \pm 0.02$ & $-0.01\pm0.10$ & - & - \\
Zn & combined & - & $4.56 \pm 0.05$ & $-0.03 \pm 0.03$ & $-0.03\pm0.10$ & - & - \\
Zn & 4722 & 4.49 & $4.56 \pm 0.05$ & $-0.03 \pm 0.03$ & $-0.03\pm0.10$ & - & - \\
Zn & 4811 & 4.46 & $4.56 \pm 0.05$ & $-0.03 \pm 0.03$ & $-0.03\pm0.10$ & - & - \\
Rb & 7800 & 2.60 & $2.52 \pm 0.10$ & - & $-0.08\pm0.28$ & - & - \\
Sr & 6550 & 3.30 & $2.87 \pm 0.07$ & - & $0.50\pm0.37$ & - & - \\
Y & combined & 2.14 & $2.21 \pm 0.05$ & $-0.23 \pm 0.05$ & $-0.02\pm0.18$ & - & - \\
Y & 4855 & 2.13 & $2.21 \pm 0.05$ & $-0.23 \pm 0.05$ & $-0.02\pm0.18$ & - & - \\
Y & 4884 & 2.09 & $2.21 \pm 0.05$ & $-0.23 \pm 0.05$ & $-0.02\pm0.18$ & - & - \\
Zr & combined & - & $2.58 \pm 0.04$ & - & $0.14\pm0.30$ & - & - \\
Zr & 4739 & 2.31 & $2.58 \pm 0.04$ & - & $0.14\pm0.30$ & - & - \\
Zr & 4772 & 2.48 & $2.58 \pm 0.04$ & - & $0.14\pm0.30$ & - & - \\
Zr & 4806 & 2.43 & $2.58 \pm 0.04$ & - & $0.14\pm0.30$ & - & - \\
Zr & 4828 & 2.66 & $2.58 \pm 0.04$ & - & $0.14\pm0.30$ & - & - \\
Zr & 5681 & 3.05 & $2.58 \pm 0.04$ & - & $0.14\pm0.30$ & - & - \\
Mo & combined & - & $1.88 \pm 0.08$ & - & $0.82\pm0.42$ & - & - \\
Mo & 5858 & 2.65 & $1.88 \pm 0.08$ & - & $0.82\pm0.42$ & - & - \\
Mo & 6619 & 1.92 & $1.88 \pm 0.08$ & - & $0.82\pm0.42$ & - & - \\
Ru & combined & - & $1.75 \pm 0.08$ & - & $1.09\pm0.49$ & - & - \\
Ru & 4739 & 2.31 & $1.75 \pm 0.08$ & - & $1.09\pm0.49$ & - & - \\
Ru & 4739 & 2.31 & $1.75 \pm 0.08$ & - & $1.09\pm0.49$ & - & - \\
Ba & combined & 2.17 & $2.18 \pm 0.09$ & $-0.14 \pm 0.04$ & $-0.00\pm0.16$ & - & - \\
La & combined & - & $1.10 \pm 0.04$ & - & $0.36\pm0.21$ & - & - \\
La & 4749 & 1.27 & $1.10 \pm 0.04$ & - & $0.36\pm0.21$ & - & - \\
La & 4804 & 1.23 & $1.10 \pm 0.04$ & - & $0.36\pm0.21$ & - & - \\
La & 5806 & 1.13 & $1.10 \pm 0.04$ & - & $0.36\pm0.21$ & - & - \\
Ce & 4774 & 2.14 & $1.58 \pm 0.04$ & $-0.01 \pm 0.01$ & $0.12\pm0.14$ & $-0.11 \pm -$ & $-0.05 \pm 0.34$ \\
Nd & combined & - & $1.42 \pm 0.04$ & - & $0.37\pm0.24$ & - & - \\
Nd & 4811 & 1.62 & $1.42 \pm 0.04$ & - & $0.37\pm0.24$ & - & - \\
Nd & 5812 & 1.40 & $1.42 \pm 0.04$ & - & $0.37\pm0.24$ & - & - \\
Sm & combined & - & $0.96 \pm 0.04$ & - & $0.19\pm0.25$ & - & - \\
Sm & 4720 & 1.36 & $0.96 \pm 0.04$ & - & $0.19\pm0.25$ & - & - \\
Sm & 4848 & 1.66 & $0.96 \pm 0.04$ & - & $0.19\pm0.25$ & - & - \\
Eu & 6645 & 0.57 & $0.52 \pm 0.04$ & - & $0.13\pm0.21$ & - & - \\
\hline
\end{tabular}
\end{table*}

\begin{table*}
\caption{Reference values for Sun from GALAH DR3 (this work, from skyflats), literature, and APOGEE DR16 VESTA \citep{SDSSDR16}. The literature is a combination of IAU Solar values \citep{Prsa2016}, ages from \citet{Bonanno2002}, $\mathrm{M\_{bol,\odot}}$ from \citet{Mamajek2012}, velocity estimates ($v_\text{mic}$ and $v_\text{broad}$) from \citet{Jofre2018a}, and abundances from \citet{Asplund2009}. [M/H] is the pseudo-iron abundance sme.feh for GALAH DR3 and \textsc{m\_h} from APOGEE DR16. For APOGEE DR16 we use the a quadratic sum of $v_\text{macro}$ and $v \sin i$ as $v_\text{broad}$ value. We use values from the \hyperref{https://data.sdss.org/sas/dr16/apogee/spectro/aspcap/r12/l33/allStar-r12-l33.html}{SDSS website}, computed via [X/Fe] = [X/M] - [Fe/M] for the Vesta abundances of O, Na, V, and Ce.}\label{tab:solar_reference_values1}
\centering
\begin{tabular}{cc|ccc}
\hline
Parameter & Unit & GALAH DR3 & Literature & APOGEE DR16 \\
\hline
$T_\text{eff}$ & [K] & $5779\pm69$ & $5772\pm-$ & $5712\pm115$ \\
$\log g$ & [dex] & $4.42\pm0.18$ & $4.438\pm-$ & $4.40\pm0.08$ \\
$\mathrm{[M/H]}$ & [dex] & $0.01\pm0.06$ & $0.00\pm-$ & $0.00\pm0.01$ \\
$\mathrm{[Fe/H]}$ & [dex] & $0.00\pm0.04$ & $0.00\pm-$ & $-0.00\pm0.01$ \\
Mass & [$\mathrm{M_\odot}$] & $0.97\pm-$ & $1.00\pm-$ & $-$ \\
Age & [Gyr] & $5.83\pm-$ & $4.57\pm0.11$ & $-$ \\
$\mathrm{M_{bol,\odot}}$ & $\mathrm{[mag]}$ & $-$ & $4.7554 \pm 0.0004$ & $-$ \\
$L_\text{bol}$ & $[\mathrm{L_{bol,\odot}}]$ & $1.01\pm-$ & $1.00\pm-$ & $-$ \\
$v_\text{mic}$ & [km/s] & $1.16\pm-$ & $1.74\pm-$ & $0.94\pm-$ \\
$v_\text{broad}$ & [km/s] & $6.52\pm2.06$ & $-$ & $5.85\pm-$ \\
\hline
\end{tabular}
\end{table*}

\begin{table*}
\caption{Reference values for Arcturus from GALAH DR3 (this work), Ramirez+11 \citep{Ramirez2011}, and APOGEE DR16 \citep{SDSSDR16}. [M/H] is the pseudo-iron abundance sme.feh for GALAH DR3, not reported by \citet{Ramirez2011} and \textsc{m\_h} from APOGEE DR16. For APOGEE DR16 we use the reported $v_\text{macro}$ as $v_\text{broad}$ value, because their was no $v \sin i$ fitted.} \label{tab:arcturus_reference_values}
\centering
\begin{tabular}{ccccc}
\hline
Parameter & Unit & GALAH DR3 & Ramirez+11 & APOGEE DR16 \\
\hline
$T_\text{eff}$ & [K] & $4289\pm69$ & $4286\pm30$ & $4292\pm76$ \\
$\log g$ & [dex] & $1.65\pm0.18$ & $1.66\pm0.05$ & $1.75\pm0.06$ \\
$\mathrm{[M/H]}$ & [dex] & $-0.53\pm0.06$ & $-$ & $-0.53\pm0.01$ \\
$\mathrm{[Fe/H]}$ & [dex] & $-0.55\pm0.04$ & $-0.52\pm0.04$ & $-0.55\pm0.01$ \\
Mass & [$\mathrm{M_\odot}$] & $0.96\pm-$ & $1.08\pm0.06$ & $-$ \\
Age & [Gyr] & $9.42\pm-$ & $7.1\pm_{1.2}^{1.5}$ & $-$ \\
$L_\text{bol}$ & $[\mathrm{L_{bol,\odot}}]$ & $179.87\pm-$ & $196.94\pm-$ & $-$ \\
$v_\text{mic}$ & [km/s] & $1.57\pm-$ & $1.74\pm-$ & $1.43\pm-$ \\
$v_\text{broad}$ & [km/s] & $6.20\pm2.05$ & $-$ & $4.04\pm-$ \\
$\mathrm{[\alpha/Fe]}$ & [dex] & $0.28\pm0.01$ & $-$ & $0.23\pm0.01$ \\
$\mathrm{[Li/Fe]}$ & [dex] & $-$ & $-$ & $-$ \\
$\mathrm{[C/Fe]}$ & [dex] & $-$ & $0.43\pm0.07$ & $0.18\pm0.01$ \\
$\mathrm{[O/Fe]}$ & [dex] & $0.55\pm0.05$ & $0.50\pm0.03$ & $0.24\pm0.01$ \\
$\mathrm{[Na/Fe]}$ & [dex] & $0.27\pm0.02$ & $0.11\pm0.03$ & $-0.03\pm0.05$ \\
$\mathrm{[Mg/Fe]}$ & [dex] & $0.48\pm0.03$ & $0.37\pm0.03$ & $0.25\pm0.01$ \\
$\mathrm{[Al/Fe]}$ & [dex] & $0.35\pm0.02$ & $0.34\pm0.03$ & $0.14\pm0.02$ \\
$\mathrm{[Si/Fe]}$ & [dex] & $0.36\pm0.03$ & $0.33\pm0.04$ & $0.20\pm0.01$ \\
$\mathrm{[K/Fe]}$ & [dex] & $0.03\pm0.04$ & $0.20\pm0.07$ & $0.16\pm0.04$ \\
$\mathrm{[Ca/Fe]}$ & [dex] & $0.14\pm0.03$ & $0.11\pm0.04$ & $0.10\pm0.02$ \\
$\mathrm{[Sc/Fe]}$ & [dex] & $0.14\pm0.02$ & $0.15\pm0.08$ & $-$ \\
$\mathrm{[Ti/Fe]}$ & [dex] & $0.26\pm0.02$ & $0.27\pm0.05$ & $-$ \\
$\mathrm{[Ti2/Fe]}$ & [dex] & $0.19\pm0.02$ & $0.21\pm0.04$ & $0.48\pm0.06$ \\
$\mathrm{[V/Fe]}$ & [dex] & $-$ & $0.20\pm0.05$ & $-0.07\pm0.05$ \\
$\mathrm{[Cr/Fe]}$ & [dex] & $-0.11\pm0.03$ & $-0.05\pm0.04$ & $-0.03\pm0.04$ \\
$\mathrm{[Mn/Fe]}$ & [dex] & $-0.19\pm0.03$ & $-$ & $-0.09\pm0.02$ \\
$\mathrm{[Co/Fe]}$ & [dex] & $0.09\pm0.01$ & $0.09\pm0.04$ & $0.15\pm0.04$ \\
$\mathrm{[Ni/Fe]}$ & [dex] & $0.13\pm0.02$ & $0.06\pm0.03$ & $0.10\pm0.02$ \\
$\mathrm{[Cu/Fe]}$ & [dex] & $0.19\pm0.01$ & $-$ & $0.29\pm0.04$ \\
$\mathrm{[Zn/Fe]}$ & [dex] & $0.05\pm0.03$ & $0.22\pm0.06$ & $-$ \\
$\mathrm{[Rb/Fe]}$ & [dex] & $-$ & $-$ & $-$ \\
$\mathrm{[Sr/Fe]}$ & [dex] & $-$ & $-$ & $-$ \\
$\mathrm{[Y/Fe]}$ & [dex] & $-0.40\pm0.05$ & $-$ & $-$ \\
$\mathrm{[Zr/Fe]}$ & [dex] & $-$ & $-$ & $-$ \\
$\mathrm{[Mo/Fe]}$ & [dex] & $0.03\pm0.03$ & $-$ & $-$ \\
$\mathrm{[Ru/Fe]}$ & [dex] & $-$ & $-$ & $-$ \\
$\mathrm{[Ba/Fe]}$ & [dex] & $0.04\pm0.04$ & $-$ & $-$ \\
$\mathrm{[La/Fe]}$ & [dex] & $-$ & $-$ & $-$ \\
$\mathrm{[Ce/Fe]}$ & [dex] & $-0.28\pm0.00$ & $-$ & $-0.14\pm0.05$ \\
$\mathrm{[Nd/Fe]}$ & [dex] & $-$ & $-$ & $-$ \\
$\mathrm{[Sm/Fe]}$ & [dex] & $-0.05\pm0.02$ & $-$ & $-$ \\
$\mathrm{[Eu/Fe]}$ & [dex] & $0.20\pm0.00$ & $-$ & $-$ \\
\hline
\end{tabular}
\end{table*}

% main_catalogue_schema with \label{tab:main_catalog_schema}
\begin{table*}
\caption{Table schema of version 2 the GALAH DR3 main catalog for all spectra (\texttt{GALAH\_DR3\_main\_allspec\_v2}). All columns are part of the extended main catalog (\texttt{allspec}) and only a subset of the listed columns are included in the clean version (\texttt{allstar} with only one entry per star). For table schemas of other catalogs (including version 1), we refer the reader to the FITS headers and the table schema website: \url{https://datacentral.org.au/services/schema/}.} \label{tab:main_catalog_schema}
\centering
\begin{tabular}{lccc}
\hline
Column Name & Units & Description & Data Type \\
\hline
star\_id &  & 2MASS identifier & string \\
sobject\_id &  & GALAH identifier & integer \\
dr2\_source\_id &  & Gaia DR2 source\_id & integer \\
dr3\_source\_id &  & Gaia DR3 source\_id & integer \\
survey\_name &  & Name of survey as part of GALAH+DR3 & string \\
field\_id &  & GALAH fco field & integer \\
flag\_repeat &  & Repeat observation flag, indicating if used for clean catalog & integer \\
wg4\_field &  & GALAH WG4 field & string \\
wg4\_pipeline &  & SME pipeline version free/lbol/seis & string \\
flag\_sp &  & Stellar parameter quality flag & integer \\
teff & K & Spectroscopic effective temperature (used for fitting) & float \\
e\_teff & K & Uncertainty teff & float \\
irfm\_teff & K & IRFM temperature (not used for synthesis) & float \\
irfm\_ebv & mag & E(B-V) used for IRFM teff estimation & float \\
irfm\_ebv\_ref &  & Reference irfm\_ebv & string \\
cov\_e\_teff & K & SME covariance fitting uncertainty teff & float \\
init\_teff & K & SME initial teff & float \\
logg & log(cm.s**-2) & Surface gravity (not fitted via spectra if wg4\_pipeline not free) & float \\
e\_logg & log(cm.s**-2) & Uncertainty logg & float \\
cov\_e\_logg & log(cm.s**-2) & MonteCarlo uncertainty logg & float \\
init\_logg & log(cm.s**-2) & SME initial logg & float \\
fe\_h &  & Fe atomic abundance from Fe lines (final, 1D-NLTE) & float \\
e\_fe\_h &  & Uncertainty fe\_h & float \\
cov\_e\_fe\_h &  & SME covariance fitting uncertainty fe\_h & float \\
flag\_fe\_h &  & Quality flag fe\_h & integer \\
fe\_h\_atmo &  & sme.feh from stellar parameter run, fitted from H, Ti, Sc, Fe & float \\
e\_fe\_h\_atmo &  & Uncertainty fe\_h\_atmo & float \\
cov\_e\_fe\_h\_atmo &  & SME covariance fitting uncertainty sme.feh & float \\
init\_fe\_h\_atmo &  & SME initial sme.feh & float \\
vmic & km s-1 & Microturbulence velocity (from empirical relation) & float \\
vbroad & km s-1 & Broadening velocity (fitted sme.vsini with sme.vmac=0) & float \\
e\_vbroad & km s-1 & Uncertainty of vbroad & float \\
cov\_e\_vbroad & km s-1 & SME covariance fitting uncertainty sme.vsini & float \\
init\_vbroad & km s-1 & SME initial broadening velocity & float \\
mass & solMass & Stellar parameter fitting product of stellar mass & float \\
lbol & solLum & Stellar parameter fitting product of bolometric luminosity & float \\
age & Gyr & Stellar parameter fitting product of stellar age & float \\
chi2\_sp &  & Chi2 value of stellar parameter fitting & float \\
alpha\_fe &  & Combined, weighted alpha-process element abundance & float \\
e\_alpha\_fe &  & Uncertainty of alpha\_fe & float \\
nr\_alpha\_fe &  & Bitmask of used measurements for alpha\_fe & float \\
flag\_alpha\_fe &  & Quality flag of measurements for alpha\_fe & integer \\
flux\_A\_Fe &  & Normalised maximum absorption strength of in iron lines & float \\
chi\_A\_Fe &  & Chi2 value of iron abundance fitting & float \\
ind\_X1234\_fe & dex & Individual uncalibrated measurmenet of line/combo X1234 & float \\
ind\_cov\_e\_X1234 & dex & SME covariance fitting uncertainty ind\_X1234\_fe & float \\
ind\_flag\_X1234 &  & Quality flag fit for ind\_X1234\_fe & integer \\
X\_fe & dex & Neutral/ionised X atomic abundance (final, 1D-LTE or NLTE) & float \\
e\_X\_fe & dex & Uncertainty X\_fe & float \\
nr\_X\_fe &  & Bitmask of used X ind lines & integer \\
flag\_X\_fe &  & Quality flag of X\_fe & integer \\
ra\_dr2 & deg & Right Ascension Gaia DR2 & float \\
dec\_dr2 & deg & Declination Gaia DR2 & float \\
parallax\_dr2 & mas & propagated from Gaia DR2 & float \\
parallax\_error\_dr2 & mas & propagated from Gaia DR2 & float \\
r\_est\_dr2 & pc & propagated from 2018AJ....156...58B & float \\
r\_lo\_dr2 & pc & propagated from 2018AJ....156...58B & float \\
r\_hi\_dr2 & pc & propagated from 2018AJ....156...58B & float \\
r\_len\_dr2 & pc & propagated from 2018AJ....156...58B & float \\
rv\_galah &  & Best-method radial velocity from GALAH spectra & float \\
e\_rv\_galah &  & Uncertainty of rv\_galah & float \\
\hline
\end{tabular}
\end{table*}

\begin{table*}
\caption{Continuation of Table~\ref{tab:main_catalog_schema}}
\centering
\begin{tabular}{lccc}
\hline
Column Name & Units & Description & Data Type \\
\hline
rv\_gaia\_dr2 & km s-1 & propagated from Gaia DR2 & float \\
e\_rv\_gaia\_dr2 & km s-1 & propagated from Gaia DR2 & float \\
red\_flag &  & eduction pipeline quality flag & integer \\
ebv & mag & SFD extinction value & float \\
snr\_c1\_iraf &  & Average SNR/px CCD1 & float \\
snr\_c2\_iraf &  & Average SNR/px CCD2 & float \\
snr\_c3\_iraf &  & Average SNR/px CCD3 & float \\
snr\_c4\_iraf &  & Average SNR/px CCD4 & float \\
flag\_guess &  & GUESS reduction pipeline quality flag & integer \\
rv\_guess & km s-1 & Reduction pipeline best radial velocity & float \\
e\_rv\_guess & km s-1 & Reduction pipeline uncertainty radial velocity & float \\
teff\_guess & K & Reduction pipeline best teff & float \\
logg\_guess & log(cm.s**-2) & Reduction pipeline best logg & float \\
feh\_guess &  & Reduction pipeline best fe\_h & float \\
rv\_5854 & km s-1 & Local best fit to RV when fitting A(Ba5854) & float \\
rv\_6708 & km s-1 & Local best fit to RV when fitting A(Li6708) & float \\
rv\_6722 & km s-1 & Local best fit to RV when fitting A(Si6722) & float \\
v\_jk & mag & V magnitude estimated from 2MASS J and Ks mag & float \\
j\_m & mag & propagated from 2MASS & float \\
j\_msigcom & mag & propagated from 2MASS & float \\
h\_m & mag & propagated from 2MASS & float \\
h\_msigcom & mag & propagated from 2MASS & float \\
ks\_m & mag & propagated from 2MASS & float \\
ks\_msigcom & mag & propagated from 2MASS & float \\
ph\_qual\_tmass &  & propagated from 2MASS ph\_qual & string \\
w2mpro & mag & propagated from AllWISE & float \\
w2mpro\_error & mag & propagated from AllWISE & float \\
ph\_qual\_wise &  & propagated from AllWISE ph\_qual & string \\
a\_ks & mag & Used Ks band extinction & float \\
e\_a\_ks & mag & Uncertainty of a\_ks & float \\
bc\_ks & mag & Used Bolometric Correction for Ks band & float \\
ruwe\_dr2 &  & propagated from Gaia DR2 & float \\
\hline
\end{tabular}
\end{table*}

%%%%%%%%%%%%%%%%%%%%%%%%%%%%%%%%%%%%%%%%%%%%%%%%%%

% Don't change these lines
\bsp	% typesetting comment
\label{lastpage}
\end{document}